\documentclass[12pt]{article}

  \usepackage{graphicx}
  \usepackage{epsfig}
  \usepackage{amssymb}
  \usepackage{subfigure}
    \usepackage{feynmf}%%%{feynmp}
  \usepackage{cite} 

\begin{document}
\begin{fmffile}{VERTICES}

\newcommand{\be}{\begin{equation}}
\newcommand{\ee}{\end{equation}}
\newcommand{\nn}{\nonumber}
\newcommand{\bea}{\begin{eqnarray}}
\newcommand{\eea}{\end{eqnarray}}
\newcommand{\bfig}{\begin{figure}}
\newcommand{\efig}{\end{figure}}
\newcommand{\bc}{\begin{center}}
\newcommand{\ec}{\end{center}}
\newcommand{\bd}{\begin{displaymath}}
\newcommand{\ed}{\end{displaymath}}

\begin{titlepage}
\nopagebreak
{\flushright{
        \begin{minipage}{5cm}
        Freiburg-THEP 05/08\\
        PITHA 05/11\\
        UCLA/05/TEP/24\\
        ZU-TH 15/05\\
        {\tt hep-ph/0508254}\\
        \end{minipage}        }

}
%\vspace*{-1.5cm}                        
\vskip 1.0cm
\begin{center}
\boldmath
{\Large \bf 
Decays of Scalar and Pseudoscalar Higgs Bosons\\
into Fermions: Two-loop QCD Corrections\\
to the Higgs-Quark-Antiquark Amplitude\\
}
\vskip 1.cm
{\large  W.~Bernreuther$\rm \, ^{a}$
%%\footnote{Email: 
%%{\tt breuther@physik.rwth-aachen.de}}
},
{\large  R.~Bonciani$\rm \, ^{b}$
%%\footnote{Email: 
%%{\tt Roberto.Bonciani@physik.uni-freiburg.de}}
},
{\large T.~Gehrmann$\rm \, ^{c}$
%%\footnote{Email: 
%%{\tt Thomas.Gehrmann@physik.unizh.ch}}
}, \\
{\large R.~Heinesch$\rm \, ^{a}$
%%\footnote{Email: 
%%{\tt heinesch@physik.rwth-aachen.de}}
}, 
{\large P.~Mastrolia$\rm \, ^{d}$
%%\footnote{Email: 
%%{\tt mastrolia@physics.ucla.edu}}
}, 
%%\\ 
and
{\large E.~Remiddi$\rm \, ^{e}$
%%\footnote{Email: 
%%{\tt Ettore.Remiddi@bo.infn.it}}
}
\vskip .7cm
{\it $\rm ^a$ Institut f\"ur Theoretische Physik, RWTH Aachen,
D-52056 Aachen, Germany} 
\vskip .3cm
{\it $\rm ^b$ Fakult\"at f\"ur Mathematik und Physik, Albert-Ludwigs-Universit\"at
Freiburg, \\ D-79104 Freiburg, Germany} 
\vskip .3cm
{\it $\rm ^c$ Institut f\"ur Theoretische Physik, 
Universit\"at Z\"urich, CH-8057 Z\"urich, Switzerland}
\vskip .3cm
{\it $\rm ^d$ Department of Physics and Astronomy, UCLA,
Los Angeles, CA 90095-1547} 
\vskip .3cm
{\it $\rm ^e$ Dipartimento di Fisica dell'Universit\`a di Bologna, and 
INFN, Sezione di Bologna, I-40126 Bologna, Italy } 
\end{center}
\vskip .4cm

\begin{abstract} 
%If Higgs bosons will be discovered then the next important task will be
%the experimental determination of its couplings to the other known fundamental particles,
%in particular to heavy quarks $Q=c,b,t$. Therefore predictions of
%Higgs boson decays into heavy quark-antiquark pairs should be made as precisely
%as possible, not only for the partial decay rates but also for differential distributions.

As a first step in the aim of arriving at a differential description
of neutral Higgs boson decays into heavy quarks, $h \to Q {\bar Q}X$, to
second order in the QCD coupling $\alpha_S$, 
we have computed the $hQ{\bar Q}$ amplitude at the two-loop level in QCD for
a general neutral Higgs boson which has both scalar and pseudoscalar couplings to quarks.
This amplitude is given in terms of a scalar and a pseudoscalar vertex form 
factor, for which we present closed analytic expressions   in terms of 
one-dimensional harmonic polylogarithms of maximum weight 4. The results hold for
arbitrary four-momentum squared, $q^2$, of the Higgs boson and of the heavy
quark mass, $m$. Moreover 
we derive the approximate expressions of these form factors near threshold and in the
asymptotic regime $m^2/q^2 \ll 1$.

\flushright{
        \begin{minipage}{12.3cm}
\hspace*{18.5mm} Heavy quark form factors.
        \end{minipage}        }
\end{abstract}
\vfill
\end{titlepage}

\section{Introduction \label{Intro}}
One of the key issues of present-day particle physics is
the search for Higgs bosons, that is, the search for  an answer to the  question whether or not
the Higgs particle of the standard model (SM), or
several types of Higgs bosons as predicted by SM extensions, do exist. 
If Higgs particles will be found at the Tevatron and/or at the LHC, then the
next important goal will be the determination  of its, respectively their,
 couplings to fermions and gauge  bosons.
This task may be achieved to some extent at the LHC, but it is expected that
it can be  carried out in detail only at a high-energy, high-luminosity
linear $e^+ e^-$ collider. 
There is a vast literature on the phenomenology of (non)standard Higgs boson(s)
(cf. \cite{Djouadi:2005gi,Djouadi:2005gj} for a recent review), and on the interplay 
of hadron and lepton colliders, in particular as far as Higgs physics is
concerned \cite{Weiglein:2004hn}. 
\par
In this paper we are concerned with the
couplings of neutral Higgs bosons to quarks (and leptons).
Apart from the possibility of getting information
about the top Yukawa coupling in 
associated production of $t{\bar t} h$, the Yukawa couplings 
of neutral Higgs bosons must be determined in their decays
to $b {\bar b}, \, c{\bar c},$ and $\tau^+\tau^-$ (indirect information on the 
Yukawa couplings can also be obtained from the branching fractions of the
channels $h \to \gamma \gamma, \, g g$). 
If heavy non-standard Higgs bosons exist, experimental investigation of
the channel $h \to t {\bar t}$ may also be feasible. The fermionic
decay rates have been calculated quite precisely, both for the SM and for some of its
extensions, notably the minimal supersymmetric extension (MSSM). As far as QCD corrections
to the  decay rates of scalar and pseudoscalar Higgs bosons into heavy 
quarks, $h \to Q {\bar Q}$, 
are concerned, the order $\alpha_S$ contributions 
were determined a long time ago 
\cite{Braaten:1980yq,Sakai:1980fa,Inami:1980qp,Drees:1990dq}, and 
the order $\alpha_S^2$ corrections were computed,
employing quark mass expansions to various orders, by 
\cite{Gorishnii:1983cu,Gorishnii:1991zr,Surguladze:1994em,Surguladze:1994gc,
Larin:1995sq,Chetyrkin:1995pd,Harlander:1997xa,Harlander:2003ai}. 
(The hadronic decay rate, i.e., with the gluon channel included, 
is known to order $\alpha_S^3$ for $m_q=0$ ($q\neq t$) and $m_h \ll m_t$, 
see \cite{Chetyrkin:1997vj, Chetyrkin:1998mw}).
Results on closed quark loop
insertions to the scalar \cite{Melnikov:1995yp, Hoang:1997ca} 
and pseudoscalar \cite{Hoang:1997ca}
vertex functions at two loops are known in the literature already.
The SM electroweak corrections to $h\to f {\bar f}$ were determined in
\cite{Fleischer:1980ub, Bardin:1990zj, Kniehl:1991ze, Dabelstein:1991ky,
  Djouadi:1997rj, Durand:1994pw, Ghinculov:1994se} 
and the radiative corrections that arise within the MSSM in
\cite{Dabelstein:1995js,Eberl:1999he,Heinemeyer:2000fa,Guasch:2003cv}.  
\par
However, not only the partial decay rates but also  differential distributions
and, in the case of tau leptons and top quarks, also the spin polarizations and correlations
of the final state 
fermion-antifermion pairs should be predicted as precisely as possible,
since these observables contain important information about the properties of the decaying boson.
For instance, neutral Higgs boson(s) may exist that are CP-violating admixtures of 
scalar and pseudoscalar states. In order to find out whether $h$ has quantum numbers
$J^{PC} = 0^{++}, \, 0^{-+}$, or has undefined CP parity one may employ suitable 
spin-spin correlations and polarization asymmetries which were proposed 
and computed, including
radiative corrections, in
 \cite{Bernreuther:1993df,Bernreuther:1997af,Bernreuther:1998qv}. 
These apply, for light
and intermediate mass Higgs bosons, to $h \to \tau^+ \tau^-$, while in the case of heavy
Higgs bosons the channel $h \to t {\bar t}$ becomes also relevant.
\par
We envisage a completely differential
description of the (on-shell  or off-shell) 
decays
%\footnote{This includes the resonant production of Higgs bosons at a
%  future muon collider, $\mu^+ \mu^- \to h \to Q {\bar Q} X$.} 
of neutral Higgs
bosons to heavy quarks, $h \to Q {\bar Q} X$, to second order in the
QCD coupling $\alpha_S$, where $Q=c,b,t$
(this includes the resonant production of Higgs bosons at a
future muon collider, $\mu^+ \mu^- \to h \to Q {\bar Q} X$). 
Such a differential description  is necessary, as already indicated above,
  for the computation of
many observables, especially if phase space cuts are to be taken into account.
To this order in the perturbative expansion,
the amplitudes for 
$h\to Q {\bar Q},$  
$h\to Q {\bar Q}g$, 
and
$h\to Q {\bar Q} \, i\,j$, where $i\, j=gg,\, q{\bar q},\,  Q {\bar Q}$ are 
required within the SM -- and, in addition, a suitable jet calculus, which 
is yet to be worked out for massive
quarks at next-to-next-to-leading order (NNLO) in the QCD coupling (for 
 the case of massless quarks  a NNLO subtraction method was recently developed
\cite{Gehrmann-DeRidder:2005cm}, cf. also references cited therein).
We consider the most general neutral Higgs boson state, i.e., a state which 
can couple both to scalar and pseudoscalar fermion currents. This
contains  scalar and pseudoscalar Higgs particles
as special cases. As a first step in this project we present in this paper
 the $ h \to Q {\bar Q}$  amplitudes to order $\alpha_S^2$ with full quark
mass dependence for arbitrary
squared four-momentum of the spin zero particle. Of course this yields
as a special case also the second order QED corrections for the
amplitudes $h \to f {\bar f}$, $f=$ quark or lepton. 
\par
The paper is organized as follows: In Section 2 we define the scalar and 
pseudoscalar vertex amplitudes for on-shell $Q, \bar Q$ states. These
amplitudes are determined by one form factor each. Then we specify the 
renormalization scheme employed in this paper. In Sections 3 and 4
we give the renormalized one- and two-loop 
scalar and pseudoscalar form factors in the spacelike
and in the timelike regions, respectively. In Section 5 and 6 we perform the
threshold and asymptotic expansions, respectively, of the form factors. 
We conclude in Section 7.

\section{The Higgs-Quark-Antiquark Amplitude \label{Hqqa}}

We consider a general neutral Higgs boson $h$ that couples both to
scalar and pseudoscalar fermion currents. This includes  scalar or pseudoscalar
Higgs particles  as special cases. The Yukawa couplings of $h$ to a massive quark $Q$ read:
\begin{equation}
{\cal L}_Y \; = \; -\frac{m_0}{v}\left[ a_Q {\bar Q} Q  \, + \, b_Q  {\bar Q}i \gamma_5 Q \right] h \, ,  
\label{yukint}
\end{equation}
where $Q$ and $h$ denote bare fields, $m_0$ is the bare mass of $Q$, 
$v = (\sqrt 2 G_F)^{-1/2}$ is the SM Higgs vacuum 
expectation value with $G_F$ being Fermi's constant, 
and $a_Q,b_Q$ are dimensionless ``reduced'' Yukawa couplings whose values depend on the 
parameters of the specific model under consideration.

%%%%%%%%%%%%%%%%%%%% one-loop Vertex %%%%%%%%%%%%%%%%%%%%%%%%%%%%%%%%%%%%%
\bfig
\bc
\subfigure[]{
\begin{fmfgraph*}(30,30)
\fmfleft{i}
\fmfright{o1,o2}
\fmfforce{0.8w,0.93h}{v2}
\fmfforce{0.8w,0.07h}{v1}
\fmfforce{0.2w,0.5h}{v5}
%\fmfforce{0.66w,0.5h}{v55}
\fmf{double}{v1,o1}
\fmf{double}{v2,o2}
\fmf{dashes}{i,v5}
\fmflabel{$p_2$}{o1}
\fmflabel{$p_1$}{o2}
\fmf{double,tension=.3}{v2,v5}
\fmf{double,tension=.3}{v1,v5}
\end{fmfgraph*}}
%
%%%%%%%%%%%%%%%%%%%%%%%
%
\hspace{20mm}
\subfigure[]{
\begin{fmfgraph*}(30,30)
\fmfleft{i}
\fmfright{o1,o2}
\fmfforce{0.8w,0.93h}{v2}
\fmfforce{0.8w,0.07h}{v1}
\fmfforce{0.2w,0.5h}{v5}
\fmf{double}{v1,o1}
\fmf{double}{v2,o2}
\fmf{dashes}{i,v5}
\fmflabel{$p_2$}{o1}
\fmflabel{$p_1$}{o2}
\fmf{double,tension=.3}{v2,v5}
\fmf{double,tension=.3}{v1,v5}
\fmf{gluon,tension=0}{v1,v2}
\end{fmfgraph*}}
%
%%%%%%%%%%%%%%%%%%%%%%%
%
\vspace*{8mm}
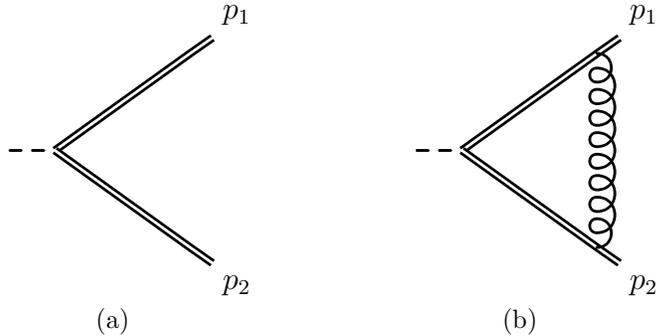
\caption{\label{fig1} Tree-level and one-loop diagrams that contribute
 to the heavy-quark scalar and pseudoscalar form factors (\ref{decomp}). 
The curly 
line represents a gluon, the double-line denotes the  heavy quark of 
mass $m$. The external quarks are on their mass-shell: 
$p_1^2 = p_2^2 = m^{2}$.}
\ec
\efig
%%%%%%%%%%%%%%%%%%%%%%%%%%%%%%%%%%%%%%%%%%%%%%%%%%%%%%%%%%%%%%%%%%%%%%%%

%%%%%%%%%%%%%%%%%%%% two-loop Vertex %%%%%%%%%%%%%%%%%%%%%%%%%%%%%%%
\bfig
\bc
\subfigure[]{
\begin{fmfgraph*}(30,30)
\fmfleft{i}
\fmfright{o1,o2}
\fmf{double}{v1,o1}
\fmf{double}{v2,o2}
\fmf{dashes}{i,v5}
\fmflabel{$p_{2}$}{o1}
\fmflabel{$p_{1}$}{o2}
\fmf{double,tension=.4}{v2,v3}
\fmf{double,tension=.3}{v3,v5}
\fmf{double,tension=.4}{v1,v4}
\fmf{double,tension=.3}{v4,v5}
\fmf{gluon,tension=0}{v1,v2}
\fmf{gluon,tension=0}{v4,v3}
\end{fmfgraph*} }
%
%%%%%%%%%%%%%%%%%%%%%%%
%
%
\hspace{8mm}
\subfigure[]{
\begin{fmfgraph*}(30,30)
\fmfleft{i}
\fmfright{o1,o2}
\fmf{double}{v1,o1}
\fmf{double}{v2,o2}
\fmf{dashes}{i,v5}
\fmflabel{$p_{2}$}{o1}
\fmflabel{$p_{1}$}{o2}
\fmf{double,tension=.3}{v2,v3}
\fmf{double,tension=.3}{v3,v5}
\fmf{double,tension=.3}{v1,v4}
\fmf{double,tension=.3}{v4,v5}
\fmf{gluon,tension=0}{v4,v2}
\fmf{gluon,tension=0}{v1,v3}
\end{fmfgraph*} }
%
%%%%%%%%%%%%%%%%%%%%%%%
%
\hspace{8mm}
\subfigure[]{
\begin{fmfgraph*}(30,30)
\fmfleft{i}
\fmfright{o1,o2}
\fmfforce{0.8w,0.93h}{v2}
\fmfforce{0.8w,0.07h}{v1}
\fmfforce{0.8w,0.5h}{v3}
\fmfforce{0.2w,0.5h}{v5}
\fmfforce{0.8w,0.3h}{v10}
\fmf{double}{v1,o1}
\fmf{double}{v2,o2}
\fmf{dashes}{i,v5}
\fmflabel{$p_{2}$}{o1}
\fmflabel{$p_{1}$}{o2}
\fmf{double,tension=0}{v2,v5}
\fmf{double,tension=0}{v3,v4}
\fmf{gluon,tension=.4}{v4,v1}
\fmf{double,tension=.4}{v4,v5}
\fmf{double,tension=0}{v1,v3}
\fmf{gluon,tension=0}{v3,v2}
\end{fmfgraph*} } \\
%
%%%%%%%%%%%%%%%%%%%%%%%
%
\subfigure[]{
\begin{fmfgraph*}(30,30)
\fmfleft{i}
\fmfright{o1,o2}
\fmfforce{0.8w,0.93h}{v2}
\fmfforce{0.8w,0.07h}{v1}
\fmfforce{0.2w,0.5h}{v5}
\fmfforce{0.8w,0.4h}{v9}
\fmfforce{0.5w,0.45h}{v10}
\fmfforce{0.8w,0.5h}{v11}
\fmf{double}{v1,o1}
\fmf{double}{v2,o2}
\fmf{dashes}{i,v5}
\fmflabel{$p_{2}$}{o1}
\fmflabel{$p_{1}$}{o2}
\fmf{double}{v2,v3}
\fmf{gluon,tension=.25,right}{v3,v4}
\fmf{double,tension=.25}{v3,v4}
\fmf{double}{v4,v5}
\fmf{double}{v1,v5}
\fmf{gluon}{v1,v2}
\end{fmfgraph*} }
%
%%%%%%%%%%%%%%%%%%%%%%%
%
\hspace{8mm}
\subfigure[]{
\begin{fmfgraph*}(30,30)
\fmfleft{i}
\fmfright{o1,o2}
\fmfforce{0.8w,0.93h}{v2}
\fmfforce{0.8w,0.07h}{v1}
\fmfforce{0.8w,0.5h}{v3}
\fmfforce{0.2w,0.5h}{v5}
\fmf{double}{v1,o1}
\fmf{double}{v2,o2}
\fmf{dashes}{i,v5}
\fmflabel{$p_{2}$}{o1}
\fmflabel{$p_{1}$}{o2}
\fmf{double,tension=0}{v1,v5}
\fmf{gluon,tension=0}{v4,v3}
\fmf{double,tension=.4}{v2,v4}
\fmf{double,tension=.4}{v4,v5}
\fmf{gluon,tension=0}{v1,v3}
\fmf{gluon,tension=0}{v3,v2}
\end{fmfgraph*}}
%
%%%%%%%%%%%%%%%%%%%%
%
\hspace{8mm}
\subfigure[]{
\begin{fmfgraph*}(30,30)
\fmfleft{i}
\fmfright{o1,o2}
\fmf{double}{v1,o1}
\fmf{double}{v2,o2}
\fmf{dashes}{i,v5}
\fmflabel{$p_{2}$}{o1}
\fmflabel{$p_{1}$}{o2}
\fmf{gluon,tension=.4}{v2,v3}
\fmf{double,tension=.3}{v3,v5}
\fmf{gluon,tension=.4}{v1,v4}
\fmf{double,tension=.3}{v4,v5}
\fmf{double,tension=0}{v1,v2}
\fmf{double,tension=0}{v4,v3}
\end{fmfgraph*} }
%
%%%%%%%%%%%%%%%%%%%%%%%
\\
\subfigure[]{
\begin{fmfgraph*}(30,30)
\fmfleft{i}
\fmfright{o1,o2}
\fmfforce{0.8w,0.93h}{v2}
\fmfforce{0.8w,0.07h}{v1}
\fmfforce{0.8w,0.3h}{v3}
\fmfforce{0.8w,0.7h}{v4}
\fmfforce{0.2w,0.5h}{v5}
\fmf{double}{v1,o1}
\fmf{double}{v2,o2}
\fmf{dashes}{i,v5}
\fmflabel{$p_{2}$}{o1}
\fmflabel{$p_{1}$}{o2}
\fmf{double}{v2,v5}
\fmf{gluon}{v1,v3}
\fmf{gluon}{v4,v2}
\fmf{double}{v1,v5}
\fmf{double,right}{v4,v3}
\fmf{double,right}{v3,v4}
\end{fmfgraph*} }
%
%%%%%%%%%%%%%%%%%%%%%%%
%
\hspace{8mm}
\subfigure[]{
\begin{fmfgraph*}(30,30)
\fmfleft{i}
\fmfright{o1,o2}
\fmfforce{0.8w,0.93h}{v2}
\fmfforce{0.8w,0.07h}{v1}
\fmfforce{0.8w,0.3h}{v3}
\fmfforce{0.8w,0.7h}{v4}
\fmfforce{0.2w,0.5h}{v5}
\fmf{double}{v1,o1}
\fmf{double}{v2,o2}
\fmf{dashes}{i,v5}
\fmflabel{$p_{2}$}{o1}
\fmflabel{$p_{1}$}{o2}
\fmf{double}{v2,v5}
\fmf{gluon}{v1,v3}
\fmf{gluon}{v4,v2}
\fmf{double}{v1,v5}
\fmf{plain,right}{v4,v3}
\fmf{plain,right}{v3,v4}
\end{fmfgraph*} } 
%
%%%%%%%%%%%%%%%%%%%%%%%
%
\hspace{8mm}
\subfigure[]{
\begin{fmfgraph*}(30,30)
\fmfleft{i}
\fmfright{o1,o2}
\fmfforce{0.8w,0.93h}{v2}
\fmfforce{0.8w,0.07h}{v1}
\fmfforce{0.8w,0.3h}{v3}
\fmfforce{0.8w,0.7h}{v4}
\fmfforce{0.2w,0.5h}{v5}
\fmf{double}{v1,o1}
\fmf{double}{v2,o2}
\fmf{dashes}{i,v5}
\fmflabel{$p_{2}$}{o1}
\fmflabel{$p_{1}$}{o2}
\fmf{double}{v2,v5}
\fmf{gluon}{v1,v3}
\fmf{gluon}{v4,v2}
\fmf{double}{v1,v5}
\fmf{dots,right}{v4,v3}
\fmf{dots,right}{v3,v4}
\end{fmfgraph*} }
%
%%%%%%%%%%%%%%%%%%%%%%%
%
\hspace{8mm}
\subfigure[]{
\begin{fmfgraph*}(30,30)
\fmfleft{i}
\fmfright{o1,o2}
\fmfforce{0.8w,0.93h}{v2}
\fmfforce{0.8w,0.07h}{v1}
\fmfforce{0.8w,0.3h}{v3}
\fmfforce{0.8w,0.7h}{v4}
\fmfforce{0.2w,0.5h}{v5}
\fmf{double}{v1,o1}
\fmf{double}{v2,o2}
\fmf{dashes}{i,v5}
\fmflabel{$p_{2}$}{o1}
\fmflabel{$p_{1}$}{o2}
\fmf{double}{v2,v5}
\fmf{gluon}{v1,v3}
\fmf{gluon}{v4,v2}
\fmf{double}{v1,v5}
\fmf{gluon,left}{v3,v4}
\fmf{gluon,left}{v4,v3}
\end{fmfgraph*} } 
%
%%%%%%%%%%%%%%%%%%%%
%
\vspace*{5mm}
\caption{\label{fig2} 
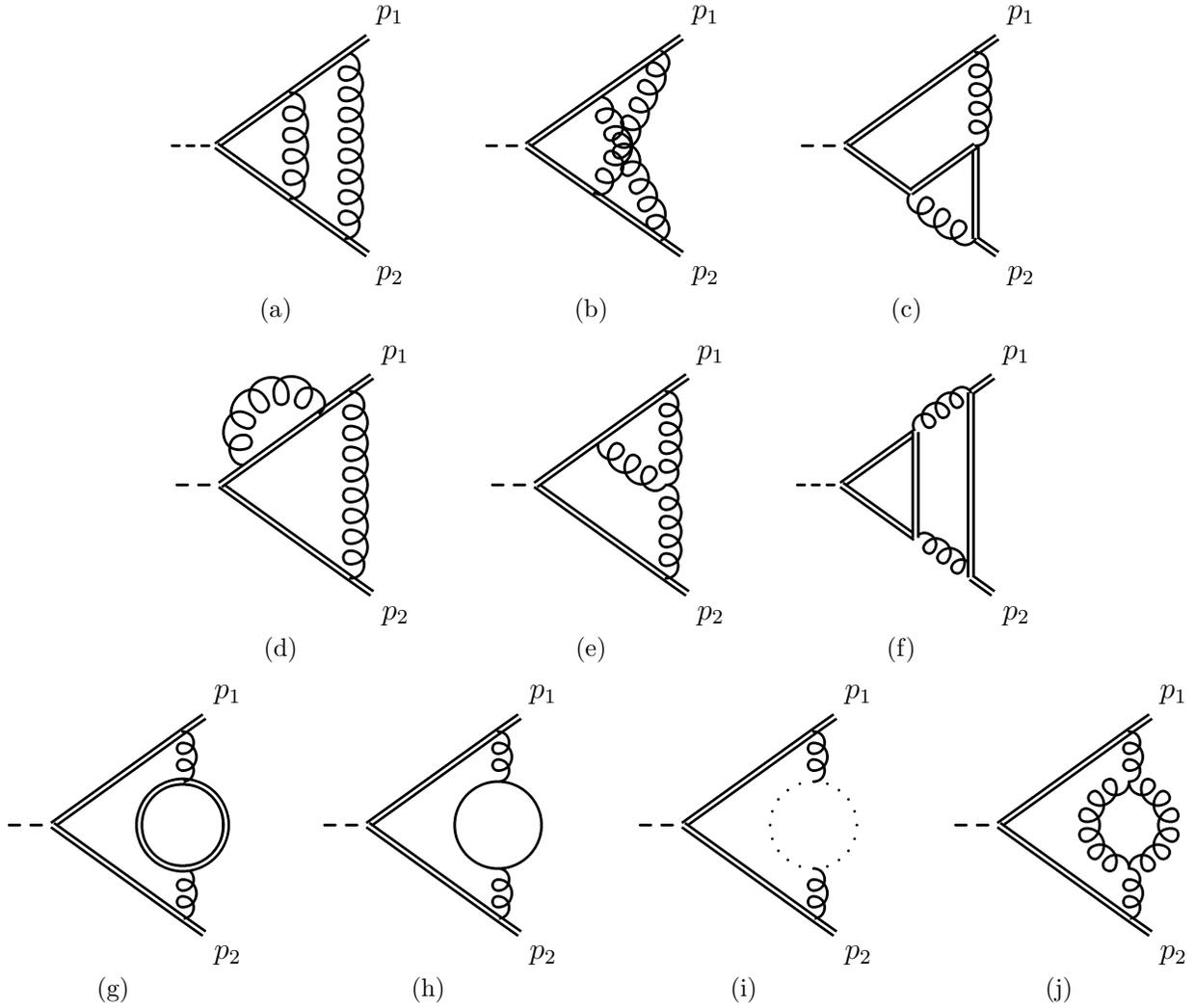
Two-loop vertex diagrams involved in the calculation of the 
form factors at order ${\mathcal O}(\alpha_{S}^{2})$. 
The notation is as in Fig. 1.  The single straight
and dotted lines  represent massless quarks and ghosts, respectively.
The external quarks  are on their mass-shell, 
$p_1^2 = p_2^2 = m^{2}$. Crossed diagrams are not drawn.
} 
\ec
\efig
%%%%%%%%%%%%%%%%%%%%%%%%%%%%%%%%%%%%%%%%%%%%%%%%%%%%%%%%%%%%%%%%%%%%%%%%

\subsection{Scalar and Pseudoscalar Form Factors \label{spffact}}

We consider the 
decay $h(q) \to Q(p_1) + {\bar Q}(p_2)$
 and aim at computing the scalar and pseudoscalar parts of
the $hQ{\bar Q}$ amplitude, for on-shell
$Q,  {\bar Q}$ and for arbitrary squared four-momentum $q^2 = s=(p_1+p_2)^2$, to second order
in the QCD coupling $\alpha_S$ and to lowest order in the Yukawa couplings. For definiteness
we work in QCD with $N_f$ massless flavors and one massive quark $Q$. The interaction
(\ref{yukint}) implies that the  $hQ{\bar Q}$ amplitude $V$ is determined by two form
factors: 
\bea
V_{c_1  c_2}(p_1,p_2) & = & i \, \bar{u}_{c_1}(p_1) \ \Gamma_{c_1 c_2}(q) \
v_{c_2}(p_2)\ , \\
& & \nn \\
\Gamma_{c_1 c_2}(q) & = & - \,  \frac{m}{v} \ 
                   \Big(  a_Q F_{S}(s) 
                         \,  + i \, b_Q F_{P}(s) \ \gamma_5 
                         \Big) \delta_{c_1 c_2} \ .
\label{decomp}
\eea
Here $F_{S}$ and  $F_{P}$ denote the renormalized scalar and pseudoscalar
form factors (whose tree-level values are equal to one), $m$ is the on-shell mass of $Q$, 
and $c_1, c_2$ are color labels.
The expansion of these form factors in powers of  
$\alpha_S/(2\pi)$, where $\alpha_S$ is  the standard
$\overline{\mathrm{MS}}$ coupling in $N_f+1$ flavor QCD  defined
at the renormalization scale $\mu$, reads
\bea
\hspace{-5mm}
F_{i} \Bigl( \epsilon,s,\frac{\mu^2}{m^2} \Bigr) & = & 
   1 + \left( \frac{\alpha_S}{2 \pi} \right) 
        F^{(1l)}_{i} \Bigl( \epsilon,s,\frac{\mu^2}{m^2} \Bigr)
     + \left( \frac{\alpha_S}{2 \pi} \right)^{2} 
        F^{(2l)}_{i} \Bigl( \epsilon,s,\frac{\mu^2}{m^2} \Bigr) \nn\\
& &  + {\mathcal O} 
     \Bigg( \! \left( \frac{\alpha_S}{2 \pi} \right) ^{3} \! \Bigg)
     , \qquad i=S,P \ .
\label{b00020} 
\eea
We regularize both the ultraviolet (UV) and infrared
divergences using dimensional regularization in
$D=4 - 2\epsilon$ dimensions. We
 use a renormalization scheme where the  wave function and the mass of $Q$
are defined in the on-shell scheme, see below. The UV-renormalized
form factors (\ref{b00020}) still depend on $\epsilon$ because 
they are infrared divergent. 
\par
The contributions to $\Gamma$ are, for the renormalization scheme employed here, 
the one-particle-irreducible 
vertex functions shown in Figs.~\ref{fig1} and \ref{fig2}, and counterterm diagrams which are
not displayed. We computed all of them in Feynman gauge.

The contribution to each form factor can be extracted from $V_{c_1  c_2}$
by means of a suitable projection. In particular one has,
\bea
 F_S &=& - \ \frac{v}{2 \,m \, a_Q \, (s-4m^2)} \ {\rm Tr} \ [ \ I_D \ (\hat p_1 + m) \
   \Gamma \ (\hat p_2 - m) \ ] \ ,\\
 F_P &=&   - \ \frac{i \, v}{2 \,m \, b_Q \, s} \ {\rm Tr} \ [\ \gamma_5 \ (\hat p_1 + m) \ \Gamma
   \ (\hat p_2 - m) \ ] \ ,
\eea
where $\hat p = p_\mu \gamma^\mu$ and $I_D$ is the unit matrix in Dirac space.

In what follows  $N_f$ denotes the
number of light quarks  running in the
loops Fig.~\ref{fig2} (h), and $C_F =(N_c^2-1)/(2N_c)$, $C_A =
N_c$, $T_R =1/2$, where  $N_c$ is the number of colors. Notice that the scalar
and pseudoscalar triangle diagrams analogous to Fig.~\ref{fig2} (f), with light quarks
circulating in the triangles, give zero contribution because a trace over an 
odd number of $\gamma$ matrices is involved if the masses
of these quarks are neglected in the triangle.

\subsection{Renormalization \label{renorm}}

Before specifying our renormalization scheme, a remark on the prescription for
the matrix $\gamma_5$ within  dimensional regularization is in order.
We use an anticommuting $\gamma_5$ in $D$ dimensions with  $\gamma_5^2=1$.
This is known to be an adequate prescription in the case of diagrams containing only
 open fermion lines, because it does not lead to spurious anomalies which must be canceled
by performing an additional renormalization of the respective current, which  in our case
is the pseudoscalar current. As to the diagrams Figs.~\ref{fig1} and
\ref{fig2}, an exception to this 
rule is the pseudoscalar triangle diagram Fig.~\ref{fig2} (f). However, this diagram
 is both UV- and IR-finite and may therefore be
evaluated in 4 dimensions.  It was
computed already in \cite{Bernreuther:2005rw}.

We use renormalized perturbation theory with $\alpha_S = g_s^2/(4\,\pi)$
being defined as the $\overline{\mathrm{MS}}$ coupling in $N_f+1$ flavor QCD, 
while we define the mass
$m$ and the wave-function of the heavy quark $Q$ in the on-shell (OS)
scheme. 

For the renormalization of the $hQ{\bar Q}$ vertex function to
two-loop order we need the coupling
renormalization $Z_g^{\overline{\mathrm{MS}}}$
and the gluon wave function $Z_3^{\overline{\mathrm{MS}}}$ to one-loop, the latter
one in the Feynman gauge. These renormalization constants are well-known and will not
be reproduced here (c.f., e.g., \cite{Bernreuther:2004ih}).

The renormalization constants of the heavy-quark mass and wave-function
are defined in the OS-scheme. We need these constants to  two-loop order; they 
were computed in \cite{Broad,Mel}.
Using the results of \cite{Mel},
we express the one- and two-loop contributions in terms of the renormalized  
$\overline{\mathrm{MS}}$ coupling $\alpha_S$. The quark wave function renormalization
constant $Z_2^{{\mathrm OS}}$ expressed in this way may be found in 
\cite{Bernreuther:2004ih}. Here we give only the mass renormalization
constant  $Z_m^{{\mathrm OS}}$ to two-loops ($m_0=Z_m^{\mathrm OS}m$): We
have $Z_m^{{\mathrm OS}} = Z_{m,{\mathrm OS}}^{(1l)} + Z_{m,{\mathrm OS}}^{(2l)}$
where

\bea
%%%%%%%%%%%%%%%%%%%%%%%%%%%%%%%%%%%%%%%%%%%%%%%%%%%%%%%%%%%%%%%%%%%%%%%%
Z_{m,{\mathrm OS}}^{(1l)} 
& = & - \ \frac{\alpha_S}{2 \pi}  \, C(\epsilon) \, 
\left( \frac{\mu^{2}}{m^2} \right)^{\epsilon} \
\frac{C_{F}}{2} \frac{(3-2 \epsilon)}{\epsilon \, (1-2 \epsilon)} \ ,  
\label{c003} \\
%%%%%%%%%%%%%%%%%%%%%%%%%%%%%%%%%%%%%%%%%%%%%%%%%%%%%%%%%%%%%%%%%%%%%%%%
&& \nn \\
Z_{m,{\mathrm OS}}^{(2l)} 
& = & 
\! \! \! \! \left( \frac{\alpha_S}{2 \pi}
\right)^2 C^2(\epsilon) \, \Biggl\{
\left( \frac{\mu^2}{m^2} \right)^{2 \epsilon} \times 
\nn\\
\! \! \! \! & & \! \! \! \!
  \Bigg[
     C_F \Bigg( 
               \frac{9}{32 \epsilon^2}
             + \frac{45}{64 \epsilon}
             + \frac{199}{128}
             - \frac{3}{4} \zeta_3
             + 3 \zeta_2 \ln(2) 
             - \frac{15}{8} \zeta_2
          \Bigg)
\nn\\
\! \! \! \! & & \! \! \! \!
   + C_A  \Bigg(
             - \frac{11}{32 \epsilon^2}
             - \frac{91}{64 \epsilon}
             - \frac{605}{128}
             + \frac{3}{8} \zeta_3
             - \frac{3}{2} \zeta_2 \ln(2) 
             + \frac{1}{2} \zeta_2
          \Bigg)
\nn\\
\! \! \! \! & & \! \! \! \!
   + T_R N_f \Bigg(
               \frac{1}{8 \epsilon^2}
             + \frac{7}{16 \epsilon}
             + \frac{45}{32}
             + \frac{1}{2} \zeta_2
            \Bigg)
\nn\\
\! \! \! \! & & \! \! \! \!
   + T_R \Bigg(
               \frac{1}{8 \epsilon^2}
             + \frac{7}{16 \epsilon}
             + \frac{69}{32}
             - \zeta_2
          \Bigg)
  \Bigg]
\nn\\
\! \! \! \! & & \! \! \! \!
   + \left( \frac{\mu^2}{m^2} \right)^{\epsilon} \Biggl[
     C_{F} C_{A} \Biggl( 
     \frac{33}{12 \epsilon^2} 
   + \frac{11}{3 \epsilon}
   + \frac{22}{3} \Biggr)  \nn\\
& & \qquad \qquad
   - C_{F} T_{R} (N_f+1) \Biggl( 
     \frac{1}{\epsilon^2} 
   + \frac{4}{3 \epsilon}
   + \frac{8}{3} \Biggr)
\Biggr] \Biggr\}
\, ,
\label{c003bis} 
\eea
%%%%%%%%%%%%%%%%%%%%%%%%%%%%%%%%%%%%%%%%%%%%%%%%%%%%%%%%%%%%%%%%%%%%%%%%
and 
\begin{equation}
C(\epsilon) = (4 \pi)^{\epsilon} \, \Gamma \left( 1 + \epsilon \right) 
\end{equation}
is an overall factor associated with each loop integral  in $D$ dimensions.
Here and in the following the Riemann zeta function of integer argument is 
denoted by $\zeta_n$.

Further, we need the renormalization constant $Z_{\mathrm{1F}}$ for the
$Q\bar{Q}$-gluon vertex to one loop. This constant is fixed by the
Slavnov-Taylor identity
$Z_{\mathrm{1F}}
=  Z_g^{\overline{\mathrm{MS}}} \; Z_2^{\mathrm{OS}} \;
\sqrt{Z_3^{\overline{\mathrm{MS}}}}$. 

Finally, we come to the overall renormalization of the $hQ{\bar Q}$ vertex function.
Because of the interaction (\ref{yukint}) the renormalization of the Yukawa couplings
is fixed by the quark mass and wave function renormalization constants. 
Moreover, as stated above, in our prescription for $\gamma_5$ no additional counterterms
are needed for the pseudoscalar vertex to order $\alpha_S^2$. Thus, the bare and
renormalized $hQ{\bar Q}$ vertex functions are related by
\begin{equation}
\Gamma \; = \; Z_m^{\mathrm{OS}} Z_2^{\mathrm{OS}}\, \Gamma_B \, .
\end{equation}

The counterterm diagram contributions to $\Gamma$ are, as far as their topological
structure is concerned, identical to those given in 
\cite{Bernreuther:2004ih}, where the vector and axial
vector parts of the  $Z^* \to Q {\bar Q}$ amplitude to order $\alpha_S^2$  were 
computed, and will therefore not be exhibited here.

\section{Renormalized Form Factors for Spacelike $s<0$ \label{FFrenorm}}

The renormalized form factors at the one- and two-loop level 
$F^{(1l)}_{i}$ and
$F^{(2l)}_{i} (i=S,P)$ are computed with the technique that was applied
in \cite{RoPieRem2, Bernreuther:2004ih, Bernreuther:2005rw}
to the computation of the two-loop photon vertex in QED and to the two-loop
vector and axial vector vertices in QCD, respectively. Performing the $D$-dimensional
$\gamma$ algebra using FORM \cite{FORM}, we obtain the form factors expressed
in terms of scalar integrals, which in the two-loop case amounts to
several hundreds. They  can
be expressed as a combination of only 35 master integrals (MIs) 
by means of the by-now standard Laporta algorithm \cite{Lap},
which makes use of integration-by-parts identities \cite{Chet}, 
Lorentz invariance \cite{Rem3} and general symmetry relations. This 
{\it reduction} algorithm is performed exactly in $D=(4-2 \epsilon)$ 
dimensions \cite{DimReg}. Once the expression in terms of the MIs is 
found, one can expand the result in powers of $\epsilon$ around $\epsilon=0$ 
($D=4$) and use the values of the MIs given in \cite{RoPieRem1,RoPieRem3},
which were evaluated with the differential equations method 
%\cite{Kot,Rem1,Rem2}.
\cite{DEmethod}.
The expressions are given  as a Laurent series in $\epsilon$ 
where both UV- and IR divergences, regularized by the same parameter $D$, show
up as poles in $\epsilon$. After adding the counterterm diagrams, we obtain
$F^{(1l)}_{i}$ and
$F^{(2l)}_{i}$, and the remaining poles in $\epsilon$ are due to IR divergences.
\par
Notice that the sum of the diagrams in
Figs.~\ref{fig1} and \ref{fig2} and of the counterterm diagrams
 has a non-trivial dependence on the renormalization scale
$\mu$: 
the counterterms with renormalization constants defined in the OS-scheme
carry an overall factor $(\mu^2/m^2)^{2 \epsilon}$; while those with 
renormalization constants defined in the
in the $\overline{\mathrm{MS}}$-scheme carry just one
factor $(\mu^2/m^2)^{\epsilon}$, coming from the loop integration.
Due to the mismatch between the $\epsilon$-dependence of
the constants in the OS-scheme and of those in the 
$\overline{\mathrm{MS}}$-scheme, the $\epsilon$-expansion of the result 
generates terms proportional 
to $\ln{(\mu^2/m^2)}$ and $\ln^2{(\mu^2/m^2)}$.
In this section we give
the one- and two-loop renormalized form factors choosing the
renormalization scale at the value of the heavy quark mass, $\mu=m$. For the terms
proportional to $\ln{(\mu^2/m^2)}$ and $\ln^2{(\mu^2/m^2)}$ we refer the
reader to Section~\ref{munotm}. 
\par
The one-loop
form  factors are presented below,
including the term of order  $\epsilon$ 
while the two-loop contributions are given to order $(\epsilon)^0$.
They are  expressed in terms of 1-dimensional harmonic
polylogarithms $\mbox{H}(\vec{a};x)$ up to weight 4 \cite{Polylog,Polylog3}, 
which are functions of the
dimensionless variable $x$ defined by
\be
x \; = \; \frac{\sqrt{-s+4m^2} - \sqrt{-s} }{\sqrt{-s+4m^2} + \sqrt{-s}} \ .
\ee
We give our results firstly in the kinematical region in which $q^2=s$ is
spacelike ($0\leq x \leq 1$), where the form factors are real. 
In Section \ref{sec_analytic} we
shall perform the analytical continuation to the physical region above
threshold, $s > 4m^2$, $-1 < x \leq 0$, and explicitly decompose 
the form factors into real
and imaginary parts.

\subsection{One-Loop Renormalized Form Factors}

The one-loop renormalized form factors are obtained by adding the contributions
of diagram of Fig.~\ref{fig1} (b) and its counterterm.

\subsubsection{The Scalar Case}

By setting $\mu=m$ and defining
\be
F_{S}^{(1l)}(s,\epsilon) = 
C(\epsilon) \ C_F 
\Bigg(
      \frac{1}{\epsilon}a_{-1} + a_0 + \epsilon \ a_1 
\Bigg)
\, ,
\label{1lrenF0}
\ee
one finds:
\bea
%%%%%%%%%%%%%%%%%%%%%%%%%%%%%%%%%%%%%%%%%%%%%%%%%%%%%
a_{-1} = &&
-1 - \frac{\left( 1 + x^2 \right) \,H(0,x)}{1 - x^2}
\ ; \\
&& \nn \\
%%%%%%%%%%%%%%%%%%%%%%%%%%%%%%%%%%%%%%%%%%%%%%%%%%%%%
a_{0} = &&
  -\left( \frac{1 - x^2 - \zeta_2 - x^2\,\zeta_2}{1 - x^2}
      \right)  
+ \frac{4\,x\,H(0,x)}{1 - x^2} 
\nn \\
&& 
+ 
   \frac{2\,\left( 1 + x^2 \right) \,H(-1,0,x)}{1 - x^2} 
- 
   \frac{\left( 1 + x^2 \right) \,H(0,0,x)}{1 - x^2}
\ ; \\
&& \nn \\
%%%%%%%%%%%%%%%%%%%%%%%%%%%%%%%%%%%%%%%%%%%%%%%%%%%%%
a_{1} = &&
  \frac{2\,\left( -1 + x^2 - 2\,x\,\zeta_2 + \zeta_3 + 
        x^2\,\zeta_3 \right) }{1 - x^2} 
\nn \\
&& 
- 
   \frac{2\,\left( 1 + x^2 \right) \,\zeta_2\,H(-1,x)}{1 - x^2} 
\nn \\
&& 
+ 
   \frac{\left( -1 + 6\,x - x^2 + \zeta_2 + x^2\,\zeta_2 \right)
        \,H(0,x)}{1 - x^2} 
\nn \\
&& 
- \frac{8\,x\,H(-1,0,x)}{1 - x^2} 
+ 
   \frac{4\,x\,H(0,0,x)}{1 - x^2} 
\nn \\
&& 
- 
   \frac{4\,\left( 1 + x^2 \right) \,H(-1,-1,0,x)}{1 - x^2} 
+ 
   \frac{2\,\left( 1 + x^2 \right) \,H(-1,0,0,x)}{1 - x^2} 
\nn \\
&& 
+ 
   \frac{2\,\left( 1 + x^2 \right) \,H(0,-1,0,x)}{1 - x^2} 
- 
   \frac{\left( 1 + x^2 \right) \,H(0,0,0,x)}{1 - x^2}
\ .
%%%%%%%%%%%%%%%%%%%%%%%%%%%%%%%%%%%%%%%%%%%%%%%%%%%%%
\label{1lrenF0coeff}
\eea

\subsubsection{The Pseudoscalar Case}

Using an analogous expansion for the pseudoscalar form factor,
\be
F_{P}^{(1l)}(s,\epsilon) = 
C(\epsilon) \ C_F 
\Bigg(
      \frac{1}{\epsilon} \bar a_{-1} + \bar a_0 + \epsilon \ \bar a_1 
\Bigg)
\, ,
\label{1lrenF5}
\ee
one finds:

\bea
%%%%%%%%%%%%%%%%%%%%%%%%%%%%%%%%%%%%%%%%%%%%%%%%%%%%%
\bar a_{-1} = && a_{-1}
\ ; \\
&& \nn \\
%%%%%%%%%%%%%%%%%%%%%%%%%%%%%%%%%%%%%%%%%%%%%%%%%%%%%
\bar a_{0} = && a_{0} - \frac{4 x \,H(0,x)}{1 - x^2} 
\ ; \\
&& \nn \\
%%%%%%%%%%%%%%%%%%%%%%%%%%%%%%%%%%%%%%%%%%%%%%%%%%%%%
\bar a_{1} = &&
  \frac{-4\,\left( 1 + x^2 \right) \,H(-1,-1,0,x)}{1 - x^2} + 
   \frac{2\,\left( 1 + x^2 \right) \,H(-1,0,0,x)}{1 - x^2} \nn \\ && + 
   \frac{2\,\left( 1 + x^2 \right) \,H(0,-1,0,x)}{1 - x^2} - 
   \frac{\left( 1 + x^2 \right) \,H(0,0,0,x)}{1 - x^2} \nn \\ && - 
   \frac{2\,\left( 1 + x^2 \right) \,H(-1,x)\,\zeta_2}{1 - x^2} \nn \\ && + 
   \frac{H(0,x)\,\left( -1 - 2\,x - x^2 + \zeta_2 + x^2\,\zeta_2 \right) }
    {1 - x^2} \nn \\ && + \frac{2\,\left( -1 + x^2 + \zeta_3 + x^2\,\zeta_3 \right) }
    {1 - x^2}
\ .
%%%%%%%%%%%%%%%%%%%%%%%%%%%%%%%%%%%%%%%%%%%%%%%%%%%%%
\label{1lrenF5coeff}
\eea

\subsection{Two-Loop Renormalized Form Factors}

The two-loop renormalized scalar and pseudoscalar form factors are obtained
by adding the contributions 
of the diagrams in Fig.~\ref{fig2} (a)--(j), of the crossed ones,
and of their counterterms. 
The pseudoscalar contribution of Fig.~\ref{fig2} (f) was
already computed in \cite{Bernreuther:2005rw} and it is added here for completeness.

\subsubsection{The Scalar Case}

By setting $\mu=m$ and defining

\bea
F_{S}^{(2l)}(\epsilon,s) = & \hspace*{-0.2cm} C^2(\epsilon) &
\hspace*{-0.3cm}
\Bigg\{
 C_F^2 \Bigg(
               \frac{b_{-2}}{\epsilon^2} 
             + \frac{b_{-1}}{\epsilon}   
             + \ b_{0}
       \Bigg)
 + C_F C_A \Bigg(
               \frac{c_{-2}}{\epsilon^2} 
             + \frac{c_{-1}}{\epsilon}   
             + \ c_{0}
       \Bigg)
\nn \\
&& 
 + C_F T_R N_f \Bigg(
               \frac{d_{-2}}{\epsilon^2} 
             + \frac{d_{-1}}{\epsilon}   
             + \ d_{0}
       \Bigg)
 + C_F T_R \Bigg(
               \frac{e_{-2}}{\epsilon^2} 
             + \frac{e_{-1}}{\epsilon}   
             + \ e_{0}
       \Bigg)
\nn \\
&& 
+ \mathcal{O}(\epsilon)
\Bigg\}
\, ,
\label{2lrenF0}
\eea

\noindent
one finds:

\bea
%%%%%%%%%%%%%%%%%%%%%%%%%%%%%%%%%%%%%%%%%%%%%%%%%%%%%%%%%%%%%%%%%%%%%%%%%%%%%%%%%%%%%%%%
%%%% C_F^2
%%%%%%%%%%%%%%%%%%%%%%%%%%%%%%%%%%%%%%%%%%%%%%%%%%%%%%%%%%%%%%%%%%%%%%%%%%%%%%%%%%%%%%%%
b_{-2} &=&  
  \frac{1}{2} + \frac{\left( 1 + x^2 \right) \,H(0;x)}{1 - x^2} 
+ 
   \frac{{\left( 1 + x^2 \right) }^2\,H(0,0;x)}{{\left( 1 - x^2 \right) }^2}
\ ; \\
&& \nn \\
%%%%%%%%%%%%%%%%%%%%%%%%%%%%%%%%%%%%%%%%%%%%%%%%%%%%%%%%%%%%%%%%%%%%%%%%%%%%%%%%%%%%%%%%
b_{-1} &=& 
  -\left( \frac{-1 + x^2 + \zeta_2 + x^2\,\zeta_2}{1 - x^2}
      \right)  
\nn \\
&&
- \frac{\left( -1 + 4\,x - 4\,x^3 + x^4 + \zeta_2 + 
        2\,x^2\,\zeta_2 + x^4\,\zeta_2 \right) \,H(0;x)}{{\left(
         1 - x^2 \right) }^2} 
\nn \\
&&
- 
   \frac{2\,\left( 1 + x^2 \right) \,H(-1,0;x)}{1 - x^2} 
- 
   \frac{\left( -1 + 8\,x + 8\,x^3 + x^4 \right) \,H(0,0;x)}
    {{\left( 1 - x^2 \right) }^2} 
\nn \\
&&
- 
   \frac{4\,{\left( 1 + x^2 \right) }^2\,H(-1,0,0;x)}
    {{\left( 1 - x^2 \right) }^2} 
- 
   \frac{2\,{\left( 1 + x^2 \right) }^2\,H(0,-1,0;x)}
    {{\left( 1 - x^2 \right) }^2} 
\nn \\
&&
+ 
   \frac{3\,{\left( 1 + x^2 \right) }^2\,H(0,0,0;x)}
    {{\left( 1 - x^2 \right) }^2}
\ ; \\
&& \nn \\
%%%%%%%%%%%%%%%%%%%%%%%%%%%%%%%%%%%%%%%%%%%%%%%%%%%%%%%%%%%%%%%%%%%%%%%%%%%%%%%%%%%%%%%%
b_{0} &=& 
  \frac{1}{20\,{\left( 1 - x \right) }^2\,{\left( 1 + x \right) }^3} 
\Big[
      145 + 145\,x - 290\,x^2 - 290\,x^3 + 145\,x^4 
\nn \\
&&
+ 145\,x^5 + 
      20\,\zeta_2 - 760\,x\,\zeta_2 + 
      1440\,\ln(2)\,x\,\zeta_2 + 520\,x^2\,\zeta_2 
\nn \\
&&
- 
      1440\,\ln(2)\,x^2\,\zeta_2 + 720\,x^3\,\zeta_2 - 
      1440\,\ln(2)\,x^3\,\zeta_2 
\nn \\
&&
- 540\,x^4\,\zeta_2 + 
      1440\,\ln(2)\,x^4\,\zeta_2 + 40\,x^5\,\zeta_2 - 
      118\,{\zeta_2}^2 
\nn \\
&&
+ 18\,x\,{\zeta_2}^2 - 
      352\,x^2\,{\zeta_2}^2 + 840\,x^3\,{\zeta_2}^2 + 
      226\,x^4\,{\zeta_2}^2 
\nn \\
&&
+ 362\,x^5\,{\zeta_2}^2 - 
      280\,\zeta_3 - 960\,x\,\zeta_3 + 
      200\,x^2\,\zeta_3 
\nn \\
&&
+ 200\,x^3\,\zeta_3 - 
      880\,x^4\,\zeta_3 - 200\,x^5\,\zeta_3
\Big]
\nn \\
&&
- 
   \frac{\left( -5 + 49\,x - 53\,x^2 + x^3 \right) \,\zeta_2\,H(-1;x)}
    {\left( 1 - x \right) \,{\left( 1 + x \right) }^2} 
\nn \\
&&
- 
   \frac{1}{{\left( 1 - x \right) }^2\,{\left( 1 + x \right) }^3} 
\Big[
\      -3 + 7\,x + 10\,x^2 - 10\,x^3 - 7\,x^4 + 3\,x^5 
\nn \\
&&
- 
        2\,\zeta_2 - 10\,x\,\zeta_2 + 47\,x^2\,\zeta_2 - 
        27\,x^3\,\zeta_2 - 45\,x^4\,\zeta_2 
\nn \\
&&
+ 
        5\,x^5\,\zeta_2 + 8\,\zeta_3 + 24\,x\,\zeta_3 - 
        56\,x^2\,\zeta_3 + 56\,x^3\,\zeta_3 
\nn \\
&&
- 
        24\,x^4\,\zeta_3 - 8\,x^5\,\zeta_3 
\Big]  \,H(0;x)
\nn \\
&&
+ 
   \frac{4\,{\left( 1 + x^2 \right) }^2\,\zeta_3\,H(1;x)}
    {{\left( 1 - x^2 \right) }^2} 
\nn \\
&&
+ 
   \frac{\left( 15 - 42\,x + 42\,x^3 - 15\,x^4 + 4\,\zeta_2 + 
        8\,x^2\,\zeta_2 + 4\,x^4\,\zeta_2 \right) \,H(-1,0;x)}
      {{\left( 1 - x^2 \right) }^2} 
\nn \\
&&
- 
   \frac{2\,\left( -7 - 7\,x + 16\,x^2 - 20\,x^3 + 5\,x^4 + 5\,x^5 \right) \,
      \zeta_2\,H(0,-1;x)}{{\left( 1 - x \right) }^2\,
      {\left( 1 + x \right) }^3} 
\nn \\
&&
+ 
   \frac{1}{{\left( 1 - x \right) }^2\,{\left( 1 + x \right) }^3} 
   \Big[ 3 - 12\,x + 36\,x^2 - 22\,x^3 - 47\,x^4 + 26\,x^5 - 
        3\,\zeta_2 
\nn \\
&&
+ x\,\zeta_2 - 12\,x^2\,\zeta_2 + 
        32\,x^3\,\zeta_2 + 9\,x^4\,\zeta_2 + 
        13\,x^5\,\zeta_2 
   \Big] \,H(0,0;x)
\nn \\
&&
+ 
   \frac{4}{{\left( 1 - x \right) }^2\,{\left( 1 + x \right) }^3} 
   \Big[ 
        -2 + 2\,x + 4\,x^2 - 4\,x^3 - 2\,x^4 + 2\,x^5 + 
        \zeta_2 
\nn \\
&&
+ 5\,x\,\zeta_2 - 18\,x^2\,\zeta_2 + 
        22\,x^3\,\zeta_2 - 3\,x^4\,\zeta_2 + x^5\,\zeta_2
    \Big] \,H(1,0;x)
\nn \\
&&
+ 
   \frac{4\,\left( 1 + x^2 \right) \,H(-1,-1,0;x)}{1 - x^2} 
\nn \\
&&
+ 
   \frac{\left( 5 + 58\,x - 2\,x^2 + 58\,x^3 + 9\,x^4 \right) \,H(-1,0,0;x)}
    {{\left( 1 - x^2 \right) }^2} 
\nn \\
&&
+ 
   \frac{2\,\left( 3 - 22\,x + 20\,x^2 - 22\,x^3 + 5\,x^4 \right) \,
      H(0,-1,0;x)}{{\left( 1 - x^2 \right) }^2} 
\nn \\
&&
- 
   \frac{\left( -1 + 31\,x + 65\,x^2 - 9\,x^3 + 10\,x^5 \right) \,H(0,0,0;x)}
    {{\left( 1 - x \right) }^2\,{\left( 1 + x \right) }^3} 
\nn \\
&&
- 
   \frac{4\,\left( 1 - 6\,x + 2\,x^2 - 6\,x^3 + x^4 \right) \,H(0,1,0;x)}
    {{\left( 1 - x^2 \right) }^2} 
\nn \\
&&
+ 
   \frac{2\,\left( 5 - 16\,x + 6\,x^2 - 16\,x^3 + 5\,x^4 \right) \,H(1,0,0;x)}
    {{\left( 1 - x^2 \right) }^2} 
\nn \\
&&
+ 
   \frac{16\,{\left( 1 + x^2 \right) }^2\,H(-1,-1,0,0;x)}
    {{\left( 1 - x^2 \right) }^2} 
+ 
   \frac{8\,{\left( 1 + x^2 \right) }^2\,H(-1,0,-1,0;x)}
    {{\left( 1 - x^2 \right) }^2} 
\nn \\
&&
- 
   \frac{12\,{\left( 1 + x^2 \right) }^2\,H(-1,0,0,0;x)}
    {{\left( 1 - x^2 \right) }^2} 
+ 
   \frac{4\,{\left( 1 + x^2 \right) }^2\,H(0,-1,-1,0;x)}
    {{\left( 1 - x^2 \right) }^2} 
\nn \\
&&
- 
   \frac{2\,\left( 3 - 5\,x + 36\,x^2 - 16\,x^3 + 15\,x^4 + 7\,x^5 \right) \,
      H(0,-1,0,0;x)}{{\left( 1 - x \right) }^2\,{\left( 1 + x \right) }^3} 
\nn \\
&&
- 
   \frac{2\,\left( -1 + 3\,x + 10\,x^2 + 10\,x^3 + 7\,x^4 + 11\,x^5 \right) \,
      H(0,0,-1,0;x)}{{\left( 1 - x \right) }^2\,{\left( 1 + x \right) }^3} 
\nn \\
&&
+ 
   \frac{\left( 7 + 7\,x + 16\,x^2 + 52\,x^3 + 27\,x^4 + 27\,x^5 \right) \,
      H(0,0,0,0;x)}{{\left( 1 - x \right) }^2\,{\left( 1 + x \right) }^3} 
\nn \\
&&
+ 
   \frac{4\,\left( -1 + 2\,x^2 + 3\,x^4 \right) \,H(0,0,1,0;x)}
    {{\left( 1 - x^2 \right) }^2} 
\nn \\
&&
- 
   \frac{4\,\left( -1 + x \right) \,H(0,1,0,0;x)}{1 + x} 
- 
   \frac{8\,{\left( 1 + x^2 \right) }^2\,H(1,0,-1,0;x)}
    {{\left( 1 - x^2 \right) }^2} 
\nn \\
&&
+ 
   \frac{8\,\left( 1 + 3\,x - 8\,x^2 + 12\,x^3 - x^4 + x^5 \right) \,
      H(1,0,0,0;x)}{{\left( 1 - x \right) }^2\,{\left( 1 + x \right) }^3} 
\nn \\
&&
+ 
   \frac{8\,{\left( 1 + x^2 \right) }^2\,H(1,0,1,0;x)}
    {{\left( 1 - x^2 \right) }^2}
\ ; 
%%%%%%%%%%%%%%%%%%%%%%%%%%%%%%%%%%%%%%%%%%%%%%%%%%%%%%%%%%%%%%%%%%%%%%%%%%%%%%%%%%%%%%%%
\eea
\bea
%%%%%%%%%%%%%%%%%%%%%%%%%%%%%%%%%%%%%%%%%%%%%%%%%%%%%%%%%%%%%%%%%%%%%%%%%%%%%%%%%%%%%%%%
%%%% C_F*C_A
%%%%%%%%%%%%%%%%%%%%%%%%%%%%%%%%%%%%%%%%%%%%%%%%%%%%%%%%%%%%%%%%%%%%%%%%%%%%%%%%%%%%%%%%
c_{-2} &=&  
  \frac{11}{12} + \frac{11\,\left( 1 + x^2 \right) \,H(0;x)}
    {12\,\left( 1 - x^2 \right) }
\ ; \\
&& \nn \\
%%%%%%%%%%%%%%%%%%%%%%%%%%%%%%%%%%%%%%%%%%%%%%%%%%%%%%%%%%%%%%%%%%%%%%%%%%%%%%%%%%%%%%%%
c_{-1} &=& 
  \frac{1}{36\,{\left( 1 - x^2 \right) }^2} 
\Big[
      -49 + 98\,x^2 - 49\,x^4 + 18\,\zeta_2 - 
      72\,x^2\,\zeta_2 
\nn \\
&&
      + 54\,x^4\,\zeta_2 - 
      18\,\zeta_3 - 36\,x^2\,\zeta_3 - 18\,x^4\,\zeta_3
\Big]
\nn \\
&&
- 
   \frac{\left( 1 + x^2 \right) \,
      \left( 67 - 67\,x^2 - 18\,\zeta_2 + 54\,x^2\,\zeta_2
        \right) \,H(0;x)}{36\,{\left( 1 - x^2 \right) }^2} 
\nn \\
&&
+ 
   \frac{\left( 1 + x^2 \right) \,H(-1,0;x)}{1 - x^2} - 
   \frac{2\,x^2\,H(0,0;x)}{1 - x^2} 
\nn \\
&&
- 
   \frac{\left( 1 + x^2 \right) \,H(1,0;x)}{1 - x^2} + 
   \frac{{\left( 1 + x^2 \right) }^2\,H(0,-1,0;x)}
    {{\left( 1 - x^2 \right) }^2} 
\nn \\
&&
- 
   \frac{2\,\left( x^2 + x^4 \right) \,H(0,0,0;x)}
    {{\left( 1 - x^2 \right) }^2} - 
   \frac{{\left( 1 + x^2 \right) }^2\,H(0,1,0;x)}{{\left( 1 - x^2 \right) }^2}
\ ; \\
&& \nn \\
%%%%%%%%%%%%%%%%%%%%%%%%%%%%%%%%%%%%%%%%%%%%%%%%%%%%%%%%%%%%%%%%%%%%%%%%%%%%%%%%%%%%%%%%
c_{0} &=& 
  \frac{1}{540\,{\left( 1 - x \right) }^2\,{\left( 1 + x \right) }^3} 
  \Big[
      -4345 - 4345\,x + 8690\,x^2 + 8690\,x^3 - 4345\,x^4 
\nn \\
&&
- 4345\,x^5 + 
      2730\,\zeta_2 + 16860\,x\,\zeta_2 - 
      19440\,\ln(2)\,x\,\zeta_2 
\nn \\
&&
- 
      18900\,x^2\,\zeta_2 + 
      19440\,\ln(2)\,x^2\,\zeta_2 - 
      15840\,x^3\,\zeta_2 
\nn \\
&&
+ 
      19440\,\ln(2)\,x^3\,\zeta_2 + 
      16170\,x^4\,\zeta_2 - 
      19440\,\ln(2)\,x^4\,\zeta_2 
\nn \\
&&
- 
      1020\,x^5\,\zeta_2 - 81\,{\zeta_2}^2 - 
      81\,x\,{\zeta_2}^2 - 3132\,x^2\,{\zeta_2}^2 - 
      864\,x^3\,{\zeta_2}^2 
\nn \\
&&
- 1917\,x^4\,{\zeta_2}^2 - 
      1917\,x^5\,{\zeta_2}^2 + 4410\,\zeta_3 - 
      1530\,x\,\zeta_3 
\nn \\
&&
+ 2700\,x^2\,\zeta_3 + 
      2700\,x^3\,\zeta_3 - 7110\,x^4\,\zeta_3 - 
      1170\,x^5\,\zeta_3
   \Big]
\nn \\
&&
+ 
   \frac{1}{6\,{\left( 1 - x^2 \right) }^2}
   \Big[ -25\,\zeta_2 + 162\,x\,\zeta_2 - 
        306\,x^2\,\zeta_2 + 162\,x^3\,\zeta_2 
\nn \\
&&
+ 
        7\,x^4\,\zeta_2 + 12\,\zeta_3 + 
        24\,x^2\,\zeta_3 + 12\,x^4\,\zeta_3 
   \Big] \,H(-1;x) 
\nn \\
&&
+ 
   \frac{1}{54\,{\left( 1 - x \right) }^2\,{\left( 1 + x \right) }^3} 
   \Big[ 
        -121 + 1133\,x + 1254\,x^2 - 1254\,x^3 - 1133\,x^4 
\nn \\
&&
+ 
        121\,x^5 - 99\,\zeta_2 + 117\,x\,\zeta_2 + 
        999\,x^2\,\zeta_2 + 729\,x^3\,\zeta_2 
\nn \\
&&
- 
        1764\,x^4\,\zeta_2 + 18\,x^5\,\zeta_2 + 
        351\,\zeta_3 + 351\,x\,\zeta_3 + 
        594\,x^2\,\zeta_3 
\nn \\
&&
- 702\,x^3\,\zeta_3 - 
        405\,x^4\,\zeta_3 - 405\,x^5\,\zeta_3 
   \Big] \,H(0;x)
\nn \\
&&
- 
   \frac{\left( 1 + x^2 \right) \,
      \left( -\zeta_2 + x^2\,\zeta_2 + 2\,\zeta_3 + 
        2\,x^2\,\zeta_3 \right) \,H(1;x)}{{\left( 1 - x^2 \right) }^2}
\nn \\
&&
    - \frac{\left( -53 + 102\,x - 102\,x^3 + 53\,x^4 - 72\,\zeta_2 - 
        72\,x^2\,\zeta_2 \right) \,H(-1,0;x)}{18\,
      {\left( 1 - x^2 \right) }^2} 
\nn \\
&&
+ 
   \frac{\left( -5 - 5\,x + 20\,x^2 - 16\,x^3 + 7\,x^4 + 7\,x^5 \right) \,
      \zeta_2\,H(0,-1;x)}{{\left( 1 - x \right) }^2\,
      {\left( 1 + x \right) }^3} 
\nn \\
&&
+ 
   \frac{1}{18\,{\left( 1 - x \right) }^2\,{\left( 1 + x \right) }^3} 
   \Big[ -67 + 74\,x + 78\,x^2 - 24\,x^3 - 11\,x^4 - 50\,x^5 
\nn \\
&&
+ 
        18\,\zeta_2 + 18\,x\,\zeta_2 + 
        54\,x^2\,\zeta_2 + 18\,x^3\,\zeta_2 + 
        18\,x^4\,\zeta_2 
\nn \\
&&
+ 18\,x^5\,\zeta_2 
   \Big] \,H(0,0;x)
\nn \\
&&
+ 
   \frac{{\left( 1 + x^2 \right) }^2\,\zeta_2\,H(0,1;x)}
    {{\left( 1 - x^2 \right) }^2} 
\nn \\
&&
+ 
   \frac{2\,\left( 1 + x^2 \right) \,
      \left( 1 - x^2 - 4\,\zeta_2 + 2\,x^2\,\zeta_2 \right) \,
      H(1,0;x)}{{\left( 1 - x^2 \right) }^2} 
\nn \\
&&
- 
   \frac{52\,\left( 1 + x^2 \right) \,H(-1,-1,0;x)}
    {3\,\left( 1 - x^2 \right) } 
\nn \\
&&
+ 
   \frac{\left( 61 + 175\,x - 35\,x^2 + 79\,x^3 \right) \,H(-1,0,0;x)}
    {6\,\left( 1 - x \right) \,{\left( 1 + x \right) }^2} 
\nn \\
&&
+ 
   \frac{6\,\left( 1 + x^2 \right) \,H(-1,1,0;x)}{1 - x^2} 
\nn \\
&&
+ 
   \frac{2\,\left( 1 + 10\,x + 25\,x^2 + 34\,x^3 \right) \,H(0,-1,0;x)}
    {3\,\left( 1 - x \right) \,{\left( 1 + x \right) }^2} 
\nn \\
&&
- 
   \frac{\left( 11 - 2\,x - 53\,x^2 - 74\,x^3 + 62\,x^4 \right) \,H(0,0,0;x)}
    {6\,\left( 1 - x \right) \,{\left( 1 + x \right) }^3} 
\nn \\
&&
- 
   \frac{4\,x\,\left( 1 + x + 2\,x^2 \right) \,H(0,1,0;x)}
    {\left( 1 - x \right) \,{\left( 1 + x \right) }^2} 
\nn \\
&&
+ 
   \frac{6\,\left( 1 + x^2 \right) \,H(1,-1,0;x)}{1 - x^2} 
+ 
   \frac{4\,\left( -3 + x^2 \right) \,H(1,0,0;x)}{1 - x^2} 
\nn \\
&&
- 
   \frac{2\,\left( 1 + x^2 \right) \,H(1,1,0;x)}{1 - x^2} 
- 
   \frac{4\,{\left( 1 + x^2 \right) }^2\,H(-1,0,-1,0;x)}
    {{\left( 1 - x^2 \right) }^2} 
\nn \\
&&
+ 
   \frac{2\,\left( 3 + 4\,x^2 + x^4 \right) \,H(-1,0,0,0;x)}
    {{\left( 1 - x^2 \right) }^2} 
%\nn \\
%&&
+ 
   \frac{4\,{\left( 1 + x^2 \right) }^2\,H(-1,0,1,0;x)}
    {{\left( 1 - x^2 \right) }^2} 
\nn \\
&&
- 
   \frac{10\,{\left( 1 + x^2 \right) }^2\,H(0,-1,-1,0;x)}
    {{\left( 1 - x^2 \right) }^2} 
\nn \\
&&
+ 
   \frac{2\,\left( 1 + x + x^2 + 15\,x^3 + 7\,x^4 + 7\,x^5 \right) \,
      H(0,-1,0,0;x)}{{\left( 1 - x \right) }^2\,{\left( 1 + x \right) }^3} 
\nn \\
&&
+ 
   \frac{6\,{\left( 1 + x^2 \right) }^2\,H(0,-1,1,0;x)}
    {{\left( 1 - x^2 \right) }^2} 
\nn \\
&&
+ 
   \frac{2\,\left( 1 + x + 22\,x^2 + 2\,x^3 + 11\,x^4 + 11\,x^5 \right) \,
      H(0,0,-1,0;x)}{{\left( 1 - x \right) }^2\,{\left( 1 + x \right) }^3} 
\nn \\
&&
- 
   \frac{3\,x^2\,\left( 1 + 7\,x + 4\,x^2 + 4\,x^3 \right) \,H(0,0,0,0;x)}
    {{\left( 1 - x \right) }^2\,{\left( 1 + x \right) }^3} 
\nn \\
&&
- 
   \frac{2\,\left( 1 + 8\,x^2 + 7\,x^4 \right) \,H(0,0,1,0;x)}
    {{\left( 1 - x^2 \right) }^2} 
+ 
   \frac{6\,{\left( 1 + x^2 \right) }^2\,H(0,1,-1,0;x)}
    {{\left( 1 - x^2 \right) }^2} 
\nn \\
&&
- 
   \frac{2\,\left( 3 + 3\,x - 8\,x^2 + 16\,x^3 + x^4 + x^5 \right) \,
      H(0,1,0,0;x)}{{\left( 1 - x \right) }^2\,{\left( 1 + x \right) }^3} 
\nn \\
&&
- 
   \frac{2\,{\left( 1 + x^2 \right) }^2\,H(0,1,1,0;x)}
    {{\left( 1 - x^2 \right) }^2} 
+ 
   \frac{4\,{\left( 1 + x^2 \right) }^2\,H(1,0,-1,0;x)}
    {{\left( 1 - x^2 \right) }^2} 
\nn \\
&&
+ 
   \frac{2\,\left( -5 - 4\,x^2 + x^4 \right) \,H(1,0,0,0;x)}
    {{\left( 1 - x^2 \right) }^2} 
- 
   \frac{4\,{\left( 1 + x^2 \right) }^2\,H(1,0,1,0;x)}
    {{\left( 1 - x^2 \right) }^2}
\ ; 
%%%%%%%%%%%%%%%%%%%%%%%%%%%%%%%%%%%%%%%%%%%%%%%%%%%%%%%%%%%%%%%%%%%%%%%%%%%%%%%%%%%%%%%%
\eea
\bea
%%%%%%%%%%%%%%%%%%%%%%%%%%%%%%%%%%%%%%%%%%%%%%%%%%%%%%%%%%%%%%%%%%%%%%%%%%%%%%%%%%%%%%%%
%%%% C_F*T_R*N_f
%%%%%%%%%%%%%%%%%%%%%%%%%%%%%%%%%%%%%%%%%%%%%%%%%%%%%%%%%%%%%%%%%%%%%%%%%%%%%%%%%%%%%%%%
d_{-2} &=&  
  - \frac{1}{3}  
  - \frac{\left( 1 + x^2 \right) \,H(0;x)}{3\,\left( 1 - x^2 \right) }
\ ; \\
%%%%%%%%%%%%%%%%%%%%%%%%%%%%%%%%%%%%%%%%%%%%%%%%%%%%%%%%%%%%%%%%%%%%%%%%%%%%%%%%%%%%%%%%
d_{-1} &=& 
  \frac{5}{9} + \frac{5\,\left( 1 + x^2 \right) \,H(0;x)}
    {9\,\left( 1 - x^2 \right) }
\ ; \\
%%%%%%%%%%%%%%%%%%%%%%%%%%%%%%%%%%%%%%%%%%%%%%%%%%%%%%%%%%%%%%%%%%%%%%%%%%%%%%%%%%%%%%%%
d_{0} &=& 
  - \frac{ -49 + 49\,x^2 - 6\,\zeta_2 - 72\,x\,\zeta_2 + 
        66\,x^2\,\zeta_2 + 36\,\zeta_3 + 36\,x^2\,\zeta_3
         }{27\,\left( 1 - x^2 \right) } 
\nn \\
&&
+ 
   \frac{4\,\left( 1 + x^2 \right) \,\zeta_2\,H(-1;x)}
    {3\,\left( 1 - x^2 \right) } 
\nn \\
&&
+ 
   \frac{2\,\left( 14 - 96\,x + 14\,x^2 + 9\,\zeta_2 + 
        9\,x^2\,\zeta_2 \right) \,H(0;x)}{27\,\left( 1 - x^2 \right) }
\nn \\
&&
    - \frac{4\,\left( 5 - 12\,x + 5\,x^2 \right) \,H(-1,0;x)}
    {9\,\left( 1 - x^2 \right) } 
+ 
   \frac{2\,\left( 5 - 12\,x + 5\,x^2 \right) \,H(0,0;x)}
    {9\,\left( 1 - x^2 \right) } 
\nn \\
&&
+ 
   \frac{8\,\left( 1 + x^2 \right) \,H(-1,-1,0;x)}
    {3\,\left( 1 - x^2 \right) } 
- 
   \frac{4\,\left( 1 + x^2 \right) \,H(-1,0,0;x)}
    {3\,\left( 1 - x^2 \right) } 
\nn \\
&&
- 
   \frac{4\,\left( 1 + x^2 \right) \,H(0,-1,0;x)}
    {3\,\left( 1 - x^2 \right) } 
+ 
   \frac{2\,\left( 1 + x^2 \right) \,H(0,0,0;x)}{3\,\left( 1 - x^2 \right) }
\ ; 
%%%%%%%%%%%%%%%%%%%%%%%%%%%%%%%%%%%%%%%%%%%%%%%%%%%%%%%%%%%%%%%%%%%%%%%%%%%%%%%%%%%%%%%%
\eea
\bea
%%%%%%%%%%%%%%%%%%%%%%%%%%%%%%%%%%%%%%%%%%%%%%%%%%%%%%%%%%%%%%%%%%%%%%%%%%%%%%%%%%%%%%%%
%%%% C_F*T_R
%%%%%%%%%%%%%%%%%%%%%%%%%%%%%%%%%%%%%%%%%%%%%%%%%%%%%%%%%%%%%%%%%%%%%%%%%%%%%%%%%%%%%%%%
e_{-2} &=& 0 
\ ; 
\\
%%%%%%%%%%%%%%%%%%%%%%%%%%%%%%%%%%%%%%%%%%%%%%%%%%%%%%%%%%%%%%%%%%%%%%%%%%%%%%%%%%%%%%%%
e_{-1} &=& 0 
\ ; \\
%%%%%%%%%%%%%%%%%%%%%%%%%%%%%%%%%%%%%%%%%%%%%%%%%%%%%%%%%%%%%%%%%%%%%%%%%%%%%%%%%%%%%%%%
e_{0} &=& 
 -\frac{1}{{135\,{\left( 1 - x \right) }^3\,\left( 1 + x \right) }^4} 
 \Big[ -2035 - 1435\,x + 5505\,x^2 + 4905\,x^3 
\nn \\
&&
- 4905\,x^4 - 
        5505\,x^5 + 1435\,x^6 + 2035\,x^7 + 810\,\zeta_2 + 
        270\,x\,\zeta_2 
\nn \\
&&
- 9810\,x^2\,\zeta_2 + 
        9450\,x^3\,\zeta_2 - 810\,x^4\,\zeta_2 + 
        5490\,x^5\,\zeta_2 
\nn \\
&&
- 4590\,x^6\,\zeta_2 - 
        810\,x^7\,\zeta_2 + 432\,x\,{\zeta_2}^2 + 
        2160\,x^2\,{\zeta_2}^2 
\nn \\
&&
+ 2592\,x^3\,{\zeta_2}^2 + 
        2592\,x^4\,{\zeta_2}^2 + 2160\,x^5\,{\zeta_2}^2 + 
        432\,x^6\,{\zeta_2}^2 
\nn \\
&&
+ 8640\,x^2\,\zeta_3 + 
        8640\,x^3\,\zeta_3 - 8640\,x^4\,\zeta_3 - 
        8640\,x^5\,\zeta_3 
 \Big]
\nn \\
&&
+ 
   \frac{2}{27\,{\left( 1 - x^2 \right) }^3} 
 \Big[ 28 + 144\,x + 212\,x^2 - 768\,x^3 + 212\,x^4 + 144\,x^5 + 
        28\,x^6 
\nn \\
&&
+ 9\,\zeta_2 - 9\,x^2\,\zeta_2 - 
        9\,x^4\,\zeta_2 + 9\,x^6\,\zeta_2 - 
        162\,x\,\zeta_3 - 648\,x^2\,\zeta_3 
\nn \\
&&
+ 
        756\,x^3\,\zeta_3 - 648\,x^4\,\zeta_3 - 
        162\,x^5\,\zeta_3 \Big] \,H(0;x)
\nn \\
&&
- \frac{32\,x\,H(-1,0;x)}{1 - x^2} 
\nn \\
&&
- 
   \frac{2}{9\,{\left( 1 - x \right) }^3\,{\left( 1 + x \right) }^4} 
\Big[ -5 - 23\,x + 101\,x^2 + 531\,x^3 - 27\,x^4 
\nn \\
&&
- 353\,x^5 - 
        229\,x^6 + 5\,x^7 + 18\,x\,\zeta_2 + 90\,x^2\,\zeta_2 + 
        468\,x^3\,\zeta_2 
\nn \\
&&
+ 468\,x^4\,\zeta_2 + 
        90\,x^5\,\zeta_2 + 18\,x^6\,\zeta_2 
\Big] \,H(0,0;x)
\nn \\
&&
+ 
   \frac{8\,x\,H(0,1;x)}{1 - x^2} 
\nn \\
&&
+ 
   \frac{8\,x\,\left( 2 + 4\,x + 2\,x^2 + \zeta_2 + 
        6\,x\,\zeta_2 + x^2\,\zeta_2 \right) \,H(1,0;x)}{
      \left( 1 - x \right) \,{\left( 1 + x \right) }^3} 
\nn \\
&&
+ 
   \frac{128\,x^2\,H(0,-1,0;x)}{{\left( 1 - x^2 \right) }^2} 
- 
   \frac{2\,\left( -1 + 48\,x^2 + x^4 \right) \,H(0,0,0;x)}
    {3\,{\left( 1 - x^2 \right) }^2} 
\nn \\
&&
- 
   \frac{64\,x^2\,H(0,1,0;x)}{{\left( 1 - x^2 \right) }^2} 
+ 
   \frac{128\,x^2\,H(1,0,0;x)}{{\left( 1 - x^2 \right) }^2} 
- 
   \frac{256\,x^3\,H(0,0,-1,0;x)}{{\left( 1 - x^2 \right) }^3} 
\nn \\
&&
- 
   \frac{2\,x\,\left( 1 + x \right) \,H(0,0,0,0;x)}
    {{\left( 1 - x \right) }^3} 
- 
   \frac{12\,x\,\left( 1 + x \right) \,H(0,0,0,1;x)}
    {{\left( 1 - x \right) }^3} 
\nn \\
&&
- 
   \frac{8\,x\,\left( 1 + 6\,x + x^2 \right) \,H(0,0,1,0;x)}
    {\left( 1 - x \right) \,{\left( 1 + x \right) }^3} 
\nn \\
&&
+ 
   \frac{8\,x\,\left( 1 + 4\,x - 26\,x^2 + 4\,x^3 + x^4 \right) \,
      H(0,1,0,0;x)}{{\left( 1 - x^2 \right) }^3} 
\nn \\
&&
+ 
   \frac{8\,x\,\left( 1 + 6\,x + x^2 \right) \,H(1,0,0,0;x)}
    {\left( 1 - x \right) \,{\left( 1 + x \right) }^3}
\ . 
%%%%%%%%%%%%%%%%%%%%%%%%%%%%%%%%%%%%%%%%%%%%%%%%%%%%%%%%%%%%%%%%%%%%%%%%%%%%%%%%%%%%%%%%
\eea

\subsubsection{The Pseudoscalar Case}

As done in the scalar case, by choosing $\mu=m$ and defining

\bea
F_{P}^{(2l)}(\epsilon,s) = & \hspace*{-0.2cm} C^2(\epsilon) &
\hspace*{-0.3cm}
\Bigg\{
 C_F^2 \Bigg(
               \frac{\bar b_{-2}}{\epsilon^2} 
             + \frac{\bar b_{-1}}{\epsilon}   
             + \ \bar b_{0}
       \Bigg)
 + C_F C_A \Bigg(
               \frac{\bar c_{-2}}{\epsilon^2} 
             + \frac{\bar c_{-1}}{\epsilon}   
             + \ \bar c_{0}
       \Bigg)
\nn \\
&& 
 + C_F T_R N_f \Bigg(
               \frac{\bar d_{-2}}{\epsilon^2} 
             + \frac{\bar d_{-1}}{\epsilon}   
             + \ \bar d_{0}
       \Bigg)
 + C_F T_R \Bigg(
               \frac{\bar e_{-2}}{\epsilon^2} 
             + \frac{\bar e_{-1}}{\epsilon}   
             + \ \bar e_{0}
       \Bigg)
\nn \\
&& 
+ \mathcal{O}(\epsilon)
\Bigg\}
\, ,
\label{2lrenF5}
\eea

\noindent
one finds:

\bea
%%%%%%%%%%%%%%%%%%%%%%%%%%%%%%%%%%%%%%%%%%%%%%%%%%%%%%%%%%%%%%%%%%%%%%%%%%%%%%%%%%%%%%%%
%%%% C_F^2
%%%%%%%%%%%%%%%%%%%%%%%%%%%%%%%%%%%%%%%%%%%%%%%%%%%%%%%%%%%%%%%%%%%%%%%%%%%%%%%%%%%%%%%%
\bar b_{-2} &=&  b_{-2}
\ ; \\
&& \nn \\
%%%%%%%%%%%%%%%%%%%%%%%%%%%%%%%%%%%%%%%%%%%%%%%%%%%%%%%%%%%%%%%%%%%%%%%%%%%%%%%%%%%%%%%%
\bar b_{-1} &=&  b_{-1}
 + \frac{4\,x\,H(0;x)}{1 - x^2} + 
   \frac{8\,x\,\left( 1 + x^2 \right) \,H(0,0;x)}{{\left( 1 - x^2 \right) }^2}
\ ; \\
&& \nn \\
%%%%%%%%%%%%%%%%%%%%%%%%%%%%%%%%%%%%%%%%%%%%%%%%%%%%%%%%%%%%%%%%%%%%%%%%%%%%%%%%%%%%%%%%
\bar b_{0} &=& 
  \frac{4\,\left( 1 + x^2 \right) \,H(-1,-1,0;x)}{1 - x^2} \nn \\ && + 
   \frac{\left( 5 + 14\,x + 46\,x^2 + 14\,x^3 + 9\,x^4 \right) \,H(-1,0,0;x)}
    {{\left( 1 - x^2 \right) }^2} \nn \\ && + 
   \frac{2\,\left( 3 - 14\,x - 4\,x^2 - 14\,x^3 + 5\,x^4 \right) \,
      H(0,-1,0;x)}{{\left( 1 - x^2 \right) }^2} \nn \\ && + 
   \frac{\left( 1 - 9\,x - 9\,x^2 + 71\,x^3 - 16\,x^4 + 10\,x^5 \right) \,
      H(0,0,0;x)}{{\left( 1 - x \right) }^3\,{\left( 1 + x \right) }^2} \nn \\
   && - 
   \frac{4\,\left( 1 + x^2 \right) \,H(0,1,0;x)}{{\left( 1 + x \right) }^2} + 
   \frac{2\,\left( 5 - 12\,x - 18\,x^2 - 12\,x^3 + 5\,x^4 \right) \,
      H(1,0,0;x)}{{\left( 1 - x^2 \right) }^2} \nn \\ && + 
   \frac{16\,{\left( 1 + x^2 \right) }^2\,H(-1,-1,0,0;x)}
    {{\left( 1 - x^2 \right) }^2} + 
   \frac{8\,{\left( 1 + x^2 \right) }^2\,H(-1,0,-1,0;x)}
    {{\left( 1 - x^2 \right) }^2} \nn \\ && - 
   \frac{12\,{\left( 1 + x^2 \right) }^2\,H(-1,0,0,0;x)}
    {{\left( 1 - x^2 \right) }^2} + 
   \frac{4\,{\left( 1 + x^2 \right) }^2\,H(0,-1,-1,0;x)}
    {{\left( 1 - x^2 \right) }^2} \nn \\ && + 
   \frac{2\,\left( -3 + 11\,x - 12\,x^2 + 8\,x^3 + x^4 + 7\,x^5 \right) \,
      H(0,-1,0,0;x)}{{\left( 1 - x \right) }^3\,{\left( 1 + x \right) }^2} \nn
   \\ && + 
   \frac{2\,\left( 1 - 5\,x - 10\,x^2 + 10\,x^3 - 15\,x^4 + 11\,x^5 \right) \,
      H(0,0,-1,0;x)}{{\left( 1 - x \right) }^3\,{\left( 1 + x \right) }^2} \nn
   \\ && - 
   \frac{\left( -7 + 7\,x - 24\,x^2 + 44\,x^3 - 27\,x^4 + 27\,x^5 \right) \,
      H(0,0,0,0;x)}{{\left( 1 - x \right) }^3\,{\left( 1 + x \right) }^2} \nn
   \\ && + 
   \frac{4\,\left( -1 + 2\,x^2 + 3\,x^4 \right) \,H(0,0,1,0;x)}
    {{\left( 1 - x^2 \right) }^2} \nn \\ && + 
   \frac{4\,\left( 1 - 4\,x + 2\,x^2 - 4\,x^3 + x^4 \right) \,H(0,1,0,0;x)}
    {{\left( 1 - x \right) }^3\,\left( 1 + x \right) } \nn \\ && - 
   \frac{8\,{\left( 1 + x^2 \right) }^2\,H(1,0,-1,0;x)}
    {{\left( 1 - x^2 \right) }^2} \nn \\ && + 
   \frac{8\,\left( 1 + 2\,x + 2\,x^2 - 2\,x^3 + x^4 \right) \,H(1,0,0,0;x)}
    {{\left( 1 - x^2 \right) }^2} \nn \\ && + 
   \frac{8\,{\left( 1 + x^2 \right) }^2\,H(1,0,1,0;x)}
    {{\left( 1 - x^2 \right) }^2} \nn \\ && + 
   \frac{\left( 5 - 18\,x + 54\,x^2 - 18\,x^3 + x^4 \right) \,H(-1;x)\,
      \zeta_2}{{\left( 1 - x^2 \right) }^2} \nn \\ && + 
   \frac{2\,\left( 7 - 7\,x + 8\,x^2 + 4\,x^3 - 5\,x^4 + 5\,x^5 \right) \,
      H(0,-1;x)\,\zeta_2}{{\left( 1 - x \right) }^3\,
      {\left( 1 + x \right) }^2} \nn \\ && + 
   \frac{4\,H(1,0;x)\,\left( -2 + 2\,x^4 + \zeta_2 + 4\,x\,\zeta_2 + 
        2\,x^2\,\zeta_2 - 4\,x^3\,\zeta_2 + x^4\,\zeta_2 \right) }{{\left(
         1 - x^2 \right) }^2} \nn \\ && + 
   \frac{H(-1,0;x)\,\left( 15 + 2\,x - 2\,x^3 - 15\,x^4 + 4\,\zeta_2 + 
        8\,x^2\,\zeta_2 + 4\,x^4\,\zeta_2 \right) }{{\left( 1 - x^2 \right)
         }^2} \nn \\ && 
- \frac{1} {{\left( 1 - x \right)}^3\,{\left( 1 + x \right) }^2} 
\Big[ -3 + 14\,x + 6\,x^2 - 8\,x^3 - 35\,x^4 + 26\,x^5 + 3\,\zeta_2 
\nn \\ && - 
        7\,x\,\zeta_2 + 4\,x^2\,\zeta_2 + 24\,x^3\,\zeta_2 - 
        17\,x^4\,\zeta_2 + 13\,x^5\,\zeta_2 \Big] \, H(0,0;x)
\nn \\ && + 
   \frac{4\,{\left( 1 + x^2 \right) }^2\,H(1;x)\,\zeta_3}
    {{\left( 1 - x^2 \right) }^2} \nn \\ && - 
   \frac{1}{20\, {\left( 1 - x \right) }^3\,{\left( 1 + x \right) }^2} 
\Big[-145 + 145\,x + 290\,x^2 - 290\,x^3 - 145\,x^4 + 145\,x^5 \nn \\ &&- 
      20\,\zeta_2 + 240\,x\,\zeta_2 - 1600\,x^2\,\zeta_2 + 
      1640\,x^3\,\zeta_2 - 300\,x^4\,\zeta_2 + 40\,x^5\,\zeta_2 \nn \\ &&- 
      480\,x\,\ln(2)\,\zeta_2 + 1440\,x^2\,\ln(2)\,\zeta_2 - 
      1440\,x^3\,\ln(2)\,\zeta_2 \nn \\ && + 480\,x^4\,\ln(2)\,\zeta_2 + 
      118\,{\zeta_2}^2 - 254\,x\,{\zeta_2}^2 + 48\,x^2\,{\zeta_2}^2 + 
      536\,x^3\,{\zeta_2}^2 \nn \\ &&- 498\,x^4\,{\zeta_2}^2 + 
      362\,x^5\,{\zeta_2}^2 + 280\,\zeta_3 + 280\,x^2\,\zeta_3 - 
      280\,x^3\,\zeta_3 \nn \\ &&- 80\,x^4\,\zeta_3 - 200\,x^5\,\zeta_3
\Big]
\nn \\ && + 
   \frac{H(0;x)}{{\left( 1 - x \right) }^3\,{\left( 1 + x \right) }^2}
\,\Big[ 3 - x - 2\,x^2 - 2\,x^3 - x^4 + 3\,x^5 + 
        2\,\zeta_2 - 6\,x\,\zeta_2 \nn \\ &&- 11\,x^2\,\zeta_2 + 
        73\,x^3\,\zeta_2 - 15\,x^4\,\zeta_2 + 5\,x^5\,\zeta_2 - 
        8\,\zeta_3 - 8\,x\,\zeta_3 \nn \\ && + 8\,x^2\,\zeta_3 + 8\,x^3\,\zeta_3 - 
        8\,x^4\,\zeta_3 - 8\,x^5\,\zeta_3 \Big]
\ ; 
%%%%%%%%%%%%%%%%%%%%%%%%%%%%%%%%%%%%%%%%%%%%%%%%%%%%%%%%%%%%%%%%%%%%%%%%%%%%%%%%%%%%%%%%
\eea
\bea
%%%%%%%%%%%%%%%%%%%%%%%%%%%%%%%%%%%%%%%%%%%%%%%%%%%%%%%%%%%%%%%%%%%%%%%%%%%%%%%%%%%%%%%%
%%%% C_F*C_A
%%%%%%%%%%%%%%%%%%%%%%%%%%%%%%%%%%%%%%%%%%%%%%%%%%%%%%%%%%%%%%%%%%%%%%%%%%%%%%%%%%%%%%%%
\bar c_{-2} &=&  c_{-2}
\ ; \\
&& \nn \\
%%%%%%%%%%%%%%%%%%%%%%%%%%%%%%%%%%%%%%%%%%%%%%%%%%%%%%%%%%%%%%%%%%%%%%%%%%%%%%%%%%%%%%%%
\bar c_{-1} &=&  c_{-1}
\ ; \\
&& \nn \\
%%%%%%%%%%%%%%%%%%%%%%%%%%%%%%%%%%%%%%%%%%%%%%%%%%%%%%%%%%%%%%%%%%%%%%%%%%%%%%%%%%%%%%%%
\bar c_{0} &=& 
  \frac{-52\,\left( 1 + x^2 \right) \,H(-1,-1,0;x)}
    {3\,\left( 1 - x^2 \right) } \nn \\ && - 
   \frac{\left( -61 - 6\,x + 66\,x^2 - 6\,x^3 + 79\,x^4 \right) \,H(-1,0,0;x)}
    {6\,{\left( 1 - x^2 \right) }^2} \nn \\ && + 
   \frac{6\,\left( 1 + x^2 \right) \,H(-1,1,0;x)}{1 - x^2} \nn \\ && - 
   \frac{2\,\left( -1 - 15\,x - 15\,x^2 - 15\,x^3 + 34\,x^4 \right) \,
      H(0,-1,0;x)}{3\,{\left( 1 - x^2 \right) }^2} \nn \\ && - 
   \frac{\left( 11 - 35\,x + 45\,x^2 + 189\,x^3 - 128\,x^4 + 62\,x^5 \right)
        \,H(0,0,0;x)}{6\,{\left( 1 - x \right) }^3\,{\left( 1 + x \right) }^2}
    \nn \\ && - \frac{4\,x\,\left( 1 + x + 2\,x^2 \right) \,H(0,1,0;x)}
    {\left( 1 - x \right) \,{\left( 1 + x \right) }^2} \nn \\ && + 
   \frac{6\,\left( 1 + x^2 \right) \,H(1,-1,0;x)}{1 - x^2} \nn \\ && - 
   \frac{4\,\left( 3 - 2\,x - 4\,x^2 - 2\,x^3 + x^4 \right) \,H(1,0,0;x)}
    {{\left( 1 - x^2 \right) }^2} \nn \\ && - 
   \frac{2\,\left( 1 + x^2 \right) \,H(1,1,0;x)}{1 - x^2} \nn \\ && - 
   \frac{4\,{\left( 1 + x^2 \right) }^2\,H(-1,0,-1,0;x)}
    {{\left( 1 - x^2 \right) }^2} \nn \\ && + 
   \frac{2\,\left( 3 + 4\,x^2 + x^4 \right) \,H(-1,0,0,0;x)}
    {{\left( 1 - x^2 \right) }^2} \nn \\ && + 
   \frac{4\,{\left( 1 + x^2 \right) }^2\,H(-1,0,1,0;x)}
    {{\left( 1 - x^2 \right) }^2} \nn \\ && - 
   \frac{10\,{\left( 1 + x^2 \right) }^2\,H(0,-1,-1,0;x)}
    {{\left( 1 - x^2 \right) }^2} \nn \\ && - 
   \frac{2\,\left( -1 + x - 5\,x^2 + 11\,x^3 - 7\,x^4 + 7\,x^5 \right) \,
      H(0,-1,0,0;x)}{{\left( 1 - x \right) }^3\,{\left( 1 + x \right) }^2} \nn \\ && + 
   \frac{6\,{\left( 1 + x^2 \right) }^2\,H(0,-1,1,0;x)}
    {{\left( 1 - x^2 \right) }^2} \nn \\ && - 
   \frac{2\,\left( -1 + x - 14\,x^2 + 10\,x^3 - 11\,x^4 + 11\,x^5 \right) \,
      H(0,0,-1,0;x)}{{\left( 1 - x \right) }^3\,{\left( 1 + x \right) }^2} \nn \\ && + 
   \frac{x^2\,\left( -7 + 17\,x - 12\,x^2 + 12\,x^3 \right) \,H(0,0,0,0;x)}
    {{\left( 1 - x \right) }^3\,{\left( 1 + x \right) }^2} \nn \\ && - 
   \frac{2\,\left( 1 + 8\,x^2 + 7\,x^4 \right) \,H(0,0,1,0;x)}
    {{\left( 1 - x^2 \right) }^2} \nn \\ && + 
   \frac{6\,{\left( 1 + x^2 \right) }^2\,H(0,1,-1,0;x)}
    {{\left( 1 - x^2 \right) }^2} \nn \\ && + 
   \frac{2\,\left( -3 + 3\,x + 8\,x^3 - x^4 + x^5 \right) \,H(0,1,0,0;x)}
    {{\left( 1 - x \right) }^3\,{\left( 1 + x \right) }^2} \nn \\ && - 
   \frac{2\,{\left( 1 + x^2 \right) }^2\,H(0,1,1,0;x)}
    {{\left( 1 - x^2 \right) }^2} \nn \\ && + 
   \frac{4\,{\left( 1 + x^2 \right) }^2\,H(1,0,-1,0;x)}
    {{\left( 1 - x^2 \right) }^2} \nn \\ && + 
   \frac{2\,\left( -5 - 4\,x^2 + x^4 \right) \,H(1,0,0,0;x)}
    {{\left( 1 - x^2 \right) }^2} \nn \\ && - 
   \frac{4\,{\left( 1 + x^2 \right) }^2\,H(1,0,1,0;x)}
    {{\left( 1 - x^2 \right) }^2} \nn \\ && - 
   \frac{\left( 5 - 5\,x + 4\,x^2 + 8\,x^3 - 7\,x^4 + 7\,x^5 \right) \,
      H(0,-1;x)\,\zeta_2}{{\left( 1 - x \right) }^3\,
      {\left( 1 + x \right) }^2} \nn \\ && + 
   \frac{{\left( 1 + x^2 \right) }^2\,H(0,1;x)\,\zeta_2}
    {{\left( 1 - x^2 \right) }^2} \nn \\ && - 
   \frac{H(-1,0;x)\,\left( -53 - 54\,x + 54\,x^3 + 53\,x^4 - 72\,\zeta_2 - 
        72\,x^2\,\zeta_2 \right) }{18\,{\left( 1 - x^2 \right) }^2} \nn \\ && + 
   \frac{2\,\left( 1 + x^2 \right) \,H(1,0;x)\,
      \left( 1 - x^2 - 4\,\zeta_2 + 2\,x^2\,\zeta_2 \right) }{{\left( 1 - 
         x^2 \right) }^2} \nn \\ && 
- \frac{H(0,0;x)}{18\,{\left( 1 - x \right) }^3\,{\left( 1 + x \right) }^2} \,
      \Big[ 67 - 148\,x + 36\,x^2 - 90\,x^3 + 185\,x^4 - 50\,x^5 - 
        18\,\zeta_2 \nn \\ && + 18\,x\,\zeta_2 - 126\,x^2\,\zeta_2 - 
        54\,x^3\,\zeta_2 - 18\,x^4\,\zeta_2 + 18\,x^5\,\zeta_2 \Big]
\nn \\ && - 
   \frac{\left( 1 + x^2 \right) \,H(1;x)\,
      \left( -\zeta_2 + x^2\,\zeta_2 + 2\,\zeta_3 + 2\,x^2\,\zeta_3
        \right) }{{\left( 1 - x^2 \right) }^2} 
   \nn \\ && + 
   \frac{H(-1;x)}{{6\,\left( 1 - x^2 \right) }^2} 
\,\Big[ -25\,\zeta_2 + 54\,x\,\zeta_2 - 
        162\,x^2\,\zeta_2 + 54\,x^3\,\zeta_2 + 7\,x^4\,\zeta_2 \nn \\ && + 
        12\,\zeta_3 + 24\,x^2\,\zeta_3 + 12\,x^4\,\zeta_3 \Big]
   \nn \\ && - 
   \frac{1}{540\,{\left( 1 - x \right) }^3\,{\left( 1 + x \right) }^2} 
   \Big[
      4345 - 4345\,x - 8690\,x^2 + 8690\,x^3 + 4345\,x^4 \nn \\ && - 4345\,x^5 - 
      2730\,\zeta_2 - 1320\,x\,\zeta_2 + 25200\,x^2\,\zeta_2 - 
      23580\,x^3\,\zeta_2 \nn \\ && + 3450\,x^4\,\zeta_2 - 1020\,x^5\,\zeta_2 + 
      6480\,x\,\ln(2)\,\zeta_2 - 19440\,x^2\,\ln(2)\,\zeta_2 \nn \\ &&+ 
      19440\,x^3\,\ln(2)\,\zeta_2 - 6480\,x^4\,\ln(2)\,\zeta_2 + 
      81\,{\zeta_2}^2 - 81\,x\,{\zeta_2}^2 \nn \\ &&- 108\,x^2\,{\zeta_2}^2 - 
      4104\,x^3\,{\zeta_2}^2 + 1917\,x^4\,{\zeta_2}^2 - 
      1917\,x^5\,{\zeta_2}^2 - 4410\,\zeta_3 \nn \\ && + 4950\,x\,\zeta_3 - 
      2700\,x^2\,\zeta_3 + 2700\,x^3\,\zeta_3 + 630\,x^4\,\zeta_3 - 
      1170\,x^5\,\zeta_3\Big]
\nn \\ && - 
   \frac{H(0;x)}{54\,{\left( 1 - x \right) }^3\,{\left( 1 + x \right) }^2}
\,\Big[ 121 - 283\,x + 162\,x^2 + 162\,x^3 - 283\,x^4 + 
        121\,x^5 \nn \\ && + 99\,\zeta_2 - 315\,x\,\zeta_2 - 135\,x^2\,\zeta_2 + 
        2241\,x^3\,\zeta_2 - 612\,x^4\,\zeta_2 \nn \\ && + 18\,x^5\,\zeta_2 - 
        351\,\zeta_3 + 351\,x\,\zeta_3 - 162\,x^2\,\zeta_3 - 
        270\,x^3\,\zeta_3 \nn \\ && + 405\,x^4\,\zeta_3 - 405\,x^5\,\zeta_3 \Big]
\ ; 
%%%%%%%%%%%%%%%%%%%%%%%%%%%%%%%%%%%%%%%%%%%%%%%%%%%%%%%%%%%%%%%%%%%%%%%%%%%%%%%%%%%%%%%%
\eea
\bea
%%%%%%%%%%%%%%%%%%%%%%%%%%%%%%%%%%%%%%%%%%%%%%%%%%%%%%%%%%%%%%%%%%%%%%%%%%%%%%%%%%%%%%%%
%%%% C_F*T_R*N_f
%%%%%%%%%%%%%%%%%%%%%%%%%%%%%%%%%%%%%%%%%%%%%%%%%%%%%%%%%%%%%%%%%%%%%%%%%%%%%%%%%%%%%%%%
\bar d_{-2} &=& d_{-2}
\ ; \\
&& \nn \\
%%%%%%%%%%%%%%%%%%%%%%%%%%%%%%%%%%%%%%%%%%%%%%%%%%%%%%%%%%%%%%%%%%%%%%%%%%%%%%%%%%%%%%%%
\bar d_{-1} &=& d_{-1}
\ ; \\
&& \nn \\
%%%%%%%%%%%%%%%%%%%%%%%%%%%%%%%%%%%%%%%%%%%%%%%%%%%%%%%%%%%%%%%%%%%%%%%%%%%%%%%%%%%%%%%%
\bar d_{0} &=& 
  \frac{-20\,\left( 1 + x^2 \right) \,H(-1,0;x)}
    {9\,\left( 1 - x^2 \right) } + 
   \frac{10\,\left( 1 + x^2 \right) \,H(0,0;x)}{9\,\left( 1 - x^2 \right) } \nn \\ && + 
   \frac{8\,\left( 1 + x^2 \right) \,H(-1,-1,0;x)}
    {3\,\left( 1 - x^2 \right) } - 
   \frac{4\,\left( 1 + x^2 \right) \,H(-1,0,0;x)}
    {3\,\left( 1 - x^2 \right) } \nn \\ && - 
   \frac{4\,\left( 1 + x^2 \right) \,H(0,-1,0;x)}
    {3\,\left( 1 - x^2 \right) } + 
   \frac{2\,\left( 1 + x^2 \right) \,H(0,0,0;x)}
    {3\,\left( 1 - x^2 \right) } \nn \\ && + 
   \frac{4\,\left( 1 + x^2 \right) \,H(-1;x)\,\zeta_2}
    {3\,\left( 1 - x^2 \right) } + 
   \frac{2\,\left( 1 + x^2 \right) \,H(0;x)\,\left( 14 + 9\,\zeta_2 \right) }
    {27\,\left( 1 - x^2 \right) } \nn \\ && - 
   \frac{-49 + 49\,x^2 - 6\,\zeta_2 + 66\,x^2\,\zeta_2 + 36\,\zeta_3 + 
      36\,x^2\,\zeta_3}{27\,\left( 1 - x^2 \right) }
\ ; 
%%%%%%%%%%%%%%%%%%%%%%%%%%%%%%%%%%%%%%%%%%%%%%%%%%%%%%%%%%%%%%%%%%%%%%%%%%%%%%%%%%%%%%%%
\eea
\bea
%%%%%%%%%%%%%%%%%%%%%%%%%%%%%%%%%%%%%%%%%%%%%%%%%%%%%%%%%%%%%%%%%%%%%%%%%%%%%%%%%%%%%%%%
%%%% C_F*T_R
%%%%%%%%%%%%%%%%%%%%%%%%%%%%%%%%%%%%%%%%%%%%%%%%%%%%%%%%%%%%%%%%%%%%%%%%%%%%%%%%%%%%%%%%
\bar e_{-2} &=& e_{-2} = 0
\ ; \\
&& \nn \\ 
%%%%%%%%%%%%%%%%%%%%%%%%%%%%%%%%%%%%%%%%%%%%%%%%%%%%%%%%%%%%%%%%%%%%%%%%%%%%%%%%%%%%%%%%
\bar e_{-1} &=& e_{-1} = 0
\ ; \\
&& \nn \\ 
%%%%%%%%%%%%%%%%%%%%%%%%%%%%%%%%%%%%%%%%%%%%%%%%%%%%%%%%%%%%%%%%%%%%%%%%%%%%%%%%%%%%%%%%
\bar e_{0} &=& 
  \frac{2\,\left( 1 - 2\,x + 26\,x^2 - 2\,x^3 + x^4 \right) \,H(0,0,0;x)}
    {3\,{\left( 1 - x \right) }^3\,\left( 1 + x \right) } - 
   \frac{2\,x\,H(0,0,0,0;x)}{1 - x^2} \nn \\ && - \frac{12\,x\,H(0,0,0,1;x)}{1 - x^2} - 
   \frac{8\,x\,H(0,0,1,0;x)}{1 - x^2} + \frac{8\,x\,H(0,1,0,0;x)}{1 - x^2} \nn \\ && + 
   \frac{8\,x\,H(1,0,0,0;x)}{1 - x^2} + 
   \frac{8\,x\,H(1,0;x)\,\zeta_2}{1 - x^2} \nn \\ && + 
   \frac{2\,H(0,0;x)}{9\, {\left( 1 - x \right) }^2\,{\left( 1 + x \right) }^4} 
     \Big[ 5 + 22\,x + 59\,x^2 + 116\,x^3 + 59\,x^4 + 
        22\,x^5 + 5\,x^6 \nn \\ && - 18\,x\,\zeta_2 - 36\,x^2\,\zeta_2 + 
        36\,x^4\,\zeta_2 + 18\,x^5\,\zeta_2 \Big]
   \nn \\ && + 
   \frac{2\,H(0;x)}{27\,{\left( 1 - x^2 \right) }^3}
     \Big[ 28 + 24\,x - 76\,x^2 + 48\,x^3 - 76\,x^4 + 
        24\,x^5 + 28\,x^6 + 9\,\zeta_2 \nn \\ && + 207\,x^2\,\zeta_2 + 
        432\,x^3\,\zeta_2 + 207\,x^4\,\zeta_2 + 9\,x^6\,\zeta_2 - 
        162\,x\,\zeta_3 \nn \\ && + 324\,x^3\,\zeta_3 - 162\,x^5\,\zeta_3 \Big]
   \nn \\ && + 
   \frac{1}{135\,{\left( 1 - x \right) }^2\,{\left( 1 + x \right) }^4} 
     \Big[2035 + 4190\,x - 2035\,x^2 - 8380\,x^3 - 2035\,x^4 \nn \\ && + 4190\,x^5 + 
      2035\,x^6 - 810\,\zeta_2 - 1080\,x\,\zeta_2 + 4410\,x^2\,\zeta_2 + 
      12240\,x^3\,\zeta_2 \nn \\ && + 4410\,x^4\,\zeta_2 - 1080\,x^5\,\zeta_2 - 
      810\,x^6\,\zeta_2 - 432\,x\,{\zeta_2}^2 - 864\,x^2\,{\zeta_2}^2 \nn \\ && + 
      864\,x^4\,{\zeta_2}^2 + 432\,x^5\,{\zeta_2}^2 \Big]
\ . 
%%%%%%%%%%%%%%%%%%%%%%%%%%%%%%%%%%%%%%%%%%%%%%%%%%%%%%%%%%%%%%%%%%%%%%%%%%%%%%%%%%%%%%%%
\eea

\boldmath
\subsection{Form Factors for $\mu \neq m$ \label{munotm}}
\unboldmath

At the one-loop level, 
the dependence of the form factors on the
renormalization scale $\mu$ appears in an overall factor $(\mu^2/m^2)^\epsilon$:
\be
F_{i}^{(1l)} \Bigl( \epsilon,s, \frac{\mu^2}{m^2} \Bigr) = 
C(\epsilon) \, \left( \frac{\mu^2}{m^2} \right)^{\epsilon} 
F_{i}^{(1l)}(\epsilon,s) \, , \quad (i=S,P)
\label{FF1loopmunotm}
\ee
with the expression of $F_{S}^{(1l)}(\epsilon,s)$ and
$F_{P}^{(1l)}(\epsilon,s)$ given in  
Eq.~(\ref{1lrenF0}) and Eq.~(\ref{1lrenF5}) respectively.

At the two-loop level, the 
coupling constant renormalization generates logarithms of the ratio $(\mu^2/m^2)$. 
Factoring an overall $(\mu^2/m^2)^{2 
  \epsilon}$, one has \cite{Bernreuther:2004ih}: 

\bea
\hspace{-5mm}
F_{i}^{(2l)} \Bigl( \epsilon,s, \frac{\mu^2}{m^2} \Bigr) \! \! &=& 
\left( \frac{\alpha_S}{2 \pi} \right)^2 \,
C^2(\epsilon) \, \left( \frac{\mu^2}{m^2} \right)^{2 \epsilon}   \nn\\
& & \hspace*{15mm} 
\times \Biggl\{
F_{i}^{(2l)}(\epsilon,s)
- \frac{\beta_0}{\epsilon} \left[ \left( \frac{\mu^2}{m^2} 
\right)^{- \epsilon} -1 \right]
F_{i}^{(1l)}(\epsilon,s) \Biggr\}  
\label{eq:reno1} \\
\hspace{-5mm}
& = & \left( \frac{\alpha_S}{2 \pi} \right)^2 \,
C^2(\epsilon) \, \left( \frac{\mu^2}{m^2} \right)^{2 \epsilon}  \Biggl\{
F_{i}^{(2l)}(\epsilon,s)
+ \beta_0 \,
F_{i}^{(1l)}(\epsilon,s) \ln{\left( \frac{\mu^2}{m^2} \right)}
\nn\\
& & \hspace*{45mm}
-  \epsilon \, \frac{\beta_0}{2}
F_{i}^{(1l)}(\epsilon,s) 
\ln^2{\left( \frac{\mu^2}{m^2} \right)}
\Biggr\} 
\quad (i=S,P)
\, , 
\label{eq:reno2}
\eea

\noindent
where the functions $F_{i}^{(2l)}(\epsilon,s)$ are given in 
Eq.~(\ref{2lrenF0}) and Eq.~(\ref{2lrenF5}), and $\beta_0$ is 
the first coefficient of the QCD $\beta$-function,
\be
\beta_0 = \frac{11 C_A - 4 T_R (N_f+1)}{6} \, .
\ee

\section{Analytic Continuation above Threshold \label{analytical}
\label{sec_analytic}}

For application to Higgs decay, the form factors must be analytically
continued into the timelike region, above the $Q {\bar Q}$ 
threshold, $s > 4m^2$, where they develop imaginary parts.
That can be made by the substitution $s \to s + i \epsilon$,  which implies
\be
x \rightarrow - y + i \epsilon \ , 
\ee
where
\be
y = \frac{\sqrt{s} - \sqrt{s-4m^2} }{\sqrt{s} + \sqrt{s-4m^2} } \ ,
\ee
with $y=1$ at $s=4m^2$ and $y\to0$ as $s \to \infty$.

The real and imaginary parts of the form factors are defined through the
relations:
\bea
F_{i}(\epsilon,s+i\epsilon) & = & \Re \, F_{i}(\epsilon,s) 
                         + i \pi \, \Im \, F_{i}(\epsilon,s) 
                         \qquad (i=S,P) \ .
\eea

In the following two sections we will give the real and imaginary parts of
the one- and two-loop analytically continued form factors for $\mu=m$. 
The renormalization scale dependence follows from the pattern outlined 
in Section \ref{munotm}.

\subsection{One-Loop Form Factors above Threshold}

In this section we give the
analytic continuation of the coefficients $a_i$, corresponding to the scalar
form factor defined in Eq.~(\ref{1lrenF0}), and $\bar a_i$, corresponding to
the pseudoscalar one defined in Eq.~(\ref{1lrenF5}).

\subsubsection{The Scalar Case}

\bea
%%%%%%%%%%%%%%%%%%%%%%%%%%%%%%%%%%%%%%%%%%%%%%%%%%%%%
\Re \, a_{-1} = &&
-1 - \frac{\left( 1 + y^2 \right) \,H(0;y)}{1 - y^2}
\ ; \\
&& \nn \\
%%%%%%%%%%%%%%%%%%%%%%%%%%%%%%%%%%%%%%%%%%%%%%%%%%%%%
\Re \, a_{0} = &&
  -\left( \frac{1 - y^2 - 4\,\zeta_2 - 4\,y^2\,\zeta_2}
      {1 - y^2} \right)  - \frac{4\,y\,H(0;y)}{1 - y^2} 
\nn \\
&&
- 
   \frac{\left( 1 + y^2 \right) \,H(0,0;y)}{1 - y^2} - 
   \frac{2\,\left( 1 + y^2 \right) \,H(1,0;y)}{1 - y^2}
\ ; \\
&& \nn \\
%%%%%%%%%%%%%%%%%%%%%%%%%%%%%%%%%%%%%%%%%%%%%%%%%%%%%
\Re \, a_{1} = &&
  \frac{2\,\left( -1 + y^2 + 8\,y\,\zeta_2 + \zeta_3 + 
        y^2\,\zeta_3 \right) }{1 - y^2} 
\nn \\
&&
+ 
   \frac{\left( -1 - 6\,y - y^2 + 4\,\zeta_2 + 
        4\,y^2\,\zeta_2 \right) \,H(0;y)}{1 - y^2} 
\nn \\
&&
+ 
   \frac{8\,\left( 1 + y^2 \right) \,\zeta_2\,H(1;y)}{1 - y^2} - 
   \frac{4\,y\,H(0,0;y)}{1 - y^2} - \frac{8\,y\,H(1,0;y)}{1 - y^2} 
\nn \\
&&
- 
   \frac{\left( 1 + y^2 \right) \,H(0,0,0;y)}{1 - y^2} - 
   \frac{2\,\left( 1 + y^2 \right) \,H(0,1,0;y)}{1 - y^2} 
\nn \\
&&
- 
   \frac{2\,\left( 1 + y^2 \right) \,H(1,0,0;y)}{1 - y^2} - 
   \frac{4\,\left( 1 + y^2 \right) \,H(1,1,0;y)}{1 - y^2}
\ .
%%%%%%%%%%%%%%%%%%%%%%%%%%%%%%%%%%%%%%%%%%%%%%%%%%%%%
\label{1lReF0coeff}
\eea

\bea
%%%%%%%%%%%%%%%%%%%%%%%%%%%%%%%%%%%%%%%%%%%%%%%%%%%%%
\Im \, a_{-1} = &&
- \frac{1 + y^2}{1 - y^2} 
\ ; \\
&& \nn \\
%%%%%%%%%%%%%%%%%%%%%%%%%%%%%%%%%%%%%%%%%%%%%%%%%%%%%
\Im \, a_{0} = &&
  \frac{-4\,y}{1 - y^2} - \frac{\left( 1 + y^2 \right) \,H(0;y)}{1 - y^2} - 
   \frac{2\,\left( 1 + y^2 \right) \,H(1;y)}{1 - y^2}
\ ; \\
&& \nn \\
%%%%%%%%%%%%%%%%%%%%%%%%%%%%%%%%%%%%%%%%%%%%%%%%%%%%%
\Im \, a_{1} = &&
  - \frac{1 + 6\,y + y^2 - 2\,\zeta_2 - 2\,y^2\,\zeta_2}
      {1 - y^2} 
\nn \\
&&
- \frac{4\,y\,H(0;y)}{1 - y^2} - 
   \frac{8\,y\,H(1;y)}{1 - y^2} - 
   \frac{\left( 1 + y^2 \right) \,H(0,0;y)}{1 - y^2} 
\nn \\
&&
- 
   \frac{2\,\left( 1 + y^2 \right) \,H(0,1;y)}{1 - y^2} - 
   \frac{2\,\left( 1 + y^2 \right) \,H(1,0;y)}{1 - y^2} 
\nn \\
&&
- 
   \frac{4\,\left( 1 + y^2 \right) \,H(1,1;y)}{1 - y^2}
\ .
%%%%%%%%%%%%%%%%%%%%%%%%%%%%%%%%%%%%%%%%%%%%%%%%%%%%%
\label{1lImF0coeff}
\eea

\subsubsection{The Pseudoscalar Case}

\bea
%%%%%%%%%%%%%%%%%%%%%%%%%%%%%%%%%%%%%%%%%%%%%%%%%%%%%
\Re \, \bar a_{-1} = && \Re \, a_{-1} 
\ ; \\
&& \nn \\
%%%%%%%%%%%%%%%%%%%%%%%%%%%%%%%%%%%%%%%%%%%%%%%%%%%%%
\Re \, \bar a_{0} = &&
  -\frac{\left( 1 + y^2 \right) \,H(0,0;y)}{1 - y^2}  - 
   \frac{2\,\left( 1 + y^2 \right) \,H(1,0;y)}{1 - y^2} \nn \\ &&- 
   \frac{1 - y^2 - 4\,\zeta_2 - 4\,y^2\,\zeta_2}{1 - y^2}
\ ; \\
&& \nn \\
%%%%%%%%%%%%%%%%%%%%%%%%%%%%%%%%%%%%%%%%%%%%%%%%%%%%%
\Re \, \bar a_{1} = &&
  - \frac{\left( 1 + y^2 \right) \,H(0,0,0;y)}{1 - y^2}   - 
   \frac{2\,\left( 1 + y^2 \right) \,H(0,1,0;y)}{1 - y^2} \nn \\ && - 
   \frac{2\,\left( 1 + y^2 \right) \,H(1,0,0;y)}{1 - y^2} - 
   \frac{4\,\left( 1 + y^2 \right) \,H(1,1,0;y)}{1 - y^2} \nn \\ && + 
   \frac{8\,\left( 1 + y^2 \right) \,H(1;y)\,\zeta_2}{1 - y^2} \nn \\ && + 
   \frac{H(0;y)\,\left( -1 + 2\,y - y^2 + 4\,\zeta_2 + 4\,y^2\,\zeta_2
        \right) }{1 - y^2} \nn \\ && + \frac{2\,
      \left( -1 + y^2 + \zeta_3 + y^2\,\zeta_3 \right) }{1 - y^2}
\ .
%%%%%%%%%%%%%%%%%%%%%%%%%%%%%%%%%%%%%%%%%%%%%%%%%%%%%
\label{1lReF5coeff}
\eea

\bea
%%%%%%%%%%%%%%%%%%%%%%%%%%%%%%%%%%%%%%%%%%%%%%%%%%%%%
\Im \, \bar a_{-1} = && \Im \, a_{-1} 
\ ; \\
&& \nn \\
%%%%%%%%%%%%%%%%%%%%%%%%%%%%%%%%%%%%%%%%%%%%%%%%%%%%%
\Im \, \bar a_{0} = &&
  - \frac{\left( 1 + y^2 \right) \,H(0;y)}{1 - y^2}  - 
   \frac{2\,\left( 1 + y^2 \right) \,H(1;y)}{1 - y^2}
\ ; \\
&& \nn \\
%%%%%%%%%%%%%%%%%%%%%%%%%%%%%%%%%%%%%%%%%%%%%%%%%%%%%
\Im \, \bar a_{1} = &&
  - \frac{\left( 1 + y^2 \right) \,H(0,0;y)}{1 - y^2}  - 
   \frac{2\,\left( 1 + y^2 \right) \,H(0,1;y)}{1 - y^2} \nn \\ && - 
   \frac{2\,\left( 1 + y^2 \right) \,H(1,0;y)}{1 - y^2} - 
   \frac{4\,\left( 1 + y^2 \right) \,H(1,1;y)}{1 - y^2} \nn \\ && - 
   \frac{1 - 2\,y + y^2 - 2\,\zeta_2 - 2\,y^2\,\zeta_2}{1 - y^2}
\ .
%%%%%%%%%%%%%%%%%%%%%%%%%%%%%%%%%%%%%%%%%%%%%%%%%%%%%
\label{1lImF5coeff}
\eea

Our results Eqs. (53), (54) and Eqs. (59), (60) in the scalar and 
pseudoscalar cases,
agree with those of \cite{Braaten:1980yq} and \cite{Drees:1990dq}, 
respectively.

\subsection{Two-Loop Form Factors above Threshold}

In this section we give
the analytic continuation of the coefficients $b_i$,
$c_i$, $d_i$, $e_i$, of the scalar form factor defined in Eq.~(\ref{2lrenF0}) and 
$\bar b_i$, $\bar c_i$, $\bar d_i$, $\bar e_i$ of the pseudoscalar one, 
defined in Eq.~(\ref{2lrenF5}).

\subsubsection{The Scalar Case}

\bea
%%%%%%%%%%%%%%%%%%%%%%%%%%%%%%%%%%%%%%%%%%%%%%%%%%%%%%%%%%%%%%%%%%%%%%%%%%%%%%%%%%%%%%%%
%%%% C_F^2
%%%%%%%%%%%%%%%%%%%%%%%%%%%%%%%%%%%%%%%%%%%%%%%%%%%%%%%%%%%%%%%%%%%%%%%%%%%%%%%%%%%%%%%%
\Re \, b_{-2} &=&  
  \frac{1 - 2\,y^2 + y^4 - 6\,\zeta_2 - 12\,y^2\,\zeta_2 - 
      6\,y^4\,\zeta_2}{2\,{\left( 1 - y^2 \right) }^2} 
\nn \\
&&
+ 
   \frac{\left( 1 + y^2 \right) \,H(0;y)}{1 - y^2} + 
   \frac{{\left( 1 + y^2 \right) }^2\,H(0,0;y)}{{\left( 1 - y^2 \right) }^2}
\ ; \\
&& \nn \\
%%%%%%%%%%%%%%%%%%%%%%%%%%%%%%%%%%%%%%%%%%%%%%%%%%%%%%%%%%%%%%%%%%%%%%%%%%%%%%%%%%%%%%%%
\Re \, b_{-1} &=& 
  \frac{1 - 2\,y^2 + y^4 - 4\,\zeta_2 - 24\,y\,\zeta_2 - 
      24\,y^3\,\zeta_2 + 4\,y^4\,\zeta_2}{{\left( 1 - y^2
         \right) }^2} 
\nn \\
&&
- \frac{\left( -1 - 4\,y + 4\,y^3 + y^4 + 
        10\,\zeta_2 + 20\,y^2\,\zeta_2 + 10\,y^4\,\zeta_2
        \right) \,H(0;y)}{{\left( 1 - y^2 \right) }^2} 
\nn \\
&&
- 
   \frac{12\,{\left( 1 + y^2 \right) }^2\,\zeta_2\,H(1;y)}
    {{\left( 1 - y^2 \right) }^2} - 
   \frac{\left( -1 - 8\,y - 8\,y^3 + y^4 \right) \,H(0,0;y)}
    {{\left( 1 - y^2 \right) }^2} 
\nn \\
&&
+ 
   \frac{2\,\left( 1 + y^2 \right) \,H(1,0;y)}{1 - y^2} + 
   \frac{3\,{\left( 1 + y^2 \right) }^2\,H(0,0,0;y)}
    {{\left( 1 - y^2 \right) }^2} 
\nn \\
&&
+ 
   \frac{2\,{\left( 1 + y^2 \right) }^2\,H(0,1,0;y)}
    {{\left( 1 - y^2 \right) }^2} + 
   \frac{4\,{\left( 1 + y^2 \right) }^2\,H(1,0,0;y)}
    {{\left( 1 - y^2 \right) }^2}
\ ; \\
&& \nn \\
%%%%%%%%%%%%%%%%%%%%%%%%%%%%%%%%%%%%%%%%%%%%%%%%%%%%%%%%%%%%%%%%%%%%%%%%%%%%%%%%%%%%%%%%
\Re \, b_{0} &=& 
  - \frac{1}{20\,{\left( 1 - y \right) }^3\, {\left( 1 + y \right) }^2} 
\Big[ -145 + 145\,y + 290\,y^2 - 290\,y^3 - 145\,y^4 \nn \\ && + 145\,y^5 + 
        160\,\zeta_2 - 40\,y\,\zeta_2 + 
        1440\,\ln(2)\,y\,\zeta_2 + 
        1640\,y^2\,\zeta_2 \nn \\ && + 
        1440\,\ln(2)\,y^2\,\zeta_2 + 
        2040\,y^3\,\zeta_2 - 
        1440\,\ln(2)\,y^3\,\zeta_2 \nn \\ && - 
        2280\,y^4\,\zeta_2 - 
        1440\,\ln(2)\,y^4\,\zeta_2 - 
        1520\,y^5\,\zeta_2 - 272\,{\zeta_2}^2 \nn \\ && + 
        168\,y\,{\zeta_2}^2 - 848\,y^2\,{\zeta_2}^2 + 
        480\,y^3\,{\zeta_2}^2 - 496\,y^4\,{\zeta_2}^2 \nn \\ && + 
        392\,y^5\,{\zeta_2}^2 + 280\,\zeta_3 - 
        960\,y\,\zeta_3 - 200\,y^2\,\zeta_3 + 
        200\,y^3\,\zeta_3 \nn \\ && + 880\,y^4\,\zeta_3 - 
        200\,y^5\,\zeta_3 
\Big]
\nn \\
&&
+ 
   \frac{2}{{\left( 1 - y^2 \right) }^2} 
\Big[ 15\,\zeta_2 + 48\,y\,\zeta_2 + 
        18\,y^2\,\zeta_2 + 48\,y^3\,\zeta_2 + 
        15\,y^4\,\zeta_2 \nn \\ && - 2\,\zeta_3 - 
        4\,y^2\,\zeta_3 - 2\,y^4\,\zeta_3 \Big] \,H(-1;y)
\nn \\
&&
- 
   \frac{1}{{\left( 1 - y \right) }^3\,{\left( 1 + y \right) }^2} 
\Big[ -3 - 7\,y + 10\,y^2 + 10\,y^3 - 7\,y^4 - 3\,y^5 + 
        \zeta_2 \nn \\ && + 103\,y\,\zeta_2 - 148\,y^2\,\zeta_2 - 
        45\,y^4\,\zeta_2 + 25\,y^5\,\zeta_2 + 
        8\,\zeta_3 \nn \\ && - 24\,y\,\zeta_3 - 56\,y^2\,\zeta_3 - 
        56\,y^3\,\zeta_3 - 24\,y^4\,\zeta_3 + 
        8\,y^5\,\zeta_3 \Big] \,H(0;y)
\nn \\
&&
+ 
   \frac{2\,\left( 5 - 114\,y - 54\,y^2 - 114\,y^3 + 13\,y^4 \right) \,
      \zeta_2\,H(1;y)}{{\left( 1 - y^2 \right) }^2} 
\nn \\
&&
- 
   \frac{4}{{\left( 1 - y \right) }^3\,{\left( 1 + y \right) }^2} 
\Big[ -2 - 2\,y + 4\,y^2 + 4\,y^3 - 2\,y^4 - 2\,y^5 \nn \\ && - 
        5\,\zeta_2 + 13\,y\,\zeta_2 + 30\,y^2\,\zeta_2 + 
        50\,y^3\,\zeta_2 + 3\,y^4\,\zeta_2 \nn \\ && + 
        5\,y^5\,\zeta_2 \Big] \,H(-1,0;y)
\nn \\
&&
+ 
   \frac{12\,\left( 1 + y \right) \,\zeta_2\,H(0,-1;y)}{1 - y} 
\nn \\
&&
+ 
   \frac{1}{{\left( 1 - y \right) }^3\,{\left( 1 + y \right) }^2} 
\Big[ 3 + 12\,y + 36\,y^2 + 22\,y^3 - 47\,y^4 - 26\,y^5 \nn \\ && - 
        24\,\zeta_2 + 20\,y\,\zeta_2 - 
        60\,y^2\,\zeta_2 + 124\,y^3\,\zeta_2 - 
        72\,y^4\,\zeta_2 \nn \\ && + 68\,y^5\,\zeta_2 \Big] \,H(0,0;y)
\nn \\
&&
+ 
   \frac{8\,\left( -4 - 2\,y - 23\,y^2 - 7\,y^3 - 10\,y^4 + 4\,y^5 \right) \,
      \zeta_2\,H(0,1;y)}{{\left( 1 - y \right) }^3\,
      {\left( 1 + y \right) }^2} 
\nn \\
&&
-  \frac{1}{{\left( 1 - y^2 \right) }^2}
  \Big[ 15 + 42\,y - 42\,y^3 - 15\,y^4 + 40\,\zeta_2 + 
        80\,y^2\,\zeta_2 \nn \\ && + 40\,y^4\,\zeta_2 \Big] \,H(1,0;y)
\nn \\
&&
- 
   \frac{48\,{\left( 1 + y^2 \right) }^2\,\zeta_2\,H(1,1;y)}
    {{\left( 1 - y^2 \right) }^2} 
\nn \\
&&
- 
   \frac{2\,\left( 5 + 16\,y + 6\,y^2 + 16\,y^3 + 5\,y^4 \right) \,
      H(-1,0,0;y)}{{\left( 1 - y^2 \right) }^2} 
\nn \\
&&
+ 
   \frac{4\,\left( 1 + 6\,y + 2\,y^2 + 6\,y^3 + y^4 \right) \,H(0,-1,0;y)}
    {{\left( 1 - y^2 \right) }^2} 
\nn \\
&&
+ 
   \frac{\left( 1 + 31\,y - 65\,y^2 - 9\,y^3 + 10\,y^5 \right) \,H(0,0,0;y)}
    {{\left( 1 - y \right) }^3\,{\left( 1 + y \right) }^2} 
\nn \\
&&
- 
   \frac{2\,\left( 3 + 22\,y + 20\,y^2 + 22\,y^3 + 5\,y^4 \right) \,
      H(0,1,0;y)}{{\left( 1 - y^2 \right) }^2} 
\nn \\
&&
- 
   \frac{\left( 5 - 58\,y - 2\,y^2 - 58\,y^3 + 9\,y^4 \right) \,H(1,0,0;y)}
    {{\left( 1 - y^2 \right) }^2} 
\nn \\
&&
+ 
   \frac{4\,\left( 1 + y^2 \right) \,H(1,1,0;y)}{1 - y^2} + 
   \frac{8\,{\left( 1 + y^2 \right) }^2\,H(-1,0,-1,0;y)}
    {{\left( 1 - y^2 \right) }^2} 
\nn \\
&&
+ 
   \frac{8\,\left( -1 + 3\,y + 8\,y^2 + 12\,y^3 + y^4 + y^5 \right) \,
      H(-1,0,0,0;y)}{{\left( 1 - y \right) }^3\,{\left( 1 + y \right) }^2} 
\nn \\
&&
- 
   \frac{8\,{\left( 1 + y^2 \right) }^2\,H(-1,0,1,0;y)}
    {{\left( 1 - y^2 \right) }^2} - 
   \frac{4\,\left( 1 + y \right) \,H(0,-1,0,0;y)}{1 - y} 
\nn \\
&&
- 
   \frac{4\,\left( -1 + 2\,y^2 + 3\,y^4 \right) \,H(0,0,-1,0;y)}
    {{\left( 1 - y^2 \right) }^2} 
\nn \\
&&
- 
   \frac{\left( -7 + 7\,y - 16\,y^2 + 52\,y^3 - 27\,y^4 + 27\,y^5 \right) \,
      H(0,0,0,0;y)}{{\left( 1 - y \right) }^3\,{\left( 1 + y \right) }^2} 
\nn \\
&&
- 
   \frac{2\,\left( 1 + 3\,y - 10\,y^2 + 10\,y^3 - 7\,y^4 + 11\,y^5 \right) \,
      H(0,0,1,0;y)}{{\left( 1 - y \right) }^3\,{\left( 1 + y \right) }^2} 
\nn \\
&&
- 
   \frac{2\,\left( -3 - 5\,y - 36\,y^2 - 16\,y^3 - 15\,y^4 + 7\,y^5 \right) \,
      H(0,1,0,0;y)}{{\left( 1 - y \right) }^3\,{\left( 1 + y \right) }^2} 
\nn \\
&&
+ 
   \frac{4\,{\left( 1 + y^2 \right) }^2\,H(0,1,1,0;y)}
    {{\left( 1 - y^2 \right) }^2} + 
   \frac{12\,{\left( 1 + y^2 \right) }^2\,H(1,0,0,0;y)}
    {{\left( 1 - y^2 \right) }^2} 
\nn \\
&&
+ 
   \frac{8\,{\left( 1 + y^2 \right) }^2\,H(1,0,1,0;y)}
    {{\left( 1 - y^2 \right) }^2} + 
   \frac{16\,{\left( 1 + y^2 \right) }^2\,H(1,1,0,0;y)}
    {{\left( 1 - y^2 \right) }^2}
\ ;
%%%%%%%%%%%%%%%%%%%%%%%%%%%%%%%%%%%%%%%%%%%%%%%%%%%%%%%%%%%%%%%%%%%%%%%%%%%%%%%%%%%%%%%%
\eea
\bea
%%%%%%%%%%%%%%%%%%%%%%%%%%%%%%%%%%%%%%%%%%%%%%%%%%%%%%%%%%%%%%%%%%%%%%%%%%%%%%%%%%%%%%%%
%%%% C_F*C_A
%%%%%%%%%%%%%%%%%%%%%%%%%%%%%%%%%%%%%%%%%%%%%%%%%%%%%%%%%%%%%%%%%%%%%%%%%%%%%%%%%%%%%%%%
\Re \, c_{-2} &=&  
  \frac{11}{12} + \frac{11\,\left( 1 + y^2 \right) \,H(0;y)}
    {12\,\left( 1 - y^2 \right) }
\ ; \\
&& \nn \\
%%%%%%%%%%%%%%%%%%%%%%%%%%%%%%%%%%%%%%%%%%%%%%%%%%%%%%%%%%%%%%%%%%%%%%%%%%%%%%%%%%%%%%%%
\Re \, c_{-1} &=& 
  \frac{1}{36\,{\left( 1 - y^2 \right) }^2} 
  \Big[-49 + 98\,y^2 - 49\,y^4 + 18\,\zeta_2 + 
      144\,y^2\,\zeta_2 \nn \\ && - 162\,y^4\,\zeta_2 - 
      18\,\zeta_3 - 36\,y^2\,\zeta_3 - 18\,y^4\,\zeta_3 \Big]
\nn \\
&&
+ 
   \frac{\left( 1 + y^2 \right) \,
      \left( -67 + 67\,y^2 + 18\,\zeta_2 + 
        162\,y^2\,\zeta_2 \right) \,H(0;y)}{36\,
      {\left( 1 - y^2 \right) }^2} 
\nn \\
&&
+ 
   \frac{\left( 1 + y^2 \right) \,H(-1,0;y)}{1 - y^2} - 
   \frac{2\,y^2\,H(0,0;y)}{1 - y^2} - 
   \frac{\left( 1 + y^2 \right) \,H(1,0;y)}{1 - y^2} 
\nn \\
&&
+ 
   \frac{{\left( 1 + y^2 \right) }^2\,H(0,-1,0;y)}
    {{\left( 1 - y^2 \right) }^2} - 
   \frac{2\,\left( y^2 + y^4 \right) \,H(0,0,0;y)}
    {{\left( 1 - y^2 \right) }^2} 
\nn \\
&&
- 
   \frac{{\left( 1 + y^2 \right) }^2\,H(0,1,0;y)}{{\left( 1 - y^2 \right) }^2}
\ ; \\
&& \nn \\
%%%%%%%%%%%%%%%%%%%%%%%%%%%%%%%%%%%%%%%%%%%%%%%%%%%%%%%%%%%%%%%%%%%%%%%%%%%%%%%%%%%%%%%%
\Re \, c_{0} &=& 
 - \frac{1}{540\,{\left( 1 - y \right) }^3\,{\left( 1 + y \right) }^2} 
 \Big[ 4345 - 4345\,y - 8690\,y^2 + 8690\,y^3 \nn \\ && + 4345\,y^4 - 
        4345\,y^5 - 8760\,\zeta_2 + 10200\,y\,\zeta_2 - 
        19440\,\ln(2)\,y\,\zeta_2 \nn \\ && + 
        25920\,y^2\,\zeta_2 - 
        19440\,\ln(2)\,y^2\,\zeta_2 - 
        13680\,y^3\,\zeta_2 \nn \\ && + 
        19440\,\ln(2)\,y^3\,\zeta_2 - 
        17160\,y^4\,\zeta_2 + 
        19440\,\ln(2)\,y^4\,\zeta_2 \nn \\ && + 
        3480\,y^5\,\zeta_2 + 1701\,{\zeta_2}^2 - 
        1701\,y\,{\zeta_2}^2 + 10422\,y^2\,{\zeta_2}^2 \nn \\ && - 
        19494\,y^3\,{\zeta_2}^2 + 13257\,y^4\,{\zeta_2}^2 - 
        13257\,y^5\,{\zeta_2}^2 - 4410\,\zeta_3 \nn \\ && - 
        1530\,y\,\zeta_3 - 2700\,y^2\,\zeta_3 + 
        2700\,y^3\,\zeta_3 + 7110\,y^4\,\zeta_3 \nn \\ && - 
        1170\,y^5\,\zeta_3 \Big]
\nn \\
&&
- 
   \frac{1}{{\left( 1 - y^2 \right) }^2} 
\Big[ 37\,\zeta_2 - 48\,y^2\,\zeta_2 + 
        11\,y^4\,\zeta_2 - 2\,\zeta_3 - 
        4\,y^2\,\zeta_3 \nn \\ && - 2\,y^4\,\zeta_3 \Big] \,H(-1;y)
\nn \\
&&
+ 
   \frac{1}{54\,{\left( 1 - y \right) }^3\,{\left( 1 + y \right) }^2} 
 \Big[ -121 - 1133\,y + 1254\,y^2 + 1254\,y^3 \nn \\ && - 1133\,y^4 - 
        121\,y^5 + 198\,\zeta_2 + 234\,y\,\zeta_2 - 
        378\,y^2\,\zeta_2 \nn \\ && - 162\,y^3\,\zeta_2 + 
        1908\,y^4\,\zeta_2 + 1656\,y^5\,\zeta_2 \nn \\ && + 
        351\,\zeta_3 - 351\,y\,\zeta_3 + 
        594\,y^2\,\zeta_3 + 702\,y^3\,\zeta_3 - 
        405\,y^4\,\zeta_3 \nn \\ && + 405\,y^5\,\zeta_3 \Big] \,H(0;y)
\nn \\
&&
- 
   \frac{2}{3\,{\left( 1 - y^2 \right) }^2} 
\Big[ -52\,\zeta_2 + 45\,y\,\zeta_2 + 
        81\,y^2\,\zeta_2 + 45\,y^3\,\zeta_2 \nn \\ && + 
        61\,y^4\,\zeta_2 + 3\,\zeta_3 + 
        6\,y^2\,\zeta_3 + 3\,y^4\,\zeta_3 \Big] \,H(1;y)
\nn \\
&&
+ 
   \frac{2\,\left( 1 + y^2 \right) \,
      \left( -1 + y^2 - 11\,\zeta_2 + y^2\,\zeta_2 \right) \,
      H(-1,0;y)}{{\left( 1 - y^2 \right) }^2} 
\nn \\
&&
+ 
   \frac{\left( -19 + 19\,y + 46\,y^2 + 98\,y^3 - 7\,y^4 + 7\,y^5 \right) \,
      \zeta_2\,H(0,-1;y)}{{\left( 1 - y \right) }^3\,
      {\left( 1 + y \right) }^2} 
\nn \\
&&
- 
   \frac{1}{18\,{\left( 1 - y \right) }^3\,{\left( 1 + y \right) }^2} 
\Big[ 67 + 74\,y - 78\,y^2 - 24\,y^3 + 11\,y^4 - 50\,y^5 \nn \\ && - 
        18\,\zeta_2 + 18\,y\,\zeta_2 - 
        216\,y^2\,\zeta_2 + 1152\,y^3\,\zeta_2 - 
        666\,y^4\,\zeta_2 \nn \\ && + 666\,y^5\,\zeta_2 \Big] \,H(0,0;y)
\nn \\
&&
- 
   \frac{\left( -11 + 11\,y + 14\,y^2 + 106\,y^3 - 35\,y^4 + 35\,y^5 \right)
        \,\zeta_2\,H(0,1;y)}{{\left( 1 - y \right) }^3\,
      {\left( 1 + y \right) }^2} 
\nn \\
&&
+ 
   \frac{1}{18\,{\left( 1 - y^2 \right) }^2} 
\Big[ -53 - 102\,y + 102\,y^3 + 53\,y^4 + 252\,\zeta_2 + 
        360\,y^2\,\zeta_2 \nn \\ && + 108\,y^4\,\zeta_2 \Big] \,H(1,0;y)
\nn \\
&&
- 
   \frac{2\,\left( 1 + y^2 \right) \,H(-1,-1,0;y)}{1 - y^2} - 
   \frac{4\,\left( -3 + y^2 \right) \,H(-1,0,0;y)}{1 - y^2} 
\nn \\
&&
+ 
   \frac{6\,\left( 1 + y^2 \right) \,H(-1,1,0;y)}{1 - y^2} - 
   \frac{4\,y\,\left( 1 - y + 2\,y^2 \right) \,H(0,-1,0;y)}
    {{\left( 1 - y \right) }^2\,\left( 1 + y \right) } 
\nn \\
&&
- 
   \frac{\left( 11 + 2\,y - 53\,y^2 + 74\,y^3 + 62\,y^4 \right) \,H(0,0,0;y)}
    {6\,{\left( 1 - y \right) }^3\,\left( 1 + y \right) } 
\nn \\
&&
+ 
   \frac{2\,\left( -1 + 10\,y - 25\,y^2 + 34\,y^3 \right) \,H(0,1,0;y)}
    {3\,{\left( 1 - y \right) }^2\,\left( 1 + y \right) } 
\nn \\
&&
+ 
   \frac{6\,\left( 1 + y^2 \right) \,H(1,-1,0;y)}{1 - y^2} 
\nn \\
&&
+ 
   \frac{\left( -61 + 175\,y + 35\,y^2 + 79\,y^3 \right) \,H(1,0,0;y)}
    {6\,{\left( 1 - y \right) }^2\,\left( 1 + y \right) } 
\nn \\
&&
- 
   \frac{52\,\left( 1 + y^2 \right) \,H(1,1,0;y)}
    {3\,\left( 1 - y^2 \right) } - 
   \frac{4\,{\left( 1 + y^2 \right) }^2\,H(-1,0,-1,0;y)}
    {{\left( 1 - y^2 \right) }^2} 
\nn \\
&&
- 
   \frac{2\,\left( -5 - 4\,y^2 + y^4 \right) \,H(-1,0,0,0;y)}
    {{\left( 1 - y^2 \right) }^2} + 
   \frac{4\,{\left( 1 + y^2 \right) }^2\,H(-1,0,1,0;y)}
    {{\left( 1 - y^2 \right) }^2} 
\nn \\
&&
- 
   \frac{2\,{\left( 1 + y^2 \right) }^2\,H(0,-1,-1,0;y)}
    {{\left( 1 - y^2 \right) }^2} 
\nn \\
&&
- 
   \frac{2\,\left( -3 + 3\,y + 8\,y^2 + 16\,y^3 - y^4 + y^5 \right) \,
      H(0,-1,0,0;y)}{{\left( 1 - y \right) }^3\,{\left( 1 + y \right) }^2} 
\nn \\
&&
+ 
   \frac{6\,{\left( 1 + y^2 \right) }^2\,H(0,-1,1,0;y)}
    {{\left( 1 - y^2 \right) }^2} + 
   \frac{2\,\left( 1 + 8\,y^2 + 7\,y^4 \right) \,H(0,0,-1,0;y)}
    {{\left( 1 - y^2 \right) }^2} 
\nn \\
&&
+ 
   \frac{3\,y^2\,\left( -1 + 7\,y - 4\,y^2 + 4\,y^3 \right) \,H(0,0,0,0;y)}
    {{\left( 1 - y \right) }^3\,{\left( 1 + y \right) }^2} 
\nn \\
&&
+ 
   \frac{2\,\left( -1 + y - 22\,y^2 + 2\,y^3 - 11\,y^4 + 11\,y^5 \right) \,
      H(0,0,1,0;y)}{{\left( 1 - y \right) }^3\,{\left( 1 + y \right) }^2} 
\nn \\
&&
+ 
   \frac{6\,{\left( 1 + y^2 \right) }^2\,H(0,1,-1,0;y)}
    {{\left( 1 - y^2 \right) }^2} 
\nn \\
&&
+ 
   \frac{2\,\left( -1 + y - y^2 + 15\,y^3 - 7\,y^4 + 7\,y^5 \right) \,
      H(0,1,0,0;y)}{{\left( 1 - y \right) }^3\,{\left( 1 + y \right) }^2} 
\nn \\
&&
- 
   \frac{10\,{\left( 1 + y^2 \right) }^2\,H(0,1,1,0;y)}
    {{\left( 1 - y^2 \right) }^2} + 
   \frac{4\,{\left( 1 + y^2 \right) }^2\,H(1,0,-1,0;y)}
    {{\left( 1 - y^2 \right) }^2} 
\nn \\
&&
- 
   \frac{2\,\left( 3 + 4\,y^2 + y^4 \right) \,H(1,0,0,0;y)}
    {{\left( 1 - y^2 \right) }^2} - 
   \frac{4\,{\left( 1 + y^2 \right) }^2\,H(1,0,1,0;y)}
    {{\left( 1 - y^2 \right) }^2}
\ ; 
%%%%%%%%%%%%%%%%%%%%%%%%%%%%%%%%%%%%%%%%%%%%%%%%%%%%%%%%%%%%%%%%%%%%%%%%%%%%%%%%%%%%%%%%
\eea
\bea
%%%%%%%%%%%%%%%%%%%%%%%%%%%%%%%%%%%%%%%%%%%%%%%%%%%%%%%%%%%%%%%%%%%%%%%%%%%%%%%%%%%%%%%%
%%%% C_F*T_R*N_f
%%%%%%%%%%%%%%%%%%%%%%%%%%%%%%%%%%%%%%%%%%%%%%%%%%%%%%%%%%%%%%%%%%%%%%%%%%%%%%%%%%%%%%%%
\Re \, d_{-2} &=&  
  - \frac{1}{3}  
  -  \frac{\left( 1 + y^2 \right) \,H(0;y)}{3\,\left( 1 - y^2 \right) }
\ ; \\
%%%%%%%%%%%%%%%%%%%%%%%%%%%%%%%%%%%%%%%%%%%%%%%%%%%%%%%%%%%%%%%%%%%%%%%%%%%%%%%%%%%%%%%%
\Re \, d_{-1} &=& 
  \frac{5}{9} + \frac{5\,\left( 1 + y^2 \right) \,H(0;y)}
    {9\,\left( 1 - y^2 \right) }
\ ; \\
%%%%%%%%%%%%%%%%%%%%%%%%%%%%%%%%%%%%%%%%%%%%%%%%%%%%%%%%%%%%%%%%%%%%%%%%%%%%%%%%%%%%%%%%
\Re \, d_{0} &=& 
- \frac{1}{27\,\left( 1 - y^2 \right) } 
  \Big[ -49 + 49\,y^2 + 84\,\zeta_2 + 288\,y\,\zeta_2 + 
        156\,y^2\,\zeta_2 \nn \\ && + 36\,\zeta_3 + 
        36\,y^2\,\zeta_3 \Big]
\nn \\
&&
- 
   \frac{4\,\left( -7 - 48\,y - 7\,y^2 + 9\,\zeta_2 + 
        9\,y^2\,\zeta_2 \right) \,H(0;y)}{27\,\left( 1 - y^2 \right) }
\nn \\
&&
    - \frac{16\,\left( 1 + y^2 \right) \,\zeta_2\,H(1;y)}
    {3\,\left( 1 - y^2 \right) } + 
   \frac{2\,\left( 5 + 12\,y + 5\,y^2 \right) \,H(0,0;y)}
    {9\,\left( 1 - y^2 \right) } 
\nn \\
&&
+ 
   \frac{4\,\left( 5 + 12\,y + 5\,y^2 \right) \,H(1,0;y)}
    {9\,\left( 1 - y^2 \right) } + 
   \frac{2\,\left( 1 + y^2 \right) \,H(0,0,0;y)}
    {3\,\left( 1 - y^2 \right) } 
\nn \\
&&
+ 
   \frac{4\,\left( 1 + y^2 \right) \,H(0,1,0;y)}
    {3\,\left( 1 - y^2 \right) } + 
   \frac{4\,\left( 1 + y^2 \right) \,H(1,0,0;y)}
    {3\,\left( 1 - y^2 \right) } 
\nn \\
&&
+ 
   \frac{8\,\left( 1 + y^2 \right) \,H(1,1,0;y)}{3\,\left( 1 - y^2 \right) }
\ ;
%%%%%%%%%%%%%%%%%%%%%%%%%%%%%%%%%%%%%%%%%%%%%%%%%%%%%%%%%%%%%%%%%%%%%%%%%%%%%%%%%%%%%%%%
\eea
\bea
%%%%%%%%%%%%%%%%%%%%%%%%%%%%%%%%%%%%%%%%%%%%%%%%%%%%%%%%%%%%%%%%%%%%%%%%%%%%%%%%%%%%%%%%
%%%% C_F*T_R
%%%%%%%%%%%%%%%%%%%%%%%%%%%%%%%%%%%%%%%%%%%%%%%%%%%%%%%%%%%%%%%%%%%%%%%%%%%%%%%%%%%%%%%%
\Re \, e_{-2} &=& 0 
\ ; \\
%%%%%%%%%%%%%%%%%%%%%%%%%%%%%%%%%%%%%%%%%%%%%%%%%%%%%%%%%%%%%%%%%%%%%%%%%%%%%%%%%%%%%%%%
\Re \, e_{-1} &=& 0
\ ; \\
%%%%%%%%%%%%%%%%%%%%%%%%%%%%%%%%%%%%%%%%%%%%%%%%%%%%%%%%%%%%%%%%%%%%%%%%%%%%%%%%%%%%%%%%
\Re \, e_{0} &=& 
-  \frac{1}{135\,{\left( 1 - y^2 \right) }^3} 
\Big[ -2035 - 600\,y + 4905\,y^2 - 4905\,y^4 + 600\,y^5 \nn \\ && + 
        2035\,y^6 + 1260\,\zeta_2 - 1080\,y\,\zeta_2 - 
        19980\,y^2\,\zeta_2 + 18360\,y^3\,\zeta_2 \nn \\ && + 
        19980\,y^4\,\zeta_2 - 17280\,y^5\,\zeta_2 - 
        1260\,y^6\,\zeta_2 + 783\,y\,{\zeta_2}^2 \nn \\ && - 
        3132\,y^2\,{\zeta_2}^2 + 32346\,y^3\,{\zeta_2}^2 - 
        3132\,y^4\,{\zeta_2}^2 + 783\,y^5\,{\zeta_2}^2 \nn \\ && + 
        8640\,y^2\,\zeta_3 - 8640\,y^4\,\zeta_3 \Big]
\nn \\
&&
+ 
   \frac{384\,y^2\,\zeta_2\,H(-1;y)}{{\left( 1 - y^2 \right) }^2} 
\nn \\
&&
- 
   \frac{4}{27\,{\left( 1 - y^2 \right) }^3} 
\Big[ -14 + 72\,y - 106\,y^2 - 384\,y^3 - 106\,y^4 + 72\,y^5 \nn \\ && - 
        14\,y^6 + 9\,\zeta_2 - 657\,y^2\,\zeta_2 + 
        639\,y^4\,\zeta_2 + 9\,y^6\,\zeta_2 - 
        81\,y\,\zeta_3 \nn \\ && + 324\,y^2\,\zeta_3 + 
        378\,y^3\,\zeta_3 + 324\,y^4\,\zeta_3 - 
        81\,y^5\,\zeta_3 \Big] \,H(0;y)
\nn \\
&&
- 
   \frac{16\,y\,\left( -1 + 2\,y - y^2 + \zeta_2 - 
        6\,y\,\zeta_2 + y^2\,\zeta_2 \right) \,H(-1,0;y)}{
      {\left( 1 - y \right) }^3\,\left( 1 + y \right) } 
\nn \\
&&
- 
   \frac{8\,y}{{\left( 1 - y^2 \right) }^3} 
\Big[ -1 + 2\,y^2 - y^4 + 3\,\zeta_2 - 
        12\,y\,\zeta_2 - 78\,y^2\,\zeta_2  \nn \\ && - 
        12\,y^3\,\zeta_2 + 3\,y^4\,\zeta_2 \Big] \,H(0,-1;y)
\nn \\
&&
+ 
   \frac{2}{9\,{\left( 1 - y \right) }^4\,{\left( 1 + y \right) }^3} 
\Big[ 5 - 23\,y - 101\,y^2 + 531\,y^3 + 27\,y^4 - 353\,y^5 \nn \\ && + 
        229\,y^6 + 5\,y^7 - 9\,y\,\zeta_2 + 45\,y^2\,\zeta_2 + 
        198\,y^3\,\zeta_2 - 198\,y^4\,\zeta_2 \nn \\ && - 
        45\,y^5\,\zeta_2 + 9\,y^6\,\zeta_2 \Big] H(0,0;y)
\nn \\
&&
- 
   \frac{32\,y\,H(1,0;y)}{1 - y^2} - 
   \frac{128\,y^2\,H(-1,0,0;y)}{{\left( 1 - y^2 \right) }^2} + 
   \frac{64\,y^2\,H(0,-1,0;y)}{{\left( 1 - y^2 \right) }^2} 
\nn \\
&&
- 
   \frac{2\,\left( -1 + 48\,y^2 + y^4 \right) \,H(0,0,0;y)}
    {3\,{\left( 1 - y^2 \right) }^2} - 
   \frac{128\,y^2\,H(0,1,0;y)}{{\left( 1 - y^2 \right) }^2} 
\nn \\
&&
+ 
   \frac{8\,y\,\left( 1 - 6\,y + y^2 \right) \,H(-1,0,0,0;y)}
    {{\left( 1 - y \right) }^3\,\left( 1 + y \right) } 
\nn \\
&&
+ 
   \frac{8\,y\,\left( 1 - 4\,y - 26\,y^2 - 4\,y^3 + y^4 \right) \,
      H(0,-1,0,0;y)}{{\left( 1 - y^2 \right) }^3} 
\nn \\
&&
- 
   \frac{8\,y\,\left( 1 - 6\,y + y^2 \right) \,H(0,0,-1,0;y)}
    {{\left( 1 - y \right) }^3\,\left( 1 + y \right) } + 
   \frac{12\,\left( -1 + y \right) \,y\,H(0,0,0,-1;y)}
    {{\left( 1 + y \right) }^3} 
\nn \\
&&
- 
   \frac{2\,\left( -1 + y \right) \,y\,H(0,0,0,0;y)}
    {{\left( 1 + y \right) }^3} - 
   \frac{256\,y^3\,H(0,0,1,0;y)}{{\left( 1 - y^2 \right) }^3}
\ ;
%%%%%%%%%%%%%%%%%%%%%%%%%%%%%%%%%%%%%%%%%%%%%%%%%%%%%%%%%%%%%%%%%%%%%%%%%%%%%%%%%%%%%%%%
\eea

\bea
%%%%%%%%%%%%%%%%%%%%%%%%%%%%%%%%%%%%%%%%%%%%%%%%%%%%%%%%%%%%%%%%%%%%%%%%%%%%%%%%%%%%%%%%
%%%% C_F^2
%%%%%%%%%%%%%%%%%%%%%%%%%%%%%%%%%%%%%%%%%%%%%%%%%%%%%%%%%%%%%%%%%%%%%%%%%%%%%%%%%%%%%%%%
\Im \, b_{-2} &=&  
  \frac{1 + y^2}{1 - y^2}  + 
   \frac{{\left( 1 + y^2 \right) }^2\,H(0;y)}{{\left( 1 - y^2 \right) }^2}
\ ; \\
&& \nn \\
%%%%%%%%%%%%%%%%%%%%%%%%%%%%%%%%%%%%%%%%%%%%%%%%%%%%%%%%%%%%%%%%%%%%%%%%%%%%%%%%%%%%%%%%
\Im \, b_{-1} &=& 
  \frac{1 + 4\,y - 4\,y^3 - y^4 - 4\,\zeta_2 - 
      8\,y^2\,\zeta_2 - 4\,y^4\,\zeta_2}{{\left( 1 - y^2 \right)
         }^2} 
\nn \\
&&
- \frac{\left( -1 - 8\,y - 8\,y^3 + y^4 \right) \,H(0;y)}
    {{\left( 1 - y^2 \right) }^2} + 
   \frac{2\,\left( 1 + y^2 \right) \,H(1;y)}{1 - y^2} 
\nn \\
&&
+ 
   \frac{3\,{\left( 1 + y^2 \right) }^2\,H(0,0;y)}
    {{\left( 1 - y^2 \right) }^2} + 
   \frac{2\,{\left( 1 + y^2 \right) }^2\,H(0,1;y)}
    {{\left( 1 - y^2 \right) }^2} 
\nn \\
&&
+ 
   \frac{4\,{\left( 1 + y^2 \right) }^2\,H(1,0;y)}
    {{\left( 1 - y^2 \right) }^2}
\ ; \\
&& \nn \\
%%%%%%%%%%%%%%%%%%%%%%%%%%%%%%%%%%%%%%%%%%%%%%%%%%%%%%%%%%%%%%%%%%%%%%%%%%%%%%%%%%%%%%%%
\Im \, b_{0} &=& 
  - \frac{1}{{\left( 1 - y \right) }^3\,{\left( 1 + y \right) }^2} 
\Big[ -3 - 7\,y + 10\,y^2 + 10\,y^3 - 7\,y^4 - 3\,y^5 - 
        \zeta_2 \nn \\ && + 41\,y\,\zeta_2 - 18\,y^2\,\zeta_2 + 
        18\,y^3\,\zeta_2 - 45\,y^4\,\zeta_2 + 
        5\,y^5\,\zeta_2 + 8\,\zeta_3 \nn \\ && - 24\,y\,\zeta_3 - 
        56\,y^2\,\zeta_3 - 56\,y^3\,\zeta_3 - 
        24\,y^4\,\zeta_3 + 8\,y^5\,\zeta_3 \Big]
\nn \\
&&
+ 
   \frac{4\,\left( 2 + 4\,y - 4\,y^3 - 2\,y^4 + \zeta_2 + 
        2\,y^2\,\zeta_2 + y^4\,\zeta_2 \right) \,H(-1;y)}{{\left
         ( 1 - y^2 \right) }^2} 
\nn \\
&&
+ 
   \frac{1}{{\left( 1 - y \right) }^3\,{\left( 1 + y \right) }^2} 
\Big[ 3 + 12\,y + 36\,y^2 + 22\,y^3 - 47\,y^4 - 26\,y^5 \nn \\ && - 
        10\,\zeta_2 + 6\,y\,\zeta_2 - 28\,y^2\,\zeta_2 + 
        20\,y^3\,\zeta_2 - 18\,y^4\,\zeta_2 \nn \\ && + 
        14\,y^5\,\zeta_2 \Big] \,H(0;y)
\nn \\
&&
- 
   \frac{\left( 15 + 42\,y - 42\,y^3 - 15\,y^4 + 16\,\zeta_2 + 
        32\,y^2\,\zeta_2 + 16\,y^4\,\zeta_2 \right) \,H(1;y)}
      {{\left( 1 - y^2 \right) }^2} 
\nn \\
&&
- 
   \frac{2\,\left( 5 + 16\,y + 6\,y^2 + 16\,y^3 + 5\,y^4 \right) \,H(-1,0;y)}
    {{\left( 1 - y^2 \right) }^2} 
\nn \\
&&
+ 
   \frac{4\,\left( 1 + 6\,y + 2\,y^2 + 6\,y^3 + y^4 \right) \,H(0,-1;y)}
    {{\left( 1 - y^2 \right) }^2} 
\nn \\
&&
+ 
   \frac{\left( 1 + 31\,y - 65\,y^2 - 9\,y^3 + 10\,y^5 \right) \,H(0,0;y)}
    {{\left( 1 - y \right) }^3\,{\left( 1 + y \right) }^2} 
\nn \\
&&
- 
   \frac{2\,\left( 3 + 22\,y + 20\,y^2 + 22\,y^3 + 5\,y^4 \right) \,H(0,1;y)}
    {{\left( 1 - y^2 \right) }^2} 
\nn \\
&&
- 
   \frac{\left( 5 - 58\,y - 2\,y^2 - 58\,y^3 + 9\,y^4 \right) \,H(1,0;y)}
    {{\left( 1 - y^2 \right) }^2} 
\nn \\
&&
+ 
   \frac{4\,\left( 1 + y^2 \right) \,H(1,1;y)}{1 - y^2} + 
   \frac{8\,{\left( 1 + y^2 \right) }^2\,H(-1,0,-1;y)}
    {{\left( 1 - y^2 \right) }^2} 
\nn \\
&&
+ 
   \frac{8\,\left( -1 + 3\,y + 8\,y^2 + 12\,y^3 + y^4 + y^5 \right) \,
      H(-1,0,0;y)}{{\left( 1 - y \right) }^3\,{\left( 1 + y \right) }^2} 
\nn \\
&&
- 
   \frac{8\,{\left( 1 + y^2 \right) }^2\,H(-1,0,1;y)}
    {{\left( 1 - y^2 \right) }^2} - 
   \frac{4\,\left( 1 + y \right) \,H(0,-1,0;y)}{1 - y} 
\nn \\
&&
- 
   \frac{4\,\left( -1 + 2\,y^2 + 3\,y^4 \right) \,H(0,0,-1;y)}
    {{\left( 1 - y^2 \right) }^2} 
\nn \\
&&
- 
   \frac{\left( -7 + 7\,y - 16\,y^2 + 52\,y^3 - 27\,y^4 + 27\,y^5 \right) \,
      H(0,0,0;y)}{{\left( 1 - y \right) }^3\,{\left( 1 + y \right) }^2} 
\nn \\
&&
- 
   \frac{2\,\left( 1 + 3\,y - 10\,y^2 + 10\,y^3 - 7\,y^4 + 11\,y^5 \right) \,
      H(0,0,1;y)}{{\left( 1 - y \right) }^3\,{\left( 1 + y \right) }^2} 
\nn \\
&&
- 
   \frac{2\,\left( -3 - 5\,y - 36\,y^2 - 16\,y^3 - 15\,y^4 + 7\,y^5 \right) \,
      H(0,1,0;y)}{{\left( 1 - y \right) }^3\,{\left( 1 + y \right) }^2} 
\nn \\
&&
+ 
   \frac{4\,{\left( 1 + y^2 \right) }^2\,H(0,1,1;y)}
    {{\left( 1 - y^2 \right) }^2} + 
   \frac{12\,{\left( 1 + y^2 \right) }^2\,H(1,0,0;y)}
    {{\left( 1 - y^2 \right) }^2} 
\nn \\
&&
+ 
   \frac{8\,{\left( 1 + y^2 \right) }^2\,H(1,0,1;y)}
    {{\left( 1 - y^2 \right) }^2} + 
   \frac{16\,{\left( 1 + y^2 \right) }^2\,H(1,1,0;y)}
    {{\left( 1 - y^2 \right) }^2}
\ ;
%%%%%%%%%%%%%%%%%%%%%%%%%%%%%%%%%%%%%%%%%%%%%%%%%%%%%%%%%%%%%%%%%%%%%%%%%%%%%%%%%%%%%%%%
\eea
\bea
%%%%%%%%%%%%%%%%%%%%%%%%%%%%%%%%%%%%%%%%%%%%%%%%%%%%%%%%%%%%%%%%%%%%%%%%%%%%%%%%%%%%%%%%
%%%% C_F*C_A
%%%%%%%%%%%%%%%%%%%%%%%%%%%%%%%%%%%%%%%%%%%%%%%%%%%%%%%%%%%%%%%%%%%%%%%%%%%%%%%%%%%%%%%%
\Im \, c_{-2} &=&  
  \frac{11\,\left( 1 + y^2 \right) }{12\,\left( 1 - y^2 \right) }
\ ; \\
&& \nn \\
%%%%%%%%%%%%%%%%%%%%%%%%%%%%%%%%%%%%%%%%%%%%%%%%%%%%%%%%%%%%%%%%%%%%%%%%%%%%%%%%%%%%%%%%
\Im \, c_{-1} &=& 
  \frac{\left( 1 + y^2 \right) \,
      \left( -67 + 67\,y^2 + 18\,\zeta_2 + 18\,y^2\,\zeta_2
        \right) }{36\,{\left( 1 - y^2 \right) }^2} 
\nn \\
&&
+ 
   \frac{\left( 1 + y^2 \right) \,H(-1;y)}{1 - y^2} - 
   \frac{2\,y^2\,H(0;y)}{1 - y^2} - 
   \frac{\left( 1 + y^2 \right) \,H(1;y)}{1 - y^2} 
\nn \\
&&
+ 
   \frac{{\left( 1 + y^2 \right) }^2\,H(0,-1;y)}
    {{\left( 1 - y^2 \right) }^2} - 
   \frac{2\,y^2\,\left( 1 + y^2 \right) \,H(0,0;y)}
    {{\left( 1 - y^2 \right) }^2} 
\nn \\
&&
- 
   \frac{{\left( 1 + y^2 \right) }^2\,H(0,1;y)}{{\left( 1 - y^2 \right) }^2}
\ ; \\
&& \nn \\
%%%%%%%%%%%%%%%%%%%%%%%%%%%%%%%%%%%%%%%%%%%%%%%%%%%%%%%%%%%%%%%%%%%%%%%%%%%%%%%%%%%%%%%%
\Im \, c_{0} &=& 
  -\frac{1}{54\,{\left( 1 - y \right) }^3\,{\left( 1 + y \right) }^2}
  \Big[ 121 + 1133\,y - 1254\,y^2 - 1254\,y^3 + 1133\,y^4 \nn \\ && + 
        121\,y^5 - 540\,y^2\,\zeta_2 + 540\,y^3\,\zeta_2 + 
        540\,y^4\,\zeta_2 - 540\,y^5\,\zeta_2 \nn \\ && - 
        351\,\zeta_3 + 351\,y\,\zeta_3 - 
        594\,y^2\,\zeta_3 - 702\,y^3\,\zeta_3 + 
        405\,y^4\,\zeta_3 \nn \\ && - 405\,y^5\,\zeta_3 \Big]
\nn \\
&&
- 
   \frac{2\,\left( 1 + y^2 \right) \,
      \left( 1 - y^2 + \zeta_2 + y^2\,\zeta_2 \right) \,H(-1;y)}
      {{\left( 1 - y^2 \right) }^2} 
\nn \\
&&
- 
   \frac{1}{18\,{\left( 1 - y \right) }^3\,{\left( 1 + y \right) }^2} 
\Big[ 67 + 74\,y - 78\,y^2 - 24\,y^3 + 11\,y^4 - 50\,y^5 \nn \\ && - 
        18\,\zeta_2 + 18\,y\,\zeta_2 - 
        108\,y^2\,\zeta_2 + 396\,y^3\,\zeta_2 - 
        234\,y^4\,\zeta_2 \nn \\ && + 234\,y^5\,\zeta_2 \Big] \,H(0;y)
\nn \\
&&
+ 
   \frac{1}{18\,{\left( 1 - y^2 \right) }^2} 
\Big[ -53 - 102\,y + 102\,y^3 + 53\,y^4 + 36\,\zeta_2 \nn \\ && + 
        72\,y^2\,\zeta_2 + 36\,y^4\,\zeta_2 \Big] \,H(1;y)
\nn \\
&&
- 
   \frac{2\,\left( 1 + y^2 \right) \,H(-1,-1;y)}{1 - y^2} - 
   \frac{4\,\left( -3 + y^2 \right) \,H(-1,0;y)}{1 - y^2} 
\nn \\
&&
+ 
   \frac{6\,\left( 1 + y^2 \right) \,H(-1,1;y)}{1 - y^2} - 
   \frac{4\,y\,\left( 1 - y + 2\,y^2 \right) \,H(0,-1;y)}
    {{\left( 1 - y \right) }^2\,\left( 1 + y \right) } 
\nn \\
&&
- 
   \frac{\left( 11 + 2\,y - 53\,y^2 + 74\,y^3 + 62\,y^4 \right) \,H(0,0;y)}
    {6\,{\left( 1 - y \right) }^3\,\left( 1 + y \right) } 
\nn \\
&&
+ 
   \frac{2\,\left( -1 + 10\,y - 25\,y^2 + 34\,y^3 \right) \,H(0,1;y)}
    {3\,{\left( 1 - y \right) }^2\,\left( 1 + y \right) } 
+ 
   \frac{6\,\left( 1 + y^2 \right) \,H(1,-1;y)}{1 - y^2} 
\nn \\
&&
+ 
   \frac{\left( -61 + 175\,y + 35\,y^2 + 79\,y^3 \right) \,H(1,0;y)}
    {6\,{\left( 1 - y \right) }^2\,\left( 1 + y \right) } - 
   \frac{52\,\left( 1 + y^2 \right) \,H(1,1;y)}{3\,\left( 1 - y^2 \right) } 
\nn \\
&&
- 
   \frac{4\,{\left( 1 + y^2 \right) }^2\,H(-1,0,-1;y)}
    {{\left( 1 - y^2 \right) }^2} - 
   \frac{2\,\left( -5 - 4\,y^2 + y^4 \right) \,H(-1,0,0;y)}
    {{\left( 1 - y^2 \right) }^2} 
\nn \\
&&
+ 
   \frac{4\,{\left( 1 + y^2 \right) }^2\,H(-1,0,1;y)}
    {{\left( 1 - y^2 \right) }^2} - 
   \frac{2\,{\left( 1 + y^2 \right) }^2\,H(0,-1,-1;y)}
    {{\left( 1 - y^2 \right) }^2} 
\nn \\
&&
- 
   \frac{2\,\left( -3 + 3\,y + 8\,y^2 + 16\,y^3 - y^4 + y^5 \right) \,
      H(0,-1,0;y)}{{\left( 1 - y \right) }^3\,{\left( 1 + y \right) }^2} 
\nn \\
&&
+ 
   \frac{6\,{\left( 1 + y^2 \right) }^2\,H(0,-1,1;y)}
    {{\left( 1 - y^2 \right) }^2} + 
   \frac{2\,\left( 1 + 8\,y^2 + 7\,y^4 \right) \,H(0,0,-1;y)}
    {{\left( 1 - y^2 \right) }^2} 
\nn \\
&&
+ 
   \frac{3\,y^2\,\left( -1 + 7\,y - 4\,y^2 + 4\,y^3 \right) \,H(0,0,0;y)}
    {{\left( 1 - y \right) }^3\,{\left( 1 + y \right) }^2} 
\nn \\
&&
+ 
   \frac{2\,\left( -1 + y - 22\,y^2 + 2\,y^3 - 11\,y^4 + 11\,y^5 \right) \,
      H(0,0,1;y)}{{\left( 1 - y \right) }^3\,{\left( 1 + y \right) }^2} 
\nn \\
&&
+ 
   \frac{6\,{\left( 1 + y^2 \right) }^2\,H(0,1,-1;y)}
    {{\left( 1 - y^2 \right) }^2} 
\nn \\
&&
+ 
   \frac{2\,\left( -1 + y - y^2 + 15\,y^3 - 7\,y^4 + 7\,y^5 \right) \,
      H(0,1,0;y)}{{\left( 1 - y \right) }^3\,{\left( 1 + y \right) }^2} 
\nn \\
&&
- 
   \frac{10\,{\left( 1 + y^2 \right) }^2\,H(0,1,1;y)}
    {{\left( 1 - y^2 \right) }^2} + 
   \frac{4\,{\left( 1 + y^2 \right) }^2\,H(1,0,-1;y)}
    {{\left( 1 - y^2 \right) }^2} 
\nn \\
&&
- 
   \frac{2\,\left( 3 + 4\,y^2 + y^4 \right) \,H(1,0,0;y)}
    {{\left( 1 - y^2 \right) }^2} - 
   \frac{4\,{\left( 1 + y^2 \right) }^2\,H(1,0,1;y)}
    {{\left( 1 - y^2 \right) }^2}
\ ; 
%%%%%%%%%%%%%%%%%%%%%%%%%%%%%%%%%%%%%%%%%%%%%%%%%%%%%%%%%%%%%%%%%%%%%%%%%%%%%%%%%%%%%%%%
\eea
\bea
%%%%%%%%%%%%%%%%%%%%%%%%%%%%%%%%%%%%%%%%%%%%%%%%%%%%%%%%%%%%%%%%%%%%%%%%%%%%%%%%%%%%%%%%
%%%% C_F*T_R*N_f
%%%%%%%%%%%%%%%%%%%%%%%%%%%%%%%%%%%%%%%%%%%%%%%%%%%%%%%%%%%%%%%%%%%%%%%%%%%%%%%%%%%%%%%%
\Im \, d_{-2} &=&  
- \frac{ 1 + y^2 }{3\,\left( 1 - y^2 \right) }
\ ; \\
&& \nn \\
%%%%%%%%%%%%%%%%%%%%%%%%%%%%%%%%%%%%%%%%%%%%%%%%%%%%%%%%%%%%%%%%%%%%%%%%%%%%%%%%%%%%%%%%
\Im \, d_{-1} &=& 
  \frac{5\,\left( 1 + y^2 \right) }{9\,\left( 1 - y^2 \right) }
\ ; \\
&& \nn \\
%%%%%%%%%%%%%%%%%%%%%%%%%%%%%%%%%%%%%%%%%%%%%%%%%%%%%%%%%%%%%%%%%%%%%%%%%%%%%%%%%%%%%%%%
\Im \, d_{0} &=& 
  \frac{4\,\left( 7 + 48\,y + 7\,y^2 \right) }{27\,\left( 1 - y^2 \right) } + 
   \frac{2\,\left( 5 + 12\,y + 5\,y^2 \right) \,H(0;y)}
    {9\,\left( 1 - y^2 \right) } 
\nn \\
&&
+ 
   \frac{4\,\left( 5 + 12\,y + 5\,y^2 \right) \,H(1;y)}
    {9\,\left( 1 - y^2 \right) } + 
   \frac{2\,\left( 1 + y^2 \right) \,H(0,0;y)}{3\,\left( 1 - y^2 \right) } 
\nn \\
&&
+ 
   \frac{4\,\left( 1 + y^2 \right) \,H(0,1;y)}{3\,\left( 1 - y^2 \right) } + 
   \frac{4\,\left( 1 + y^2 \right) \,H(1,0;y)}{3\,\left( 1 - y^2 \right) } 
\nn \\
&&
+ 
   \frac{8\,\left( 1 + y^2 \right) \,H(1,1;y)}{3\,\left( 1 - y^2 \right) }
\ ;
%%%%%%%%%%%%%%%%%%%%%%%%%%%%%%%%%%%%%%%%%%%%%%%%%%%%%%%%%%%%%%%%%%%%%%%%%%%%%%%%%%%%%%%%
\eea
\bea
%%%%%%%%%%%%%%%%%%%%%%%%%%%%%%%%%%%%%%%%%%%%%%%%%%%%%%%%%%%%%%%%%%%%%%%%%%%%%%%%%%%%%%%%
%%%% C_F*T_R
%%%%%%%%%%%%%%%%%%%%%%%%%%%%%%%%%%%%%%%%%%%%%%%%%%%%%%%%%%%%%%%%%%%%%%%%%%%%%%%%%%%%%%%%
\Im \, e_{-2} &=& 0 
\ ; \\
%%%%%%%%%%%%%%%%%%%%%%%%%%%%%%%%%%%%%%%%%%%%%%%%%%%%%%%%%%%%%%%%%%%%%%%%%%%%%%%%%%%%%%%%
\Im \, e_{-1} &=& 0
\ ; \\
%%%%%%%%%%%%%%%%%%%%%%%%%%%%%%%%%%%%%%%%%%%%%%%%%%%%%%%%%%%%%%%%%%%%%%%%%%%%%%%%%%%%%%%%
\Im \, e_{0} &=& 
  \frac{4}{27\,{\left( 1 - y^2 \right) }^3} 
\Big[ 14 - 72\,y + 106\,y^2 + 384\,y^3 + 106\,y^4 - 72\,y^5 + 
        14\,y^6 \nn \\ && + 216\,y^2\,\zeta_2 - 216\,y^4\,\zeta_2 + 
        81\,y\,\zeta_3 - 324\,y^2\,\zeta_3 - 
        378\,y^3\,\zeta_3 \nn \\ && - 324\,y^4\,\zeta_3 + 
        81\,y^5\,\zeta_3 \Big]
\nn \\
&&
+ 
   \frac{16\,y\,H(-1;y)}{1 - y^2} 
\nn \\
&&
+ 
   \frac{2}{9\,{\left( 1 - y \right) }^4\,{\left( 1 + y \right) }^3} 
\Big[ 5 - 23\,y - 101\,y^2 + 531\,y^3 + 27\,y^4 - 353\,y^5 \nn \\ && + 
        229\,y^6 + 5\,y^7 + 9\,y\,\zeta_2 - 45\,y^2\,\zeta_2 + 
        378\,y^3\,\zeta_2 - 378\,y^4\,\zeta_2 \nn \\ && + 
        45\,y^5\,\zeta_2 - 9\,y^6\,\zeta_2 \Big] \,H(0;y)
\nn \\
&&
- 
   \frac{32\,y\,H(1;y)}{1 - y^2} - 
   \frac{128\,y^2\,H(-1,0;y)}{{\left( 1 - y^2 \right) }^2} + 
   \frac{64\,y^2\,H(0,-1;y)}{{\left( 1 - y^2 \right) }^2} 
\nn \\
&&
- 
   \frac{2\,\left( -1 + 48\,y^2 + y^4 \right) \,H(0,0;y)}
    {3\,{\left( 1 - y^2 \right) }^2} - 
   \frac{128\,y^2\,H(0,1;y)}{{\left( 1 - y^2 \right) }^2} 
\nn \\
&&
+ 
   \frac{8\,y\,\left( 1 - 6\,y + y^2 \right) \,H(-1,0,0;y)}
    {{\left( 1 - y \right) }^3\,\left( 1 + y \right) } 
\nn \\
&&
+ 
   \frac{8\,y\,\left( 1 - 4\,y - 26\,y^2 - 4\,y^3 + y^4 \right) \,H(0,-1,0;y)}
    {{\left( 1 - y^2 \right) }^3} 
\nn \\
&&
- 
   \frac{8\,y\,\left( 1 - 6\,y + y^2 \right) \,H(0,0,-1;y)}
    {{\left( 1 - y \right) }^3\,\left( 1 + y \right) } 
\nn \\
&&
- 
   \frac{2\,\left( -1 + y \right) \,y\,H(0,0,0;y)}
    {{\left( 1 + y \right) }^3} - 
   \frac{256\,y^3\,H(0,0,1;y)}{{\left( 1 - y^2 \right) }^3}
\ ;
%%%%%%%%%%%%%%%%%%%%%%%%%%%%%%%%%%%%%%%%%%%%%%%%%%%%%%%%%%%%%%%%%%%%%%%%%%%%%%%%%%%%%%%%
\eea

\subsubsection{The Pseudoscalar Case}

\bea
%%%%%%%%%%%%%%%%%%%%%%%%%%%%%%%%%%%%%%%%%%%%%%%%%%%%%%%%%%%%%%%%%%%%%%%%%%%%%%%%%%%%%%%%
%%%% C_F^2
%%%%%%%%%%%%%%%%%%%%%%%%%%%%%%%%%%%%%%%%%%%%%%%%%%%%%%%%%%%%%%%%%%%%%%%%%%%%%%%%%%%%%%%%
\Re \, \bar b_{-2} &=&  \Re \, b_{-2}
\ ; \\
&& \nn \\
%%%%%%%%%%%%%%%%%%%%%%%%%%%%%%%%%%%%%%%%%%%%%%%%%%%%%%%%%%%%%%%%%%%%%%%%%%%%%%%%%%%%%%%%
\Re \, \bar b_{-1} &=& 
  \frac{\left( 1 + y^2 \right) \,H(0,0;y)}{1 - y^2} + 
   \frac{2\,\left( 1 + y^2 \right) \,H(1,0;y)}{1 - y^2} \nn \\ && + 
   \frac{3\,{\left( 1 + y^2 \right) }^2\,H(0,0,0;y)}
    {{\left( 1 - y^2 \right) }^2} + 
   \frac{2\,{\left( 1 + y^2 \right) }^2\,H(0,1,0;y)}
    {{\left( 1 - y^2 \right) }^2} \nn \\ && + 
   \frac{4\,{\left( 1 + y^2 \right) }^2\,H(1,0,0;y)}
    {{\left( 1 - y^2 \right) }^2} - 
   \frac{12\,{\left( 1 + y^2 \right) }^2\,H(1;y)\,\zeta_2}
    {{\left( 1 - y^2 \right) }^2} 
\nn \\ && - 
   \frac{\left( 1 + y^2 \right) \,H(0;y)\,
      \left( -1 + y^2 + 10\,\zeta_2 + 10\,y^2\,\zeta_2 \right) }{{\left(
         1 - y^2 \right) }^2}
\nn \\ && - 
   \frac{-1 + y^2 + 4\,\zeta_2 + 4\,y^2\,\zeta_2}{1 - y^2} 
\ ; \\
&& \nn \\
%%%%%%%%%%%%%%%%%%%%%%%%%%%%%%%%%%%%%%%%%%%%%%%%%%%%%%%%%%%%%%%%%%%%%%%%%%%%%%%%%%%%%%%%
\Re \, \bar b_{0} &=& 
  \frac{-2\,\left( 5 + 12\,y - 18\,y^2 + 12\,y^3 + 5\,y^4 \right) \,
      H(-1,0,0;y)}{{\left( 1 - y^2 \right) }^2} \nn \\ && + 
   \frac{4\,\left( 1 + y^2 \right) \,H(0,-1,0;y)}
    {{\left( 1 - y \right) }^2} \nn \\ && - 
   \frac{\left( -1 - 9\,y + 9\,y^2 + 71\,y^3 + 16\,y^4 + 10\,y^5 \right) \,
      H(0,0,0;y)}{{\left( 1 - y \right) }^2\,{\left( 1 + y \right) }^3} \nn \\ && - 
   \frac{2\,\left( 3 + 14\,y - 4\,y^2 + 14\,y^3 + 5\,y^4 \right) \,H(0,1,0;y)}
    {{\left( 1 - y^2 \right) }^2} \nn \\ && - 
   \frac{\left( 5 - 14\,y + 46\,y^2 - 14\,y^3 + 9\,y^4 \right) \,H(1,0,0;y)}
    {{\left( 1 - y^2 \right) }^2} \nn \\ && + 
   \frac{4\,\left( 1 + y^2 \right) \,H(1,1,0;y)}{1 - y^2} \nn \\ && + 
   \frac{8\,{\left( 1 + y^2 \right) }^2\,H(-1,0,-1,0;y)}
    {{\left( 1 - y^2 \right) }^2} \nn \\ && - 
   \frac{8\,\left( 1 - 2\,y + 2\,y^2 + 2\,y^3 + y^4 \right) \,H(-1,0,0,0;y)}
    {{\left( 1 - y^2 \right) }^2} \nn \\ && - 
   \frac{8\,{\left( 1 + y^2 \right) }^2\,H(-1,0,1,0;y)}
    {{\left( 1 - y^2 \right) }^2} \nn \\ && - 
   \frac{4\,\left( 1 + 4\,y + 2\,y^2 + 4\,y^3 + y^4 \right) \,H(0,-1,0,0;y)}
    {\left( 1 - y \right) \,{\left( 1 + y \right) }^3} \nn \\ && - 
   \frac{4\,\left( -1 + 2\,y^2 + 3\,y^4 \right) \,H(0,0,-1,0;y)}
    {{\left( 1 - y^2 \right) }^2} \nn \\ && + 
   \frac{\left( 7 + 7\,y + 24\,y^2 + 44\,y^3 + 27\,y^4 + 27\,y^5 \right) \,
      H(0,0,0,0;y)}{{\left( 1 - y \right) }^2\,{\left( 1 + y \right) }^3} \nn \\ && + 
   \frac{2\,\left( -1 - 5\,y + 10\,y^2 + 10\,y^3 + 15\,y^4 + 11\,y^5 \right)
        \,H(0,0,1,0;y)}{{\left( 1 - y \right) }^2\,{\left( 1 + y \right) }^3}
    \nn \\ && + \frac{2\,\left( 3 + 11\,y + 12\,y^2 + 8\,y^3 - y^4 + 7\,y^5 \right) \,
      H(0,1,0,0;y)}{{\left( 1 - y \right) }^2\,{\left( 1 + y \right) }^3} \nn \\ && + 
   \frac{4\,{\left( 1 + y^2 \right) }^2\,H(0,1,1,0;y)}
    {{\left( 1 - y^2 \right) }^2} \nn \\ && + 
   \frac{12\,{\left( 1 + y^2 \right) }^2\,H(1,0,0,0;y)}
    {{\left( 1 - y^2 \right) }^2} \nn \\ && + 
   \frac{8\,{\left( 1 + y^2 \right) }^2\,H(1,0,1,0;y)}
    {{\left( 1 - y^2 \right) }^2} \nn \\ && + 
   \frac{16\,{\left( 1 + y^2 \right) }^2\,H(1,1,0,0;y)}
    {{\left( 1 - y^2 \right) }^2} \nn \\ && - 
   \frac{2\,\left( -5 + 25\,y - 17\,y^2 + 13\,y^3 \right) \,H(1;y)\,\zeta_2}
    {\left( 1 - y \right) \,{\left( 1 + y \right) }^2} \nn \\ && + 
   \frac{12\,\left( 1 + 4\,y + 2\,y^2 + 4\,y^3 + y^4 \right) \,H(0,-1;y)\,
      \zeta_2}{\left( 1 - y \right) \,{\left( 1 + y \right) }^3} \nn \\ && - 
   \frac{8\,\left( 4 + 10\,y + 11\,y^2 + 5\,y^3 - 2\,y^4 + 4\,y^5 \right) \,
      H(0,1;y)\,\zeta_2}{{\left( 1 - y \right) }^2\,
      {\left( 1 + y \right) }^3} \nn \\ && - 
   \frac{48\,{\left( 1 + y^2 \right) }^2\,H(1,1;y)\,\zeta_2}
    {{\left( 1 - y^2 \right) }^2} \nn \\ && + 
   \frac{4\,H(-1,0;y)}{{\left( 1 - y^2 \right) }^2} 
\,\Big[ 2 - 2\,y^4 + 5\,\zeta_2 - 8\,y\,\zeta_2 + 
        10\,y^2\,\zeta_2 \nn \\ && + 8\,y^3\,\zeta_2 + 5\,y^4\,\zeta_2 \Big]
      \nn \\ && - 
   \frac{H(1,0;y)\,\left( 15 - 2\,y + 2\,y^3 - 15\,y^4 + 40\,\zeta_2 + 
        80\,y^2\,\zeta_2 + 40\,y^4\,\zeta_2 \right) }{{\left( 1 - y^2
         \right) }^2} \nn \\ && - 
   \frac{H(0,0;y)}{{\left( 1 - y \right) }^2\,{\left( 1 + y \right) }^3} 
      \Big[ -3 - 14\,y + 6\,y^2 + 8\,y^3 - 35\,y^4 - 26\,y^5 + 
        24\,\zeta_2 \nn \\ && + 28\,y\,\zeta_2 + 76\,y^2\,\zeta_2 + 
        108\,y^3\,\zeta_2 + 64\,y^4\,\zeta_2 + 68\,y^5\,\zeta_2 \Big]
  \nn \\ && + 
   \frac{2\,H(-1;y)}{{\left(1 - y^2 \right) }^2} 
   \,\Big[ 15\,\zeta_2 + 36\,y\,\zeta_2 - 
        54\,y^2\,\zeta_2 + 36\,y^3\,\zeta_2 + 15\,y^4\,\zeta_2 \nn \\ && - 
        2\,\zeta_3 - 4\,y^2\,\zeta_3 - 2\,y^4\,\zeta_3 \Big]
   \nn \\ && + 
   \frac{1}{20\,{\left( 1 - y \right) }^2\,{\left( 1 + y \right) }^3} 
   \Big[145 + 145\,y - 290\,y^2 - 290\,y^3 + 145\,y^4 + 145\,y^5 \nn \\ && - 
      160\,\zeta_2 - 600\,y\,\zeta_2 + 1960\,y^2\,\zeta_2 + 
      2120\,y^3\,\zeta_2 - 1800\,y^4\,\zeta_2 \nn \\ && - 1520\,y^5\,\zeta_2 - 
      480\,y\,\ln(2)\,\zeta_2 - 1440\,y^2\,\ln(2)\,\zeta_2 - 
      1440\,y^3\,\ln(2)\,\zeta_2 \nn \\ && - 480\,y^4\,\ln(2)\,\zeta_2 + 
      272\,{\zeta_2}^2 + 376\,y\,{\zeta_2}^2 + 912\,y^2\,{\zeta_2}^2 + 
      416\,y^3\,{\zeta_2}^2 \nn \\ && + 288\,y^4\,{\zeta_2}^2 + 
      392\,y^5\,{\zeta_2}^2 - 280\,\zeta_3 - 280\,y^2\,\zeta_3 - 
      280\,y^3\,\zeta_3 \nn \\ && + 80\,y^4\,\zeta_3 - 200\,y^5\,\zeta_3\Big]
   \nn \\ && + 
   \frac{H(0;y)}{{\left( 1 - y \right) }^2\,{\left( 1 + y \right) }^3}
\,\Big[ 3 + y - 2\,y^2 + 2\,y^3 - y^4 - 3\,y^5 - \zeta_2 - 
        21\,y\,\zeta_2 \nn \\ && + 16\,y^2\,\zeta_2 + 140\,y^3\,\zeta_2 + 
        33\,y^4\,\zeta_2 + 25\,y^5\,\zeta_2 - 8\,\zeta_3 + 
        8\,y\,\zeta_3 \nn \\ && + 8\,y^2\,\zeta_3 - 8\,y^3\,\zeta_3 - 
        8\,y^4\,\zeta_3 + 8\,y^5\,\zeta_3 \Big]
\ ;
%%%%%%%%%%%%%%%%%%%%%%%%%%%%%%%%%%%%%%%%%%%%%%%%%%%%%%%%%%%%%%%%%%%%%%%%%%%%%%%%%%%%%%%%
\eea
\bea
%%%%%%%%%%%%%%%%%%%%%%%%%%%%%%%%%%%%%%%%%%%%%%%%%%%%%%%%%%%%%%%%%%%%%%%%%%%%%%%%%%%%%%%%
%%%% C_F*C_A
%%%%%%%%%%%%%%%%%%%%%%%%%%%%%%%%%%%%%%%%%%%%%%%%%%%%%%%%%%%%%%%%%%%%%%%%%%%%%%%%%%%%%%%%
\Re \, \bar c_{-2} &=&  \Re \, c_{-2} 
\ ; \\
&& \nn \\
%%%%%%%%%%%%%%%%%%%%%%%%%%%%%%%%%%%%%%%%%%%%%%%%%%%%%%%%%%%%%%%%%%%%%%%%%%%%%%%%%%%%%%%%
\Re \, \bar c_{-1} &=& \Re \, c_{-1} 
\ ; \\
&& \nn \\
%%%%%%%%%%%%%%%%%%%%%%%%%%%%%%%%%%%%%%%%%%%%%%%%%%%%%%%%%%%%%%%%%%%%%%%%%%%%%%%%%%%%%%%%
\Re \, \bar c_{0} &=& 
  \frac{-2\,\left( 1 + y^2 \right) \,H(-1,-1,0;y)}{1 - y^2} \nn \\ && + 
   \frac{4\,\left( 3 + 2\,y - 4\,y^2 + 2\,y^3 + y^4 \right) \,H(-1,0,0;y)}
    {{\left( 1 - y^2 \right) }^2} \nn \\ && + 
   \frac{6\,\left( 1 + y^2 \right) \,H(-1,1,0;y)}{1 - y^2} \nn \\ && - 
   \frac{4\,y\,\left( 1 - y + 2\,y^2 \right) \,H(0,-1,0;y)}
    {{\left( 1 - y \right) }^2\,\left( 1 + y \right) } \nn \\ && + 
   \frac{\left( -11 - 35\,y - 45\,y^2 + 189\,y^3 + 128\,y^4 + 62\,y^5 \right)
        \,H(0,0,0;y)}{6\,{\left( 1 - y \right) }^2\,{\left( 1 + y \right) }^3}
    \nn \\ && + \frac{2\,\left( -1 + 15\,y - 15\,y^2 + 15\,y^3 + 34\,y^4 \right) \,
      H(0,1,0;y)}{3\,{\left( 1 - y^2 \right) }^2} \nn \\ && + 
   \frac{6\,\left( 1 + y^2 \right) \,H(1,-1,0;y)}{1 - y^2} \nn \\ && + 
   \frac{\left( -61 + 6\,y + 66\,y^2 + 6\,y^3 + 79\,y^4 \right) \,H(1,0,0;y)}
    {6\,{\left( 1 - y^2 \right) }^2} \nn \\ && - 
   \frac{52\,\left( 1 + y^2 \right) \,H(1,1,0;y)}
    {3\,\left( 1 - y^2 \right) } \nn \\ && - 
   \frac{4\,{\left( 1 + y^2 \right) }^2\,H(-1,0,-1,0;y)}
    {{\left( 1 - y^2 \right) }^2} \nn \\ && - 
   \frac{2\,\left( -5 - 4\,y^2 + y^4 \right) \,H(-1,0,0,0;y)}
    {{\left( 1 - y^2 \right) }^2} \nn \\ && + 
   \frac{4\,{\left( 1 + y^2 \right) }^2\,H(-1,0,1,0;y)}
    {{\left( 1 - y^2 \right) }^2} \nn \\ && - 
   \frac{2\,{\left( 1 + y^2 \right) }^2\,H(0,-1,-1,0;y)}
    {{\left( 1 - y^2 \right) }^2} \nn \\ && + 
   \frac{2\,\left( 3 + 3\,y + 8\,y^3 + y^4 + y^5 \right) \,H(0,-1,0,0;y)}
    {{\left( 1 - y \right) }^2\,{\left( 1 + y \right) }^3} \nn \\ && + 
   \frac{6\,{\left( 1 + y^2 \right) }^2\,H(0,-1,1,0;y)}
    {{\left( 1 - y^2 \right) }^2} \nn \\ && + 
   \frac{2\,\left( 1 + 8\,y^2 + 7\,y^4 \right) \,H(0,0,-1,0;y)}
    {{\left( 1 - y^2 \right) }^2} \nn \\ && - 
   \frac{y^2\,\left( 7 + 17\,y + 12\,y^2 + 12\,y^3 \right) \,H(0,0,0,0;y)}
    {{\left( 1 - y \right) }^2\,{\left( 1 + y \right) }^3} \nn \\ && - 
   \frac{2\,\left( 1 + y + 14\,y^2 + 10\,y^3 + 11\,y^4 + 11\,y^5 \right) \,
      H(0,0,1,0;y)}{{\left( 1 - y \right) }^2\,{\left( 1 + y \right) }^3} \nn \\ && + 
   \frac{6\,{\left( 1 + y^2 \right) }^2\,H(0,1,-1,0;y)}
    {{\left( 1 - y^2 \right) }^2} \nn \\ && - 
   \frac{2\,\left( 1 + y + 5\,y^2 + 11\,y^3 + 7\,y^4 + 7\,y^5 \right) \,
      H(0,1,0,0;y)}{{\left( 1 - y \right) }^2\,{\left( 1 + y \right) }^3} \nn \\ && - 
   \frac{10\,{\left( 1 + y^2 \right) }^2\,H(0,1,1,0;y)}
    {{\left( 1 - y^2 \right) }^2} \nn \\ && + 
   \frac{4\,{\left( 1 + y^2 \right) }^2\,H(1,0,-1,0;y)}
    {{\left( 1 - y^2 \right) }^2} \nn \\ && - 
   \frac{2\,\left( 3 + 4\,y^2 + y^4 \right) \,H(1,0,0,0;y)}
    {{\left( 1 - y^2 \right) }^2} \nn \\ && - 
   \frac{4\,{\left( 1 + y^2 \right) }^2\,H(1,0,1,0;y)}
    {{\left( 1 - y^2 \right) }^2} \nn \\ && - 
   \frac{\left( 19 + 19\,y + 2\,y^2 + 50\,y^3 + 7\,y^4 + 7\,y^5 \right) \,
      H(0,-1;y)\,\zeta_2}{{\left( 1 - y \right) }^2\,
      {\left( 1 + y \right) }^3} \nn \\ && + 
   \frac{\left( 11 + 11\,y + 34\,y^2 + 58\,y^3 + 35\,y^4 + 35\,y^5 \right) \,
      H(0,1;y)\,\zeta_2}{{\left( 1 - y \right) }^2\,
      {\left( 1 + y \right) }^3} \nn \\ && + 
   \frac{2\,\left( 1 + y^2 \right) \,H(-1,0;y)\,
      \left( -1 + y^2 - 11\,\zeta_2 + y^2\,\zeta_2 \right) }{{\left( 1 - 
         y^2 \right) }^2} \nn \\ && + 
     \frac{H(1,0;y)}{18\, {\left( 1 - y^2 \right) }^2} \,
      \Big[ -53 + 54\,y - 54\,y^3 + 53\,y^4 + 252\,\zeta_2 + 
        360\,y^2\,\zeta_2 \nn \\ && + 108\,y^4\,\zeta_2 \Big]
   \nn \\ && + 
   \frac{H(0,0;y)}{18\,{\left( 1 - y \right) }^2\,{\left( 1 + y \right) }^3} 
\,\Big[ -67 - 148\,y - 36\,y^2 - 90\,y^3 - 185\,y^4 - 
        50\,y^5 + 18\,\zeta_2 \nn \\ && + 18\,y\,\zeta_2 + 504\,y^2\,\zeta_2 + 
        864\,y^3\,\zeta_2 + 666\,y^4\,\zeta_2 + 666\,y^5\,\zeta_2 \Big]
   \nn \\ && - 
   \frac{2\,\left( 1 + y^2 \right) \,H(1;y)\,
      \left( -52\,\zeta_2 - 9\,y\,\zeta_2 + 61\,y^2\,\zeta_2 + 
        3\,\zeta_3 + 3\,y^2\,\zeta_3 \right) }{3\,
      {\left( 1 - y^2 \right) }^2} \nn \\ && - 
   \frac{H(-1;y)}{{\left( 1 - y^2 \right)}^2} 
   \Big[ 37\,\zeta_2 + 24\,y\,\zeta_2 - 48\,y^2\,\zeta_2 + 
        24\,y^3\,\zeta_2 + 11\,y^4\,\zeta_2 \nn \\ && - 2\,\zeta_3 - 
        4\,y^2\,\zeta_3 - 2\,y^4\,\zeta_3 \Big]
      \nn \\ && 
      + \frac{1}{540\,{\left( 1 - y \right) }^2\, {\left( 1 + y \right) }^3} 
      \Big[-4345 - 4345\,y + 8690\,y^2 + 8690\,y^3 - 4345\,y^4 \nn \\ && - 
      4345\,y^5 + 8760\,\zeta_2 + 12000\,y\,\zeta_2 - 
      21960\,y^2\,\zeta_2 - 15480\,y^3\,\zeta_2 \nn \\ && + 13200\,y^4\,\zeta_2 + 
      3480\,y^5\,\zeta_2 + 6480\,y\,\ln(2)\,\zeta_2 + 
      19440\,y^2\,\ln(2)\,\zeta_2 \nn \\ && + 19440\,y^3\,\ln(2)\,\zeta_2 + 
      6480\,y^4\,\ln(2)\,\zeta_2 - 1701\,{\zeta_2}^2 - 
      1701\,y\,{\zeta_2}^2 \nn \\ && - 16902\,y^2\,{\zeta_2}^2 - 
      13014\,y^3\,{\zeta_2}^2 - 13257\,y^4\,{\zeta_2}^2 - 
      13257\,y^5\,{\zeta_2}^2 \nn \\ && + 4410\,\zeta_3 + 4950\,y\,\zeta_3 + 
      2700\,y^2\,\zeta_3 + 2700\,y^3\,\zeta_3 - 630\,y^4\,\zeta_3 \nn \\ && - 
      1170\,y^5\,\zeta_3 \Big]
   \nn \\ && - 
   \frac{H(0;y)}{54\,{\left( 1 - y \right) }^2\,{\left( 1 + y \right) }^3}
   \,\Big[ 121 + 283\,y + 162\,y^2 - 162\,y^3 - 283\,y^4 - 
        121\,y^5 \nn \\ && - 198\,\zeta_2 - 630\,y\,\zeta_2 - 1350\,y^2\,\zeta_2 + 
        2862\,y^3\,\zeta_2 + 2844\,y^4\,\zeta_2 \nn \\ && + 1656\,y^5\,\zeta_2 - 
        351\,\zeta_3 - 351\,y\,\zeta_3 - 162\,y^2\,\zeta_3 + 
        270\,y^3\,\zeta_3 \nn \\ && + 405\,y^4\,\zeta_3 + 405\,y^5\,\zeta_3 \Big]
\ ; 
%%%%%%%%%%%%%%%%%%%%%%%%%%%%%%%%%%%%%%%%%%%%%%%%%%%%%%%%%%%%%%%%%%%%%%%%%%%%%%%%%%%%%%%%
\eea
\bea
%%%%%%%%%%%%%%%%%%%%%%%%%%%%%%%%%%%%%%%%%%%%%%%%%%%%%%%%%%%%%%%%%%%%%%%%%%%%%%%%%%%%%%%%
%%%% C_F*T_R*N_f
%%%%%%%%%%%%%%%%%%%%%%%%%%%%%%%%%%%%%%%%%%%%%%%%%%%%%%%%%%%%%%%%%%%%%%%%%%%%%%%%%%%%%%%%
\Re \, \bar d_{-2} &=&  \Re \, d_{-2} 
\ ; \\
&& \nn \\
%%%%%%%%%%%%%%%%%%%%%%%%%%%%%%%%%%%%%%%%%%%%%%%%%%%%%%%%%%%%%%%%%%%%%%%%%%%%%%%%%%%%%%%%
\Re \, \bar d_{-1} &=& \Re \, d_{-1} 
\ ; \\
&& \nn \\
%%%%%%%%%%%%%%%%%%%%%%%%%%%%%%%%%%%%%%%%%%%%%%%%%%%%%%%%%%%%%%%%%%%%%%%%%%%%%%%%%%%%%%%%
\Re \, \bar d_{0} &=& 
  \frac{10\,\left( 1 + y^2 \right) \,H(0,0;y)}{9\,\left( 1 - y^2 \right) } + 
   \frac{20\,\left( 1 + y^2 \right) \,H(1,0;y)}{9\,\left( 1 - y^2 \right) } \nn \\ && + 
   \frac{2\,\left( 1 + y^2 \right) \,H(0,0,0;y)}
    {3\,\left( 1 - y^2 \right) } + 
   \frac{4\,\left( 1 + y^2 \right) \,H(0,1,0;y)}
    {3\,\left( 1 - y^2 \right) } \nn \\ && + 
   \frac{4\,\left( 1 + y^2 \right) \,H(1,0,0;y)}
    {3\,\left( 1 - y^2 \right) } + 
   \frac{8\,\left( 1 + y^2 \right) \,H(1,1,0;y)}
    {3\,\left( 1 - y^2 \right) } \nn \\ && - 
   \frac{16\,\left( 1 + y^2 \right) \,H(1;y)\,\zeta_2}
    {3\,\left( 1 - y^2 \right) } - 
   \frac{4\,\left( 1 + y^2 \right) \,H(0;y)\,\left( -7 + 9\,\zeta_2 \right) }
    {27\,\left( 1 - y^2 \right) } \nn \\ && - 
   \frac{-49 + 49\,y^2 + 84\,\zeta_2 + 156\,y^2\,\zeta_2 + 36\,\zeta_3 + 
      36\,y^2\,\zeta_3}{27\,\left( 1 - y^2 \right) }
\ ;
%%%%%%%%%%%%%%%%%%%%%%%%%%%%%%%%%%%%%%%%%%%%%%%%%%%%%%%%%%%%%%%%%%%%%%%%%%%%%%%%%%%%%%%%
\eea
\bea
%%%%%%%%%%%%%%%%%%%%%%%%%%%%%%%%%%%%%%%%%%%%%%%%%%%%%%%%%%%%%%%%%%%%%%%%%%%%%%%%%%%%%%%%
%%%% C_F*T_R
%%%%%%%%%%%%%%%%%%%%%%%%%%%%%%%%%%%%%%%%%%%%%%%%%%%%%%%%%%%%%%%%%%%%%%%%%%%%%%%%%%%%%%%%
\Re \, \bar e_{-2} &=& \Re \, e_{-2} = 0
\ ; \\
&& \nn \\
%%%%%%%%%%%%%%%%%%%%%%%%%%%%%%%%%%%%%%%%%%%%%%%%%%%%%%%%%%%%%%%%%%%%%%%%%%%%%%%%%%%%%%%%
\Re \, \bar e_{-1} &=& \Re \, e_{-1} = 0
\ ; \\
&& \nn \\
%%%%%%%%%%%%%%%%%%%%%%%%%%%%%%%%%%%%%%%%%%%%%%%%%%%%%%%%%%%%%%%%%%%%%%%%%%%%%%%%%%%%%%%%
\Re \, \bar e_{0} &=& 
  \frac{2\,\left( 1 + 2\,y + 26\,y^2 + 2\,y^3 + y^4 \right) \,H(0,0,0;y)}
    {3\,\left( 1 - y \right) \,{\left( 1 + y \right) }^3} + 
   \frac{8\,y\,H(-1,0,0,0;y)}{1 - y^2} \nn \\ && + 
   \frac{8\,y\,H(0,-1,0,0;y)}{1 - y^2} - 
   \frac{8\,y\,H(0,0,-1,0;y)}{1 - y^2} - 
   \frac{12\,y\,H(0,0,0,-1;y)}{1 - y^2} \nn \\ && + 
   \frac{2\,y\,H(0,0,0,0;y)}{1 - y^2} - 
   \frac{16\,y\,H(-1,0;y)\,\zeta_2}{1 - y^2} - 
   \frac{24\,y\,H(0,-1;y)\,\zeta_2}{1 - y^2} 
   \nn \\ && + 
   \frac{2\,H(0,0;y)}   {9\,{\left( 1 - y \right) }^4\,{\left( 1 + y \right) }^2} 
     \Big[ 5 - 22\,y + 59\,y^2 - 116\,y^3 + 59\,y^4 - 
        22\,y^5 + 5\,y^6 \nn \\ && - 9\,y\,\zeta_2 + 18\,y^2\,\zeta_2 - 
        18\,y^4\,\zeta_2 + 9\,y^5\,\zeta_2 \Big]
   \nn \\ && - 
   \frac{4\,H(0;y)}{27\,{\left( 1 - y^2 \right) }^3}
     \Big[ -14 + 12\,y + 38\,y^2 + 24\,y^3 + 38\,y^4 + 
        12\,y^5 - 14\,y^6 \nn \\ && + 9\,\zeta_2 + 207\,y^2\,\zeta_2 - 
        432\,y^3\,\zeta_2 + 207\,y^4\,\zeta_2 + 9\,y^6\,\zeta_2 - 
        81\,y\,\zeta_3 \nn \\ && + 162\,y^3\,\zeta_3 - 81\,y^5\,\zeta_3 \Big]
   \nn \\ && + 
   \frac{1}{135\,{\left( 1 - y^2 \right) }^2} 
     \Big[2035 - 120\,y - 4310\,y^2 - 120\,y^3 + 2035\,y^4 - 1260\,\zeta_2 \nn \\ && + 
      540\,y\,\zeta_2 + 1440\,y^2\,\zeta_2 + 540\,y^3\,\zeta_2 - 
      1260\,y^4\,\zeta_2 - 783\,y\,{\zeta_2}^2 \nn \\ && + 783\,y^3\,{\zeta_2}^2\Big]
\ ;
%%%%%%%%%%%%%%%%%%%%%%%%%%%%%%%%%%%%%%%%%%%%%%%%%%%%%%%%%%%%%%%%%%%%%%%%%%%%%%%%%%%%%%%%
\eea

\bea
%%%%%%%%%%%%%%%%%%%%%%%%%%%%%%%%%%%%%%%%%%%%%%%%%%%%%%%%%%%%%%%%%%%%%%%%%%%%%%%%%%%%%%%%
%%%% C_F^2
%%%%%%%%%%%%%%%%%%%%%%%%%%%%%%%%%%%%%%%%%%%%%%%%%%%%%%%%%%%%%%%%%%%%%%%%%%%%%%%%%%%%%%%%
\Im \, \bar b_{-2} &=&  \Im \, b_{-2} 
\ ; \\
&& \nn \\
%%%%%%%%%%%%%%%%%%%%%%%%%%%%%%%%%%%%%%%%%%%%%%%%%%%%%%%%%%%%%%%%%%%%%%%%%%%%%%%%%%%%%%%%
\Im \, \bar b_{-1} &=& 
  \frac{\left( 1 + y^2 \right) \,H(0;y)}{1 - y^2} + 
   \frac{2\,\left( 1 + y^2 \right) \,H(1;y)}{1 - y^2} + 
   \frac{3\,{\left( 1 + y^2 \right) }^2\,H(0,0;y)}
    {{\left( 1 - y^2 \right) }^2} \nn \\ && + 
   \frac{2\,{\left( 1 + y^2 \right) }^2\,H(0,1;y)}
    {{\left( 1 - y^2 \right) }^2} + 
   \frac{4\,{\left( 1 + y^2 \right) }^2\,H(1,0;y)}
    {{\left( 1 - y^2 \right) }^2} \nn \\ && - 
   \frac{\left( 1 + y^2 \right) \,
      \left( -1 + y^2 + 4\,\zeta_2 + 4\,y^2\,\zeta_2 \right) }{{\left( 1 - 
         y^2 \right) }^2}
\ ; \\
&& \nn \\
%%%%%%%%%%%%%%%%%%%%%%%%%%%%%%%%%%%%%%%%%%%%%%%%%%%%%%%%%%%%%%%%%%%%%%%%%%%%%%%%%%%%%%%%
\Im \, \bar b_{0} &=& 
  \frac{-2\,\left( 5 + 12\,y - 18\,y^2 + 12\,y^3 + 5\,y^4 \right) \,H(-1,0;y)}
    {{\left( 1 - y^2 \right) }^2} \nn \\ && + 
   \frac{4\,\left( 1 + y^2 \right) \,H(0,-1;y)}{{\left( 1 - y \right) }^2} \nn \\ && - 
   \frac{\left( -1 - 9\,y + 9\,y^2 + 71\,y^3 + 16\,y^4 + 10\,y^5 \right) \,
      H(0,0;y)}{{\left( 1 - y \right) }^2\,{\left( 1 + y \right) }^3} \nn \\ && - 
   \frac{2\,\left( 3 + 14\,y - 4\,y^2 + 14\,y^3 + 5\,y^4 \right) \,H(0,1;y)}
    {{\left( 1 - y^2 \right) }^2} \nn \\ && - 
   \frac{\left( 5 - 14\,y + 46\,y^2 - 14\,y^3 + 9\,y^4 \right) \,H(1,0;y)}
    {{\left( 1 - y^2 \right) }^2} \nn \\ && + 
   \frac{4\,\left( 1 + y^2 \right) \,H(1,1;y)}{1 - y^2} \nn \\ && + 
   \frac{8\,{\left( 1 + y^2 \right) }^2\,H(-1,0,-1;y)}
    {{\left( 1 - y^2 \right) }^2} \nn \\ && - 
   \frac{8\,\left( 1 - 2\,y + 2\,y^2 + 2\,y^3 + y^4 \right) \,H(-1,0,0;y)}
    {{\left( 1 - y^2 \right) }^2} \nn \\ && - 
   \frac{8\,{\left( 1 + y^2 \right) }^2\,H(-1,0,1;y)}
    {{\left( 1 - y^2 \right) }^2} \nn \\ && - 
   \frac{4\,\left( 1 + 4\,y + 2\,y^2 + 4\,y^3 + y^4 \right) \,H(0,-1,0;y)}
    {\left( 1 - y \right) \,{\left( 1 + y \right) }^3} \nn \\ && - 
   \frac{4\,\left( -1 + 2\,y^2 + 3\,y^4 \right) \,H(0,0,-1;y)}
    {{\left( 1 - y^2 \right) }^2} \nn \\ && + 
   \frac{\left( 7 + 7\,y + 24\,y^2 + 44\,y^3 + 27\,y^4 + 27\,y^5 \right) \,
      H(0,0,0;y)}{{\left( 1 - y \right) }^2\,{\left( 1 + y \right) }^3} \nn \\ && + 
   \frac{2\,\left( -1 - 5\,y + 10\,y^2 + 10\,y^3 + 15\,y^4 + 11\,y^5 \right)
        \,H(0,0,1;y)}{{\left( 1 - y \right) }^2\,{\left( 1 + y \right) }^3} \nn \\ && + 
   \frac{2\,\left( 3 + 11\,y + 12\,y^2 + 8\,y^3 - y^4 + 7\,y^5 \right) \,
      H(0,1,0;y)}{{\left( 1 - y \right) }^2\,{\left( 1 + y \right) }^3} \nn \\ && + 
   \frac{4\,{\left( 1 + y^2 \right) }^2\,H(0,1,1;y)}
    {{\left( 1 - y^2 \right) }^2} \nn \\ && + 
   \frac{12\,{\left( 1 + y^2 \right) }^2\,H(1,0,0;y)}
    {{\left( 1 - y^2 \right) }^2} \nn \\ && + 
   \frac{8\,{\left( 1 + y^2 \right) }^2\,H(1,0,1;y)}
    {{\left( 1 - y^2 \right) }^2} \nn \\ && + 
   \frac{16\,{\left( 1 + y^2 \right) }^2\,H(1,1,0;y)}
    {{\left( 1 - y^2 \right) }^2} \nn \\ && + 
   \frac{4\,\left( 1 + y^2 \right) \,H(-1;y)\,
      \left( 2 - 2\,y^2 + \zeta_2 + y^2\,\zeta_2 \right) }{{\left( 1 - 
         y^2 \right) }^2} \nn \\ && 
 - \frac{H(0;y)}{{\left( 1 - y^2 \right) }^2} \,
      \Big( -3 - 11\,y + 17\,y^2 - 9\,y^3 - 26\,y^4 + 10\,\zeta_2 + 
        4\,y\,\zeta_2 \nn \\ && + 24\,y^2\,\zeta_2 - 4\,y^3\,\zeta_2 + 
        14\,y^4\,\zeta_2 \Big]
   \nn \\ && - 
   \frac{H(1;y)\,\left( 15 - 2\,y + 2\,y^3 - 15\,y^4 + 16\,\zeta_2 + 
        32\,y^2\,\zeta_2 + 16\,y^4\,\zeta_2 \right) }{{\left( 1 - y^2
         \right) }^2} \nn \\ && - 
   \frac{1}{\left( 1 - y \right) \,{\left( 1 + y \right) }^3}
   \Big[-3 - 4\,y - 2\,y^2 - 4\,y^3 - 3\,y^4 - 
      \zeta_2 + 2\,y\,\zeta_2 \nn \\ && + 4\,y^2\,\zeta_2 + 6\,y^3\,\zeta_2 + 
      5\,y^4\,\zeta_2 + 8\,\zeta_3 - 8\,y^2\,\zeta_3 + 8\,y^4\,\zeta_3\Big]
\ ;
%%%%%%%%%%%%%%%%%%%%%%%%%%%%%%%%%%%%%%%%%%%%%%%%%%%%%%%%%%%%%%%%%%%%%%%%%%%%%%%%%%%%%%%%
\eea
\bea
%%%%%%%%%%%%%%%%%%%%%%%%%%%%%%%%%%%%%%%%%%%%%%%%%%%%%%%%%%%%%%%%%%%%%%%%%%%%%%%%%%%%%%%%
%%%% C_F*C_A
%%%%%%%%%%%%%%%%%%%%%%%%%%%%%%%%%%%%%%%%%%%%%%%%%%%%%%%%%%%%%%%%%%%%%%%%%%%%%%%%%%%%%%%%
\Im \, \bar c_{-2} &=&  \Im \, c_{-2} 
\ ; \\
&& \nn \\
%%%%%%%%%%%%%%%%%%%%%%%%%%%%%%%%%%%%%%%%%%%%%%%%%%%%%%%%%%%%%%%%%%%%%%%%%%%%%%%%%%%%%%%%
\Im \, \bar c_{-1} &=& \Im \, c_{-1} 
\ ; \\
&& \nn \\
%%%%%%%%%%%%%%%%%%%%%%%%%%%%%%%%%%%%%%%%%%%%%%%%%%%%%%%%%%%%%%%%%%%%%%%%%%%%%%%%%%%%%%%%
\Im \, \bar c_{0} &=& 
  \frac{-2\,\left( 1 + y^2 \right) \,H(-1,-1;y)}{1 - y^2} \nn \\ && + 
   \frac{4\,\left( 3 + 2\,y - 4\,y^2 + 2\,y^3 + y^4 \right) \,H(-1,0;y)}
    {{\left( 1 - y^2 \right) }^2} \nn \\ && + 
   \frac{6\,\left( 1 + y^2 \right) \,H(-1,1;y)}{1 - y^2} \nn \\ && - 
   \frac{4\,y\,\left( 1 - y + 2\,y^2 \right) \,H(0,-1;y)}
    {{\left( 1 - y \right) }^2\,\left( 1 + y \right) } \nn \\ && + 
   \frac{\left( -11 - 35\,y - 45\,y^2 + 189\,y^3 + 128\,y^4 + 62\,y^5 \right)
        \,H(0,0;y)}{6\,{\left( 1 - y \right) }^2\,{\left( 1 + y \right) }^3} \nn \\ && +
    \frac{2\,\left( -1 + 15\,y - 15\,y^2 + 15\,y^3 + 34\,y^4 \right) \,
      H(0,1;y)}{3\,{\left( 1 - y^2 \right) }^2} \nn \\ && + 
   \frac{6\,\left( 1 + y^2 \right) \,H(1,-1;y)}{1 - y^2} \nn \\ && + 
   \frac{\left( -61 + 6\,y + 66\,y^2 + 6\,y^3 + 79\,y^4 \right) \,H(1,0;y)}
    {6\,{\left( 1 - y^2 \right) }^2} \nn \\ && - 
   \frac{52\,\left( 1 + y^2 \right) \,H(1,1;y)}{3\,\left( 1 - y^2 \right) } \nn \\ && - 
   \frac{4\,{\left( 1 + y^2 \right) }^2\,H(-1,0,-1;y)}
    {{\left( 1 - y^2 \right) }^2} \nn \\ && - 
   \frac{2\,\left( -5 - 4\,y^2 + y^4 \right) \,H(-1,0,0;y)}
    {{\left( 1 - y^2 \right) }^2} \nn \\ && + 
   \frac{4\,{\left( 1 + y^2 \right) }^2\,H(-1,0,1;y)}
    {{\left( 1 - y^2 \right) }^2} \nn \\ && - 
   \frac{2\,{\left( 1 + y^2 \right) }^2\,H(0,-1,-1;y)}
    {{\left( 1 - y^2 \right) }^2} \nn \\ && + 
   \frac{2\,\left( 3 + 3\,y + 8\,y^3 + y^4 + y^5 \right) \,H(0,-1,0;y)}
    {{\left( 1 - y \right) }^2\,{\left( 1 + y \right) }^3} \nn \\ && + 
   \frac{6\,{\left( 1 + y^2 \right) }^2\,H(0,-1,1;y)}
    {{\left( 1 - y^2 \right) }^2} \nn \\ && + 
   \frac{2\,\left( 1 + 8\,y^2 + 7\,y^4 \right) \,H(0,0,-1;y)}
    {{\left( 1 - y^2 \right) }^2} \nn \\ && - 
   \frac{y^2\,\left( 7 + 17\,y + 12\,y^2 + 12\,y^3 \right) \,H(0,0,0;y)}
    {{\left( 1 - y \right) }^2\,{\left( 1 + y \right) }^3} \nn \\ && - 
   \frac{2\,\left( 1 + y + 14\,y^2 + 10\,y^3 + 11\,y^4 + 11\,y^5 \right) \,
      H(0,0,1;y)}{{\left( 1 - y \right) }^2\,{\left( 1 + y \right) }^3} \nn \\ && + 
   \frac{6\,{\left( 1 + y^2 \right) }^2\,H(0,1,-1;y)}
    {{\left( 1 - y^2 \right) }^2} \nn \\ && - 
   \frac{2\,\left( 1 + y + 5\,y^2 + 11\,y^3 + 7\,y^4 + 7\,y^5 \right) \,
      H(0,1,0;y)}{{\left( 1 - y \right) }^2\,{\left( 1 + y \right) }^3} \nn \\ && - 
   \frac{10\,{\left( 1 + y^2 \right) }^2\,H(0,1,1;y)}
    {{\left( 1 - y^2 \right) }^2} \nn \\ && + 
   \frac{4\,{\left( 1 + y^2 \right) }^2\,H(1,0,-1;y)}
    {{\left( 1 - y^2 \right) }^2} \nn \\ && - 
   \frac{2\,\left( 3 + 4\,y^2 + y^4 \right) \,H(1,0,0;y)}
    {{\left( 1 - y^2 \right) }^2} \nn \\ && - 
   \frac{4\,{\left( 1 + y^2 \right) }^2\,H(1,0,1;y)}
    {{\left( 1 - y^2 \right) }^2} \nn \\ && - 
   \frac{2\,\left( 1 + y^2 \right) \,H(-1;y)\,
      \left( 1 - y^2 + \zeta_2 + y^2\,\zeta_2 \right) }{{\left( 1 - y^2
         \right) }^2} \nn \\ && + 
   \frac{H(1;y)}{18\, {\left( 1 - y^2 \right) }^2} \,
      \Big[ -53 + 54\,y - 54\,y^3 + 53\,y^4 + 36\,\zeta_2 + 
        72\,y^2\,\zeta_2 \nn \\ && + 36\,y^4\,\zeta_2 \Big]
   \nn \\ && + 
   \frac{H(0;y)}{18\, {\left( 1 - y^2 \right) }^2} 
   \,\Big[ -67 - 81\,y + 45\,y^2 - 135\,y^3 - 50\,y^4 + 
        18\,\zeta_2 \nn \\ && + 252\,y^2\,\zeta_2 + 234\,y^4\,\zeta_2 \Big]
   \nn \\ && + 
   \frac{1}{54\,{\left( 1 - y \right) }^2\, {\left( 1 + y \right) }^3}
     \Big[-121 - 283\,y - 162\,y^2 + 162\,y^3 + 283\,y^4 \nn \\ && + 121\,y^5 + 
      540\,y^2\,\zeta_2 + 540\,y^3\,\zeta_2 - 540\,y^4\,\zeta_2 - 
      540\,y^5\,\zeta_2 \nn \\ && + 351\,\zeta_3 + 351\,y\,\zeta_3 + 
      162\,y^2\,\zeta_3 - 270\,y^3\,\zeta_3 - 405\,y^4\,\zeta_3 \nn \\ && - 
      405\,y^5\,\zeta_3\Big]
\ ; 
%%%%%%%%%%%%%%%%%%%%%%%%%%%%%%%%%%%%%%%%%%%%%%%%%%%%%%%%%%%%%%%%%%%%%%%%%%%%%%%%%%%%%%%%
\eea
\bea
%%%%%%%%%%%%%%%%%%%%%%%%%%%%%%%%%%%%%%%%%%%%%%%%%%%%%%%%%%%%%%%%%%%%%%%%%%%%%%%%%%%%%%%%
%%%% C_F*T_R*N_f
%%%%%%%%%%%%%%%%%%%%%%%%%%%%%%%%%%%%%%%%%%%%%%%%%%%%%%%%%%%%%%%%%%%%%%%%%%%%%%%%%%%%%%%%
\Im \, \bar d_{-2} &=&  \Im \, d_{-2} 
\ ; \\
&& \nn \\
%%%%%%%%%%%%%%%%%%%%%%%%%%%%%%%%%%%%%%%%%%%%%%%%%%%%%%%%%%%%%%%%%%%%%%%%%%%%%%%%%%%%%%%%
\Im \, \bar d_{-1} &=& \Im \, d_{-1} 
\ ; \\
&& \nn \\
%%%%%%%%%%%%%%%%%%%%%%%%%%%%%%%%%%%%%%%%%%%%%%%%%%%%%%%%%%%%%%%%%%%%%%%%%%%%%%%%%%%%%%%%
\Im \, \bar d_{0} &=& 
  \frac{28\,\left( 1 + y^2 \right) }{27\,\left( 1 - y^2 \right) } + 
   \frac{10\,\left( 1 + y^2 \right) \,H(0;y)}{9\,\left( 1 - y^2 \right) } + 
   \frac{20\,\left( 1 + y^2 \right) \,H(1;y)}{9\,\left( 1 - y^2 \right) } \nn \\ && + 
   \frac{2\,\left( 1 + y^2 \right) \,H(0,0;y)}{3\,\left( 1 - y^2 \right) } + 
   \frac{4\,\left( 1 + y^2 \right) \,H(0,1;y)}{3\,\left( 1 - y^2 \right) } \nn \\ && + 
   \frac{4\,\left( 1 + y^2 \right) \,H(1,0;y)}{3\,\left( 1 - y^2 \right) } + 
   \frac{8\,\left( 1 + y^2 \right) \,H(1,1;y)}{3\,\left( 1 - y^2 \right) }
\ ;
%%%%%%%%%%%%%%%%%%%%%%%%%%%%%%%%%%%%%%%%%%%%%%%%%%%%%%%%%%%%%%%%%%%%%%%%%%%%%%%%%%%%%%%%
\eea
\bea
%%%%%%%%%%%%%%%%%%%%%%%%%%%%%%%%%%%%%%%%%%%%%%%%%%%%%%%%%%%%%%%%%%%%%%%%%%%%%%%%%%%%%%%%
%%%% C_F*T_R
%%%%%%%%%%%%%%%%%%%%%%%%%%%%%%%%%%%%%%%%%%%%%%%%%%%%%%%%%%%%%%%%%%%%%%%%%%%%%%%%%%%%%%%%
\Im \, \bar e_{-2} &=& \Im \, e_{-2} = 0 
\ ; \\
&& \nn \\
%%%%%%%%%%%%%%%%%%%%%%%%%%%%%%%%%%%%%%%%%%%%%%%%%%%%%%%%%%%%%%%%%%%%%%%%%%%%%%%%%%%%%%%%
\Im \, \bar e_{-1} &=& \Im \, e_{-1} = 0 
\ ; \\
&& \nn \\
%%%%%%%%%%%%%%%%%%%%%%%%%%%%%%%%%%%%%%%%%%%%%%%%%%%%%%%%%%%%%%%%%%%%%%%%%%%%%%%%%%%%%%%%
\Im \, \bar e_{0} &=& 
  \frac{2\,\left( 1 + 2\,y + 26\,y^2 + 2\,y^3 + y^4 \right) \,H(0,0;y)}
    {3\,\left( 1 - y \right) \,{\left( 1 + y \right) }^3} + 
   \frac{8\,y\,H(-1,0,0;y)}{1 - y^2} \nn \\ && + \frac{8\,y\,H(0,-1,0;y)}{1 - y^2} - 
   \frac{8\,y\,H(0,0,-1;y)}{1 - y^2} + \frac{2\,y\,H(0,0,0;y)}{1 - y^2} 
   \nn \\ && + 
   \frac{2 \,H(0;y)}{9\, {\left( 1 - y \right) }^4\,{\left( 1 + y \right) }^2} 
     \Big[ 5 - 22\,y + 59\,y^2 - 116\,y^3 + 59\,y^4 - 
        22\,y^5 + 5\,y^6 \nn \\ && + 9\,y\,\zeta_2 - 18\,y^2\,\zeta_2 + 
        18\,y^4\,\zeta_2 - 9\,y^5\,\zeta_2 \Big]
   \nn \\ && + 
   \frac{4}{27\,{\left( 1 - y \right) }^3\,\left( 1 + y \right) }
    \Big[ 14 - 40\,y + 28\,y^2 - 40\,y^3 + 14\,y^4 + 
        81\,y\,\zeta_3 \nn \\ && - 162\,y^2\,\zeta_3 + 81\,y^3\,\zeta_3 \Big]
\ .
%%%%%%%%%%%%%%%%%%%%%%%%%%%%%%%%%%%%%%%%%%%%%%%%%%%%%%%%%%%%%%%%%%%%%%%%%%%%%%%%%%%%%%%%
\eea

\noindent
The light quark loop insertions in Fig.~\ref{fig2} (h)
were studied before by \cite{Melnikov:1995yp, Hoang:1997ca}, using a small quark
mass for regularizing the double- and single logarithmic mass divergences of these
contributions. 
As we use dimensional regularization setting the light quark masses to zero, only
the  coefficient of the leading singularity  of the $C_F T_R N_f$ term in the
scalar and pseudoscalar form factors, 
in Eqs. (71-73), (83-85) and Eqs. (95-97), (107-109) respectively,
can be compared, and we agree with \cite{Melnikov:1995yp, Hoang:1997ca}.

The scalar and pseudoscalar $C_F T_R$ coefficients,
respectively in Eqs. (74-76), (86-88) and Eqs. (98-100), (110-112),
are obtained by adding the contributions coming from the diagrams in 
Fig.~\ref{fig2} (f) and (g). Our results for the latter, with the
heavy quark loop insertion, agree with the ones in \cite{Hoang:1997ca},
if one takes into account that the results in \cite{Hoang:1997ca} are listed
for an on-shell renormalization of the coupling constant,
while we use $\overline{{\rm MS}}$-renormalization for the
coupling constant.

\section{Threshold Expansions}

In this section we provide the expansions of our results in the threshold 
limit $s \sim 4m^2$ ($y \rightarrow 1$ in the transformed variable). 
We define
\be
\beta = \sqrt{1-\frac{4m^2}{s}} 
\ee
as the small expansion parameter and give the results up to the zeroth order 
in $\beta$.

\subsection{One-loop Scalar Case}

\bea
%%%%%%%%%%%%%%%%%%%%%%%%%%%%%%%%%%%%%%%%%%%%%%%%%%%%%
\Re \, a_{-1} &=& 
   0
\ ; \\
&& \nn \\
%%%%%%%%%%%%%%%%%%%%%%%%%%%%%%%%%%%%%%%%%%%%%%%%%%%%%
\Re \, a_{0} &=&
-1 + \frac{3\,\zeta_2}{\beta}
\ ; \\
&& \nn \\
%%%%%%%%%%%%%%%%%%%%%%%%%%%%%%%%%%%%%%%%%%%%%%%%%%%%%
\Re \, a_{1} &=&
  2 - \frac{6\,\zeta_2\,\left( -1 + \ln(2) + 
        \ln (\beta) \right) }{\beta}
\ ; 
%%%%%%%%%%%%%%%%%%%%%%%%%%%%%%%%%%%%%%%%%%%%%%%%%%%%%
\eea

\bea
%%%%%%%%%%%%%%%%%%%%%%%%%%%%%%%%%%%%%%%%%%%%%%%%%%%%%
\Im \, a_{-1} &=& 
- \frac{1}{2\,\beta}
\ ; \\
&& \nn \\
%%%%%%%%%%%%%%%%%%%%%%%%%%%%%%%%%%%%%%%%%%%%%%%%%%%%%
\Im \, a_{0} &=&
  \frac{-1 + \ln(2) + \ln (\beta)}{\beta}
\ ; \\
&& \nn \\
%%%%%%%%%%%%%%%%%%%%%%%%%%%%%%%%%%%%%%%%%%%%%%%%%%%%%
\Im \, a_{1} &=&
  \frac{1}{\beta}
\Big[-2 + 2\,\ln(2) - {\ln^2(2)} + \zeta_2 + 
     2\,\ln (\beta) \nn \\ && - 2\,\ln(2)\,\ln (\beta) - 
     {\ln^2(\beta)} \Big]
\ ; 
%%%%%%%%%%%%%%%%%%%%%%%%%%%%%%%%%%%%%%%%%%%%%%%%%%%%%
\eea

\subsection{One-loop Pseudoscalar Case}

\bea
%%%%%%%%%%%%%%%%%%%%%%%%%%%%%%%%%%%%%%%%%%%%%%%%%%%%%
\Re \, \bar a_{-1} &=& \Re \, a_{-1}
\ ; \\
&& \nn \\
%%%%%%%%%%%%%%%%%%%%%%%%%%%%%%%%%%%%%%%%%%%%%%%%%%%%%
\Re \, \bar a_{0} &=&
-3 + \frac{3\,\zeta_2}{\beta}
\ ; \\
&& \nn \\
%%%%%%%%%%%%%%%%%%%%%%%%%%%%%%%%%%%%%%%%%%%%%%%%%%%%%
\Re \, \bar a_{1} &=&
 2 - \frac{6\,\left( \ln(2) + \ln (\beta) \right) \,\zeta_2}{\beta}
\ ; 
%%%%%%%%%%%%%%%%%%%%%%%%%%%%%%%%%%%%%%%%%%%%%%%%%%%%%
\eea

\bea
%%%%%%%%%%%%%%%%%%%%%%%%%%%%%%%%%%%%%%%%%%%%%%%%%%%%%
\Im \, \bar a_{-1} &=& \Im \, a_{-1} 
\ ; \\
&& \nn \\
%%%%%%%%%%%%%%%%%%%%%%%%%%%%%%%%%%%%%%%%%%%%%%%%%%%%%
\Im \, \bar a_{0} &=&
\frac{\ln(2) + \ln (\beta)}{\beta}
\ ; \\
&& \nn \\
%%%%%%%%%%%%%%%%%%%%%%%%%%%%%%%%%%%%%%%%%%%%%%%%%%%%%
\Im \, \bar a_{1} &=&
  \frac{-{\ln^2(2)} - 2\,\ln(2)\,\ln (\beta) - {\ln^2(\beta)} + \zeta_2}
   {\beta}
\ ; 
%%%%%%%%%%%%%%%%%%%%%%%%%%%%%%%%%%%%%%%%%%%%%%%%%%%%%
\eea

\noindent

\subsection{Two-loop Scalar Case}

\bea
%%%%%%%%%%%%%%%%%%%%%%%%%%%%%%%%%%%%%%%%%%%%%%%%%%%%%%%%%%%%%%%%%%%%%%%%%%%%%%%%%%%%%%%%
%%%% C_F^2
%%%%%%%%%%%%%%%%%%%%%%%%%%%%%%%%%%%%%%%%%%%%%%%%%%%%%%%%%%%%%%%%%%%%%%%%%%%%%%%%%%%%%%%%
\Re \, b_{-2} &=&  
- \frac{3\,\zeta_2}{4\,{\beta}^2}
- \frac{3\,\zeta_2}{2} 
\ ; \\
&& \nn \\
%%%%%%%%%%%%%%%%%%%%%%%%%%%%%%%%%%%%%%%%%%%%%%%%%%%%%%%%%%%%%%%%%%%%%%%%%%%%%%%%%%%%%%%%
\Re \, b_{-1} &=& 
  \frac{3\,\zeta_2\,\left( -1 + \ln(2) + 
        \ln (\beta) \right) }{{\beta}^2} + 
   \frac{3\,\zeta_2\,\left( 1 + 4\,\ln(2) + 
        4\,\ln (\beta) \right) }{2}
\ ; \\
&& \nn \\
%%%%%%%%%%%%%%%%%%%%%%%%%%%%%%%%%%%%%%%%%%%%%%%%%%%%%%%%%%%%%%%%%%%%%%%%%%%%%%%%%%%%%%%%
\Re \, b_{0} &=& 
+ 
   \frac{3\,\zeta_2}{2\,{\beta}^2} 
\Big[ -4 + 8\,\ln(2) - 
        4\,{\ln^2(2)} + 3\,\zeta_2 + 
        8\,\ln (\beta) \nn \\ && - 
        8\,\ln(2)\,\ln (\beta) - 
        4\,{\ln^2(\beta)} \Big]
\nn \\
&&
-  \frac{3\,\zeta_2}{\beta} 
\nn \\
&&
+ 
   \frac{1}{4}
\Big[
      5 + 98\,\zeta_2 - 40\,\ln(2)\,\zeta_2 - 
      48\,{\ln^2(2)}\,\zeta_2 + 36\,{\zeta_2}^2 \nn \\ && - 
      44\,\zeta_3 - 88\,\zeta_2\,\ln (\beta) - 
      96\,\ln(2)\,\zeta_2\,\ln (\beta) \nn \\ && - 
      48\,\zeta_2\,{\ln^2(\beta)} \Big]
\ ;
%%%%%%%%%%%%%%%%%%%%%%%%%%%%%%%%%%%%%%%%%%%%%%%%%%%%%%%%%%%%%%%%%%%%%%%%%%%%%%%%%%%%%%%%
\eea
\bea
%%%%%%%%%%%%%%%%%%%%%%%%%%%%%%%%%%%%%%%%%%%%%%%%%%%%%%%%%%%%%%%%%%%%%%%%%%%%%%%%%%%%%%%%
%%%% C_F*C_A
%%%%%%%%%%%%%%%%%%%%%%%%%%%%%%%%%%%%%%%%%%%%%%%%%%%%%%%%%%%%%%%%%%%%%%%%%%%%%%%%%%%%%%%%
\Re \, c_{-2} &=&  
   0
\ ; \\
&& \nn \\
%%%%%%%%%%%%%%%%%%%%%%%%%%%%%%%%%%%%%%%%%%%%%%%%%%%%%%%%%%%%%%%%%%%%%%%%%%%%%%%%%%%%%%%%
\Re \, c_{-1} &=& 
   0
\ ; \\
&& \nn \\
%%%%%%%%%%%%%%%%%%%%%%%%%%%%%%%%%%%%%%%%%%%%%%%%%%%%%%%%%%%%%%%%%%%%%%%%%%%%%%%%%%%%%%%%
\Re \, c_{0} &=& 
- 
   \frac{\zeta_2\,\left( -97 + 66\,\ln(2) + 
        66\,\ln (\beta) \right) }{6\,\beta}
\nn \\
&&
+ \frac{49}{36} - \frac{\left( -19 + 32\,\ln(2) \right) \,
      \zeta_2}{2} - 5\,\zeta_3 - 
   4\,\zeta_2\,\ln (\beta) 
\ ; 
%%%%%%%%%%%%%%%%%%%%%%%%%%%%%%%%%%%%%%%%%%%%%%%%%%%%%%%%%%%%%%%%%%%%%%%%%%%%%%%%%%%%%%%%
\eea
\bea
%%%%%%%%%%%%%%%%%%%%%%%%%%%%%%%%%%%%%%%%%%%%%%%%%%%%%%%%%%%%%%%%%%%%%%%%%%%%%%%%%%%%%%%%
%%%% C_F*T_R*N_f
%%%%%%%%%%%%%%%%%%%%%%%%%%%%%%%%%%%%%%%%%%%%%%%%%%%%%%%%%%%%%%%%%%%%%%%%%%%%%%%%%%%%%%%%
\Re \, d_{-2} &=&  
  0
\ ; \\
&& \nn \\
%%%%%%%%%%%%%%%%%%%%%%%%%%%%%%%%%%%%%%%%%%%%%%%%%%%%%%%%%%%%%%%%%%%%%%%%%%%%%%%%%%%%%%%%
\Re \, d_{-1} &=& 
  0
\ ; \\
&& \nn \\
%%%%%%%%%%%%%%%%%%%%%%%%%%%%%%%%%%%%%%%%%%%%%%%%%%%%%%%%%%%%%%%%%%%%%%%%%%%%%%%%%%%%%%%%
\Re \, d_{0} &=& 
  \frac{2\,\zeta_2\,\left( -11 + 6\,\ln(2) + 
        6\,\ln (\beta) \right) }{3\,\beta}
  - \frac{5}{9} 
\ ;
%%%%%%%%%%%%%%%%%%%%%%%%%%%%%%%%%%%%%%%%%%%%%%%%%%%%%%%%%%%%%%%%%%%%%%%%%%%%%%%%%%%%%%%%
\eea
\bea
%%%%%%%%%%%%%%%%%%%%%%%%%%%%%%%%%%%%%%%%%%%%%%%%%%%%%%%%%%%%%%%%%%%%%%%%%%%%%%%%%%%%%%%%
%%%% C_F*T_R
%%%%%%%%%%%%%%%%%%%%%%%%%%%%%%%%%%%%%%%%%%%%%%%%%%%%%%%%%%%%%%%%%%%%%%%%%%%%%%%%%%%%%%%%
\Re \, e_{-2} &=& 0 
\ ; \\
&& \nn \\
%%%%%%%%%%%%%%%%%%%%%%%%%%%%%%%%%%%%%%%%%%%%%%%%%%%%%%%%%%%%%%%%%%%%%%%%%%%%%%%%%%%%%%%%
\Re \, e_{-1} &=& 0
\ ; \\
&& \nn \\
%%%%%%%%%%%%%%%%%%%%%%%%%%%%%%%%%%%%%%%%%%%%%%%%%%%%%%%%%%%%%%%%%%%%%%%%%%%%%%%%%%%%%%%%
\Re \, e_{0} &=& 
  \frac{145}{9} - \frac{53\,\zeta_2}{3} + 
   4\,\ln(2)\,\left( -1 + 4\,\zeta_2 \right) 
\ ;
%%%%%%%%%%%%%%%%%%%%%%%%%%%%%%%%%%%%%%%%%%%%%%%%%%%%%%%%%%%%%%%%%%%%%%%%%%%%%%%%%%%%%%%%
\eea

\bea
%%%%%%%%%%%%%%%%%%%%%%%%%%%%%%%%%%%%%%%%%%%%%%%%%%%%%%%%%%%%%%%%%%%%%%%%%%%%%%%%%%%%%%%%
%%%% C_F^2
%%%%%%%%%%%%%%%%%%%%%%%%%%%%%%%%%%%%%%%%%%%%%%%%%%%%%%%%%%%%%%%%%%%%%%%%%%%%%%%%%%%%%%%%
\Im \, b_{-2} &=&  
    0
\ ; \\
&& \nn \\
%%%%%%%%%%%%%%%%%%%%%%%%%%%%%%%%%%%%%%%%%%%%%%%%%%%%%%%%%%%%%%%%%%%%%%%%%%%%%%%%%%%%%%%%
\Im \, b_{-1} &=& 
- \frac{3\,\zeta_2}{2\,{\beta}^2}
+ \frac{1}{2\,\beta} - 3\,\zeta_2 
\ ; \\
&& \nn \\
%%%%%%%%%%%%%%%%%%%%%%%%%%%%%%%%%%%%%%%%%%%%%%%%%%%%%%%%%%%%%%%%%%%%%%%%%%%%%%%%%%%%%%%%
\Im \, b_{0} &=& 
+ \frac{6\,\zeta_2\,\left( -1 + \ln(2) 
+ \ln (\beta) \right) }{{\beta}^2} 
- \frac{\ln(2) + \ln (\beta)}{\beta} 
\nn \\
&&
+ 
   \zeta_2\,\left( 11 + 12\,\ln(2) + 
      12\,\ln (\beta) \right) 
\ ;
%%%%%%%%%%%%%%%%%%%%%%%%%%%%%%%%%%%%%%%%%%%%%%%%%%%%%%%%%%%%%%%%%%%%%%%%%%%%%%%%%%%%%%%%
\eea
\bea
%%%%%%%%%%%%%%%%%%%%%%%%%%%%%%%%%%%%%%%%%%%%%%%%%%%%%%%%%%%%%%%%%%%%%%%%%%%%%%%%%%%%%%%%
%%%% C_F*C_A
%%%%%%%%%%%%%%%%%%%%%%%%%%%%%%%%%%%%%%%%%%%%%%%%%%%%%%%%%%%%%%%%%%%%%%%%%%%%%%%%%%%%%%%%
\Im \, c_{-2} &=&  
 \frac{11}{24\,\beta}
\ ; \\
&& \nn \\
%%%%%%%%%%%%%%%%%%%%%%%%%%%%%%%%%%%%%%%%%%%%%%%%%%%%%%%%%%%%%%%%%%%%%%%%%%%%%%%%%%%%%%%%
\Im \, c_{-1} &=& 
- \frac{31}{72\,\beta}
\ ; \\
&& \nn \\
%%%%%%%%%%%%%%%%%%%%%%%%%%%%%%%%%%%%%%%%%%%%%%%%%%%%%%%%%%%%%%%%%%%%%%%%%%%%%%%%%%%%%%%%
\Im \, c_{0} &=& 
+ \frac{1}{54\,\beta}
\Big[
      -239 + 291\,\ln(2) - 
      99\,{\ln^2(2)} + 291\,\ln (\beta) \nn \\ && - 
      198\,\ln(2)\,\ln (\beta) - 
      99\,{\ln^2(\beta)} \Big]
\nn \\
&&
+   2\,\zeta_2 
\ ; 
%%%%%%%%%%%%%%%%%%%%%%%%%%%%%%%%%%%%%%%%%%%%%%%%%%%%%%%%%%%%%%%%%%%%%%%%%%%%%%%%%%%%%%%%
\eea
\bea
%%%%%%%%%%%%%%%%%%%%%%%%%%%%%%%%%%%%%%%%%%%%%%%%%%%%%%%%%%%%%%%%%%%%%%%%%%%%%%%%%%%%%%%%
%%%% C_F*T_R*N_f
%%%%%%%%%%%%%%%%%%%%%%%%%%%%%%%%%%%%%%%%%%%%%%%%%%%%%%%%%%%%%%%%%%%%%%%%%%%%%%%%%%%%%%%%
\Im \, d_{-2} &=&  
- \frac{1}{6\,\beta}
\ ; \\
%%%%%%%%%%%%%%%%%%%%%%%%%%%%%%%%%%%%%%%%%%%%%%%%%%%%%%%%%%%%%%%%%%%%%%%%%%%%%%%%%%%%%%%%
\Im \, d_{-1} &=& 
\frac{5}{18\,\beta}
\ ; \\
%%%%%%%%%%%%%%%%%%%%%%%%%%%%%%%%%%%%%%%%%%%%%%%%%%%%%%%%%%%%%%%%%%%%%%%%%%%%%%%%%%%%%%%%
\Im \, d_{0} &=& 
  \frac{2}{27\,\beta}
\Big[ 31 - 33\,\ln(2) + 9\,{\ln^2(2)} - 
       33\,\ln (\beta) \nn \\ && + 
       18\,\ln(2)\,\ln (\beta) + 
       9\,{\ln^2(\beta)} \Big]
\ ;
%%%%%%%%%%%%%%%%%%%%%%%%%%%%%%%%%%%%%%%%%%%%%%%%%%%%%%%%%%%%%%%%%%%%%%%%%%%%%%%%%%%%%%%%
\eea
\bea
%%%%%%%%%%%%%%%%%%%%%%%%%%%%%%%%%%%%%%%%%%%%%%%%%%%%%%%%%%%%%%%%%%%%%%%%%%%%%%%%%%%%%%%%
%%%% C_F*T_R
%%%%%%%%%%%%%%%%%%%%%%%%%%%%%%%%%%%%%%%%%%%%%%%%%%%%%%%%%%%%%%%%%%%%%%%%%%%%%%%%%%%%%%%%
\Im \, e_{-2} &=& 0 
\ ; \\
&& \nn \\
%%%%%%%%%%%%%%%%%%%%%%%%%%%%%%%%%%%%%%%%%%%%%%%%%%%%%%%%%%%%%%%%%%%%%%%%%%%%%%%%%%%%%%%%
\Im \, e_{-1} &=& 0
\ ; \\
&& \nn \\
%%%%%%%%%%%%%%%%%%%%%%%%%%%%%%%%%%%%%%%%%%%%%%%%%%%%%%%%%%%%%%%%%%%%%%%%%%%%%%%%%%%%%%%%
\Im \, e_{0} &=& 2
\ . 
%%%%%%%%%%%%%%%%%%%%%%%%%%%%%%%%%%%%%%%%%%%%%%%%%%%%%%%%%%%%%%%%%%%%%%%%%%%%%%%%%%%%%%%%
\eea

\subsection{Two-loop Pseudoscalar Case}

\bea
%%%%%%%%%%%%%%%%%%%%%%%%%%%%%%%%%%%%%%%%%%%%%%%%%%%%%%%%%%%%%%%%%%%%%%%%%%%%%%%%%%%%%%%%
%%%% C_F^2
%%%%%%%%%%%%%%%%%%%%%%%%%%%%%%%%%%%%%%%%%%%%%%%%%%%%%%%%%%%%%%%%%%%%%%%%%%%%%%%%%%%%%%%%
\Re \, \bar b_{-2} &=&  \Re \, b_{-2}
\ ; \\
&& \nn \\
%%%%%%%%%%%%%%%%%%%%%%%%%%%%%%%%%%%%%%%%%%%%%%%%%%%%%%%%%%%%%%%%%%%%%%%%%%%%%%%%%%%%%%%%
\Re \, \bar b_{-1} &=& 
  \frac{3\,\left( \ln(2) + \ln (\beta) \right) \,\zeta_2}{{\beta}^2} + 
   \frac{3\,\left( 1 + 4\,\ln(2) + 4\,\ln (\beta) \right) \,\zeta_2}{2}
\ ; \\
&& \nn \\
%%%%%%%%%%%%%%%%%%%%%%%%%%%%%%%%%%%%%%%%%%%%%%%%%%%%%%%%%%%%%%%%%%%%%%%%%%%%%%%%%%%%%%%%
\Re \, \bar b_{0} &=& 
  \frac{-9\,\zeta_2}{\beta} + \frac{3\,\zeta_2\,
      \left( -4\,{\ln^2(2)} - 8\,\ln(2)\,\ln (\beta) - 
        4\,{\ln^2(\beta)} + 3\,\zeta_2 \right) }{2\,{\beta}^2} \nn \\ && + 
   \frac{1}{4}
    \Big[29 - 26\,\zeta_2 - 24\,\ln(2)\,\zeta_2 - 
      48\,{\ln^2(2)}\,\zeta_2 - 120\,\ln (\beta)\,\zeta_2 \nn \\ && - 
      96\,\ln(2)\,\ln (\beta)\,\zeta_2 - 48\,{\ln^2(\beta)}\,\zeta_2 + 
      36\,{\zeta_2}^2 - 72\,\zeta_3 \Big]
\ ;
%%%%%%%%%%%%%%%%%%%%%%%%%%%%%%%%%%%%%%%%%%%%%%%%%%%%%%%%%%%%%%%%%%%%%%%%%%%%%%%%%%%%%%%%
\eea
\bea
%%%%%%%%%%%%%%%%%%%%%%%%%%%%%%%%%%%%%%%%%%%%%%%%%%%%%%%%%%%%%%%%%%%%%%%%%%%%%%%%%%%%%%%%
%%%% C_F*C_A
%%%%%%%%%%%%%%%%%%%%%%%%%%%%%%%%%%%%%%%%%%%%%%%%%%%%%%%%%%%%%%%%%%%%%%%%%%%%%%%%%%%%%%%%
\Re \, \bar c_{-2} &=&  
   0
\ ; \\
&& \nn \\
%%%%%%%%%%%%%%%%%%%%%%%%%%%%%%%%%%%%%%%%%%%%%%%%%%%%%%%%%%%%%%%%%%%%%%%%%%%%%%%%%%%%%%%%
\Re \, \bar c_{-1} &=& 
   0
\ ; \\
&& \nn \\
%%%%%%%%%%%%%%%%%%%%%%%%%%%%%%%%%%%%%%%%%%%%%%%%%%%%%%%%%%%%%%%%%%%%%%%%%%%%%%%%%%%%%%%%
\Re \, \bar c_{0} &=& 
  - \frac{17}{12} - \frac{\left( -47 + 72\,\ln(2) \right) \,\zeta_2}{2} - 
   12\,\ln (\beta)\,\zeta_2 \nn \\ && - 
   \frac{\left( -31 + 66\,\ln(2) + 66\,\ln (\beta) \right) \,\zeta_2}
    {6\,\beta} - 12\,\zeta_3
\ ; 
%%%%%%%%%%%%%%%%%%%%%%%%%%%%%%%%%%%%%%%%%%%%%%%%%%%%%%%%%%%%%%%%%%%%%%%%%%%%%%%%%%%%%%%%
\eea
\bea
%%%%%%%%%%%%%%%%%%%%%%%%%%%%%%%%%%%%%%%%%%%%%%%%%%%%%%%%%%%%%%%%%%%%%%%%%%%%%%%%%%%%%%%%
%%%% C_F*T_R*N_f
%%%%%%%%%%%%%%%%%%%%%%%%%%%%%%%%%%%%%%%%%%%%%%%%%%%%%%%%%%%%%%%%%%%%%%%%%%%%%%%%%%%%%%%%
\Re \, \bar d_{-2} &=&  
  0
\ ; \\
&& \nn \\
%%%%%%%%%%%%%%%%%%%%%%%%%%%%%%%%%%%%%%%%%%%%%%%%%%%%%%%%%%%%%%%%%%%%%%%%%%%%%%%%%%%%%%%%
\Re \, \bar d_{-1} &=& 
  0
\ ; \\
&& \nn \\
%%%%%%%%%%%%%%%%%%%%%%%%%%%%%%%%%%%%%%%%%%%%%%%%%%%%%%%%%%%%%%%%%%%%%%%%%%%%%%%%%%%%%%%%
\Re \, \bar d_{0} &=& 
  \frac{1}{3} + \frac{2\,\left( -5 + 6\,\ln(2) + 6\,\ln (\beta) \right) \,
      \zeta_2}{3\,\beta}
\ ;
%%%%%%%%%%%%%%%%%%%%%%%%%%%%%%%%%%%%%%%%%%%%%%%%%%%%%%%%%%%%%%%%%%%%%%%%%%%%%%%%%%%%%%%%
\eea
\bea
%%%%%%%%%%%%%%%%%%%%%%%%%%%%%%%%%%%%%%%%%%%%%%%%%%%%%%%%%%%%%%%%%%%%%%%%%%%%%%%%%%%%%%%%
%%%% C_F*T_R
%%%%%%%%%%%%%%%%%%%%%%%%%%%%%%%%%%%%%%%%%%%%%%%%%%%%%%%%%%%%%%%%%%%%%%%%%%%%%%%%%%%%%%%%
\Re \, \bar e_{-2} &=& 0 
\ ; \\
&& \nn \\
%%%%%%%%%%%%%%%%%%%%%%%%%%%%%%%%%%%%%%%%%%%%%%%%%%%%%%%%%%%%%%%%%%%%%%%%%%%%%%%%%%%%%%%%
\Re \, \bar e_{-1} &=& 0
\ ; \\
&& \nn \\
%%%%%%%%%%%%%%%%%%%%%%%%%%%%%%%%%%%%%%%%%%%%%%%%%%%%%%%%%%%%%%%%%%%%%%%%%%%%%%%%%%%%%%%%
\Re \, \bar e_{0} &=& 
  \frac{43}{3} - 3\,\zeta_2 + 12\,\ln(2)\,\zeta_2 - \frac{21\,\zeta_3}{2}
\ ;
%%%%%%%%%%%%%%%%%%%%%%%%%%%%%%%%%%%%%%%%%%%%%%%%%%%%%%%%%%%%%%%%%%%%%%%%%%%%%%%%%%%%%%%%
\eea

\bea
%%%%%%%%%%%%%%%%%%%%%%%%%%%%%%%%%%%%%%%%%%%%%%%%%%%%%%%%%%%%%%%%%%%%%%%%%%%%%%%%%%%%%%%%
%%%% C_F^2
%%%%%%%%%%%%%%%%%%%%%%%%%%%%%%%%%%%%%%%%%%%%%%%%%%%%%%%%%%%%%%%%%%%%%%%%%%%%%%%%%%%%%%%%
\Im \, \bar b_{-2} &=&  
    0
\ ; \\
&& \nn \\
%%%%%%%%%%%%%%%%%%%%%%%%%%%%%%%%%%%%%%%%%%%%%%%%%%%%%%%%%%%%%%%%%%%%%%%%%%%%%%%%%%%%%%%%
\Im \, \bar b_{-1} &=& 
  \frac{3}{2\,\beta} - 3\,\zeta_2 - \frac{3\,\zeta_2}{2\,{\beta}^2}
\ ; \\
&& \nn \\
%%%%%%%%%%%%%%%%%%%%%%%%%%%%%%%%%%%%%%%%%%%%%%%%%%%%%%%%%%%%%%%%%%%%%%%%%%%%%%%%%%%%%%%%
\Im \, \bar b_{0} &=& 
  \frac{-1 - 3\,\ln(2) - 3\,\ln (\beta)}{\beta} + 
   \frac{6\,\left( \ln(2) + \ln (\beta) \right) \,\zeta_2}{{\beta}^2} \nn
   \\ && + 
   3\,\left( 5 + 4\,\ln(2) + 4\,\ln (\beta) \right) \,\zeta_2
\ ;
%%%%%%%%%%%%%%%%%%%%%%%%%%%%%%%%%%%%%%%%%%%%%%%%%%%%%%%%%%%%%%%%%%%%%%%%%%%%%%%%%%%%%%%%
\eea
\bea
%%%%%%%%%%%%%%%%%%%%%%%%%%%%%%%%%%%%%%%%%%%%%%%%%%%%%%%%%%%%%%%%%%%%%%%%%%%%%%%%%%%%%%%%
%%%% C_F*C_A
%%%%%%%%%%%%%%%%%%%%%%%%%%%%%%%%%%%%%%%%%%%%%%%%%%%%%%%%%%%%%%%%%%%%%%%%%%%%%%%%%%%%%%%%
\Im \, \bar c_{-2} &=&  \Im \, c_{-2} 
\ ; \\
&& \nn \\
%%%%%%%%%%%%%%%%%%%%%%%%%%%%%%%%%%%%%%%%%%%%%%%%%%%%%%%%%%%%%%%%%%%%%%%%%%%%%%%%%%%%%%%%
\Im \, \bar c_{-1} &=& \Im \, c_{-1} 
\ ; \\
&& \nn \\
%%%%%%%%%%%%%%%%%%%%%%%%%%%%%%%%%%%%%%%%%%%%%%%%%%%%%%%%%%%%%%%%%%%%%%%%%%%%%%%%%%%%%%%%
\Im \, \bar c_{0} &=& 
  \frac{1}{54\,\beta} 
   \Big[ -47 + 93\,\ln(2) - 99\,{\ln^2(2)} + 93\,\ln (\beta) \nn \\ && - 
         198\,\ln(2)\,\ln (\beta) - 99\,{\ln^2(\beta)} \Big]
   + 6\,\zeta_2
\ ; 
%%%%%%%%%%%%%%%%%%%%%%%%%%%%%%%%%%%%%%%%%%%%%%%%%%%%%%%%%%%%%%%%%%%%%%%%%%%%%%%%%%%%%%%%
\eea
\bea
%%%%%%%%%%%%%%%%%%%%%%%%%%%%%%%%%%%%%%%%%%%%%%%%%%%%%%%%%%%%%%%%%%%%%%%%%%%%%%%%%%%%%%%%
%%%% C_F*T_R*N_f
%%%%%%%%%%%%%%%%%%%%%%%%%%%%%%%%%%%%%%%%%%%%%%%%%%%%%%%%%%%%%%%%%%%%%%%%%%%%%%%%%%%%%%%%
\Im \, \bar d_{-2} &=&  \Im \, d_{-2} 
\ ; \\
&& \nn \\
%%%%%%%%%%%%%%%%%%%%%%%%%%%%%%%%%%%%%%%%%%%%%%%%%%%%%%%%%%%%%%%%%%%%%%%%%%%%%%%%%%%%%%%%
\Im \, \bar d_{-1} &=& \Im \, d_{-1} 
\ ; \\
&& \nn \\
%%%%%%%%%%%%%%%%%%%%%%%%%%%%%%%%%%%%%%%%%%%%%%%%%%%%%%%%%%%%%%%%%%%%%%%%%%%%%%%%%%%%%%%%
\Im \, \bar d_{0} &=& 
  \frac{2}{27\,\beta}
    \,\Big[ 7 - 15\,\ln(2) + 9\,{\ln^2(2)} - 15\,\ln (\beta) \nn \\ && + 
       18\,\ln(2)\,\ln (\beta) + 9\,{\ln^2(\beta)} \Big]
\ ;
%%%%%%%%%%%%%%%%%%%%%%%%%%%%%%%%%%%%%%%%%%%%%%%%%%%%%%%%%%%%%%%%%%%%%%%%%%%%%%%%%%%%%%%%
\eea
\bea
%%%%%%%%%%%%%%%%%%%%%%%%%%%%%%%%%%%%%%%%%%%%%%%%%%%%%%%%%%%%%%%%%%%%%%%%%%%%%%%%%%%%%%%%
%%%% C_F*T_R
%%%%%%%%%%%%%%%%%%%%%%%%%%%%%%%%%%%%%%%%%%%%%%%%%%%%%%%%%%%%%%%%%%%%%%%%%%%%%%%%%%%%%%%%
\Im \, \bar e_{-2} &=& 0 
\ ; \\
&& \nn \\
%%%%%%%%%%%%%%%%%%%%%%%%%%%%%%%%%%%%%%%%%%%%%%%%%%%%%%%%%%%%%%%%%%%%%%%%%%%%%%%%%%%%%%%%
\Im \, \bar e_{-1} &=& 0
\ ; \\
&& \nn \\
%%%%%%%%%%%%%%%%%%%%%%%%%%%%%%%%%%%%%%%%%%%%%%%%%%%%%%%%%%%%%%%%%%%%%%%%%%%%%%%%%%%%%%%%
\Im \, \bar e_{0} &=& 3\,\zeta_2
\ . 
%%%%%%%%%%%%%%%%%%%%%%%%%%%%%%%%%%%%%%%%%%%%%%%%%%%%%%%%%%%%%%%%%%%%%%%%%%%%%%%%%%%%%%%%
\eea

\section{Asymptotic Expansions}

In this section we provide the expansions of the form factors in the limit 
$s \gg m^2$ ($y \rightarrow 0$ in the transformed variable). We define 
$r=s/m^2$ and $L = \ln{(r)}$ and keep terms up to the second order in $1/r$.

\subsection{One-loop Scalar Case}

\bea
%%%%%%%%%%%%%%%%%%%%%%%%%%%%%%%%%%%%%%%%%%%%%%%%%%%%%
\Re \, a_{-1} &=& 
 \frac{-3 + 2\,L}{r^2} - \frac{2}{r} -1 + L 
\ ; \\
&& \nn \\
%%%%%%%%%%%%%%%%%%%%%%%%%%%%%%%%%%%%%%%%%%%%%%%%%%%%%
\Re \, a_{0} &=&
   \frac{-19 + 32\,L - 2\,L^2 + 16\,\zeta_2}{2\,r^2}
\nn \\ &&  +
  \frac{2\,\left( 1 + 4\,L \right) }{r} + \frac{-2 - L^2 + 8\,\zeta_2}{2} 
\ ; \\
&& \nn \\
%%%%%%%%%%%%%%%%%%%%%%%%%%%%%%%%%%%%%%%%%%%%%%%%%%%%%
\Re \, a_{1} &=&
\frac{3 + 456\,L - 96\,L^2 + 4\,L^3 + 768\,\zeta_2 - 
      96\,L\,\zeta_2 + 48\,\zeta_3}{12\,r^2} \nn \\ && 
- \frac{2\,\left( -3\,L + 2\,L^2 - 16\,\zeta_2 \right) }{r} 
-2 + L + \frac{L^3}{6} 
- 4\,L\,\zeta_2 \nn \\ && + 2\,\zeta_3 
\ ; 
%%%%%%%%%%%%%%%%%%%%%%%%%%%%%%%%%%%%%%%%%%%%%%%%%%%%%
\eea 

\bea
%%%%%%%%%%%%%%%%%%%%%%%%%%%%%%%%%%%%%%%%%%%%%%%%%%%%%
\Im \, a_{-1} &=& 
- \frac{2}{r^2} -1 
\ ; \\
&& \nn \\
%%%%%%%%%%%%%%%%%%%%%%%%%%%%%%%%%%%%%%%%%%%%%%%%%%%%%
\Im \, a_{0} &=&
\frac{2\,\left( -8 + L \right) }{r^2} - \frac{8}{r} + L 
\ ; \\
&& \nn \\
%%%%%%%%%%%%%%%%%%%%%%%%%%%%%%%%%%%%%%%%%%%%%%%%%%%%%
\Im \, a_{1} &=&
   \frac{-38 + 16\,L - L^2 + 4\,\zeta_2}{r^2}
\nn \\ && 
 + \frac{2\,\left( -3 + 4\,L \right) }{r} + \frac{-2 - L^2 + 4\,\zeta_2}{2} 
\ ; 
%%%%%%%%%%%%%%%%%%%%%%%%%%%%%%%%%%%%%%%%%%%%%%%%%%%%%
\eea

\subsection{One-loop Pseudoscalar Case}

\bea
%%%%%%%%%%%%%%%%%%%%%%%%%%%%%%%%%%%%%%%%%%%%%%%%%%%%%
\Re \, \bar a_{-1} &=& \Re \, a_{-1} 
\ ; \\
&& \nn \\
%%%%%%%%%%%%%%%%%%%%%%%%%%%%%%%%%%%%%%%%%%%%%%%%%%%%%
\Re \, \bar a_{0} &=&
  \frac{2\,\left( 1 + 2\,L \right) }{r} + \frac{-2 - L^2 + 8\,\zeta_2}{2} 
  \nn \\ && + 
   \frac{-3 + 16\,L - 2\,L^2 + 16\,\zeta_2}{2\,r^2}
\ ; \\
&& \nn \\
%%%%%%%%%%%%%%%%%%%%%%%%%%%%%%%%%%%%%%%%%%%%%%%%%%%%%
\Re \, \bar a_{1} &=&
  -2 + L + \frac{L^3}{6} 
  - \frac{2\,\left( L + L^2 - 8\,\zeta_2 \right) }{r} 
  - 4\,L\,\zeta_2 + 2\,\zeta_3 \nn \\ && + 
   \frac{99 + 72\,L - 48\,L^2 + 4\,L^3 + 384\,\zeta_2 - 96\,L\,\zeta_2 + 
      48\,\zeta_3}{12\,r^2}
\ ; 
%%%%%%%%%%%%%%%%%%%%%%%%%%%%%%%%%%%%%%%%%%%%%%%%%%%%%
\eea 

\bea
%%%%%%%%%%%%%%%%%%%%%%%%%%%%%%%%%%%%%%%%%%%%%%%%%%%%%
\Im \, \bar a_{-1} &=& \Im \, a_{-1} 
\ ; \\
&& \nn \\
%%%%%%%%%%%%%%%%%%%%%%%%%%%%%%%%%%%%%%%%%%%%%%%%%%%%%
\Im \, \bar a_{0} &=&
  0
\ ; \\
&& \nn \\
%%%%%%%%%%%%%%%%%%%%%%%%%%%%%%%%%%%%%%%%%%%%%%%%%%%%%
\Im \, \bar a_{1} &=&
  0 
\ ; 
%%%%%%%%%%%%%%%%%%%%%%%%%%%%%%%%%%%%%%%%%%%%%%%%%%%%%
\eea

\subsection{Two-loop Scalar Case}

\bea
%%%%%%%%%%%%%%%%%%%%%%%%%%%%%%%%%%%%%%%%%%%%%%%%%%%%%%%%%%%%%%%%%%%%%%%%%%%%%%%%%%%%%%%%
%%%% C_F^2
%%%%%%%%%%%%%%%%%%%%%%%%%%%%%%%%%%%%%%%%%%%%%%%%%%%%%%%%%%%%%%%%%%%%%%%%%%%%%%%%%%%%%%%%
\Re \, b_{-2} &=&  
   \frac{5 - 5\,L + 2\,L^2 - 12\,\zeta_2}{r^2} + 
  \frac{-2\,\left( -1 + L \right) }{r} \nn \\ && + 
   \frac{1 - 2\,L + L^2 - 6\,\zeta_2}{2}
\ ; \\
&& \nn \\
%%%%%%%%%%%%%%%%%%%%%%%%%%%%%%%%%%%%%%%%%%%%%%%%%%%%%%%%%%%%%%%%%%%%%%%%%%%%%%%%%%%%%%%%
\Re \, b_{-1} &=& 
   \frac{17 - 87\,L + 37\,L^2 - 4\,L^3 - 232\,\zeta_2 + 80\,L\,\zeta_2}
    {2\,r^2} \nn \\ && +
   \frac{-6\,L + 9\,L^2 - 56\,\zeta_2}{r} \nn \\ && + 
   \frac{2 - 2\,L + L^2 - L^3 - 8\,\zeta_2 + 20\,L\,\zeta_2}{2} 
\ ; \\
&& \nn \\
%%%%%%%%%%%%%%%%%%%%%%%%%%%%%%%%%%%%%%%%%%%%%%%%%%%%%%%%%%%%%%%%%%%%%%%%%%%%%%%%%%%%%%%%
\Re \, b_{0} &=& 
   \frac{1}{120\,r^2} 
\Big[
      20235 + 6240\,L + 10080\,L^2 - 1760\,L^3 + 150\,L^4 \nn \\ && - 
      58680\,\zeta_2 + 41760\,L\,\zeta_2 - 7200\,L^2\,\zeta_2 - 
      34560\,\ln(2)\,\zeta_2 \nn \\ && + 10224\,{\zeta_2}^2 + 10320\,\zeta_3 - 
      10560\,L\,\zeta_3 \Big]
\nn \\ && + 
   \frac{1}{15\,r} 
\Big[ -30 - 360\,L - 125\,L^3 + 480\,\zeta_2 + 2580\,L\,\zeta_2 - 
      30\,L^2\,\zeta_2 \nn \\ && - 1080\,\ln(2)\,\zeta_2 + 78\,{\zeta_2}^2 + 
      210\,\zeta_3 - 240\,L\,\zeta_3 \Big]
\nn \\ &&
 + \frac{29}{4} - \frac{L^3}{6} + \frac{7\,L^4}{24} - 8\,\zeta_2 + 
   \frac{68\,{\zeta_2}^2}{5} - 
   \frac{3\,L^2\,\left( -1 + 8\,\zeta_2 \right) }{2} \nn \\ && - 14\,\zeta_3 + 
   L\,\left( -3 + \zeta_2 + 8\,\zeta_3 \right)  
\ ;
%%%%%%%%%%%%%%%%%%%%%%%%%%%%%%%%%%%%%%%%%%%%%%%%%%%%%%%%%%%%%%%%%%%%%%%%%%%%%%%%%%%%%%%%
\eea
\bea
%%%%%%%%%%%%%%%%%%%%%%%%%%%%%%%%%%%%%%%%%%%%%%%%%%%%%%%%%%%%%%%%%%%%%%%%%%%%%%%%%%%%%%%%
%%%% C_F*C_A
%%%%%%%%%%%%%%%%%%%%%%%%%%%%%%%%%%%%%%%%%%%%%%%%%%%%%%%%%%%%%%%%%%%%%%%%%%%%%%%%%%%%%%%%
\Re \, c_{-2} &=&  
- \frac{11\,\left( -3 + 2\,L \right) }{12\,r^2} + \frac{11}{6\,r}
  \frac{-11\,\left( -1 + L \right) }{12} 
\ ; \\
&& \nn \\
%%%%%%%%%%%%%%%%%%%%%%%%%%%%%%%%%%%%%%%%%%%%%%%%%%%%%%%%%%%%%%%%%%%%%%%%%%%%%%%%%%%%%%%%
\Re \, c_{-1} &=& 
\frac{-165 + 188\,L - 36\,L^2 + 12\,L^3 + 234\,\zeta_2 - 
      216\,L\,\zeta_2 - 72\,\zeta_3}{36\,r^2} 
\nn \\ && + \frac{-67 + 18\,\zeta_2}{18\,r} 
\nn \\ && + \frac{-49 + 67\,L + 18\,\zeta_2 - 18\,L\,\zeta_2 - 18\,\zeta_3}{36}
\ ; \\
&& \nn \\
%%%%%%%%%%%%%%%%%%%%%%%%%%%%%%%%%%%%%%%%%%%%%%%%%%%%%%%%%%%%%%%%%%%%%%%%%%%%%%%%%%%%%%%%
\Re \, c_{0} &=& 
   \frac{1}{2160\,r^2} 
\Big[-58815 + 264770\,L - 20010\,L^2 + 2580\,L^3 - 270\,L^4 \nn \\ && - 
      23520\,\zeta_2 - 63360\,L\,\zeta_2 + 15120\,L^2\,\zeta_2 + 
      311040\,\ln(2)\,\zeta_2 \nn \\ && - 55296\,{\zeta_2}^2 + 155160\,\zeta_3 - 
      51840\,L\,\zeta_3 \Big]
\nn \\ &&
 + \frac{1}{54\,r} 
\Big[628 + 2274\,L - 261\,L^2 + 126\,\zeta_2 - 108\,L\,\zeta_2 \nn \\ && + 
      1944\,\ln(2)\,\zeta_2 + 1296\,\zeta_3\Big]
\nn \\ &&
 + \frac{1}{540}
\Big[-4345 + 1210\,L - 1005\,L^2 + 165\,L^3 + 8760\,\zeta_2 \nn \\ && - 
      1980\,L\,\zeta_2 + 270\,L^2\,\zeta_2 - 1701\,{\zeta_2}^2 + 
      4410\,\zeta_3 \nn \\ && - 3510\,L\,\zeta_3\Big]
\ ; 
%%%%%%%%%%%%%%%%%%%%%%%%%%%%%%%%%%%%%%%%%%%%%%%%%%%%%%%%%%%%%%%%%%%%%%%%%%%%%%%%%%%%%%%%
\eea
\bea
%%%%%%%%%%%%%%%%%%%%%%%%%%%%%%%%%%%%%%%%%%%%%%%%%%%%%%%%%%%%%%%%%%%%%%%%%%%%%%%%%%%%%%%%
%%%% C_F*T_R*N_f
%%%%%%%%%%%%%%%%%%%%%%%%%%%%%%%%%%%%%%%%%%%%%%%%%%%%%%%%%%%%%%%%%%%%%%%%%%%%%%%%%%%%%%%%
\Re \, d_{-2} &=&  
\frac{-3 + 2\,L}{3\,r^2} - \frac{2}{3\,r} + \frac{-1 + L}{3} 
\ ; \\
&& \nn \\
%%%%%%%%%%%%%%%%%%%%%%%%%%%%%%%%%%%%%%%%%%%%%%%%%%%%%%%%%%%%%%%%%%%%%%%%%%%%%%%%%%%%%%%%
\Re \, d_{-1} &=& 
- \frac{5\,\left( -3 + 2\,L \right) }{9\,r^2} + \frac{10}{9\,r}
+ \frac{-5\,\left( -1 + L \right) }{9} 
\ ; \\
&& \nn \\
%%%%%%%%%%%%%%%%%%%%%%%%%%%%%%%%%%%%%%%%%%%%%%%%%%%%%%%%%%%%%%%%%%%%%%%%%%%%%%%%%%%%%%%%
\Re \, d_{0} &=& 
 \frac{477 - 2224\,L + 348\,L^2 - 12\,L^3 - 2568\,\zeta_2 + 
      144\,L\,\zeta_2 - 144\,\zeta_3}{54\,r^2} 
\nn \\ && 
+ \frac{8\,\left( -5 - 39\,L + 9\,L^2 - 63\,\zeta_2 \right) }{27\,r} 
\nn \\ && 
+ \frac{49 - 28\,L + 15\,L^2 - 3\,L^3 - 84\,\zeta_2 + 36\,L\,\zeta_2 - 
      36\,\zeta_3}{27}
\ ;
%%%%%%%%%%%%%%%%%%%%%%%%%%%%%%%%%%%%%%%%%%%%%%%%%%%%%%%%%%%%%%%%%%%%%%%%%%%%%%%%%%%%%%%%
\eea
\bea
%%%%%%%%%%%%%%%%%%%%%%%%%%%%%%%%%%%%%%%%%%%%%%%%%%%%%%%%%%%%%%%%%%%%%%%%%%%%%%%%%%%%%%%%
%%%% C_F*T_R
%%%%%%%%%%%%%%%%%%%%%%%%%%%%%%%%%%%%%%%%%%%%%%%%%%%%%%%%%%%%%%%%%%%%%%%%%%%%%%%%%%%%%%%%
\Re \, e_{-2} &=& 0 
\ ; \\
&& \nn \\
%%%%%%%%%%%%%%%%%%%%%%%%%%%%%%%%%%%%%%%%%%%%%%%%%%%%%%%%%%%%%%%%%%%%%%%%%%%%%%%%%%%%%%%%
\Re \, e_{-1} &=& 0
\ ; \\
&& \nn \\
%%%%%%%%%%%%%%%%%%%%%%%%%%%%%%%%%%%%%%%%%%%%%%%%%%%%%%%%%%%%%%%%%%%%%%%%%%%%%%%%%%%%%%%%
\Re \, e_{0} &=& 
  \frac{1}{270\,r^2}
\Big[15360 + 9380\,L - 3930\,L^2 + 840\,L^3 - 45\,L^4 
\nn \\ && + 33480\,\zeta_2 
- 19800\,L\,\zeta_2 + 540\,L^2\,\zeta_2 + 3132\,{\zeta_2}^2 
\nn \\ && - 10800\,\zeta_3 + 6480\,L\,\zeta_3 \Big]
\nn \\ &&
+ \frac{1}{540\,r} 
\Big[4640 + 4560\,L - 720\,L^2 + 45\,L^4 + 2880\,\zeta_2 \nn \\ && - 
      540\,L^2\,\zeta_2 - 3132\,{\zeta_2}^2 - 6480\,L\,\zeta_3\Big]
\nn \\ &&
+ \frac{407 - 56\,L + 15\,L^2 - 3\,L^3 - 252\,\zeta_2 + 36\,L\,\zeta_2}
    {27} 
\ ;
%%%%%%%%%%%%%%%%%%%%%%%%%%%%%%%%%%%%%%%%%%%%%%%%%%%%%%%%%%%%%%%%%%%%%%%%%%%%%%%%%%%%%%%%
\eea

\bea
%%%%%%%%%%%%%%%%%%%%%%%%%%%%%%%%%%%%%%%%%%%%%%%%%%%%%%%%%%%%%%%%%%%%%%%%%%%%%%%%%%%%%%%%
%%%% C_F^2
%%%%%%%%%%%%%%%%%%%%%%%%%%%%%%%%%%%%%%%%%%%%%%%%%%%%%%%%%%%%%%%%%%%%%%%%%%%%%%%%%%%%%%%%
\Im \, b_{-2} &=&  
\frac{5 - 4\,L}{r^2} + \frac{2}{r} + 1 - L 
\ ; \\
&& \nn \\
%%%%%%%%%%%%%%%%%%%%%%%%%%%%%%%%%%%%%%%%%%%%%%%%%%%%%%%%%%%%%%%%%%%%%%%%%%%%%%%%%%%%%%%%
\Im \, b_{-1} &=& 
 \frac{87 - 74\,L + 12\,L^2 - 32\,\zeta_2}{2\,r^2} 
- \frac{6\,\left( -1 + 3\,L \right) }{r} 
\nn \\ && 
+ 1 - L + \frac{3\,L^2}{2} - 4\,\zeta_2
\ ; \\
&& \nn \\
%%%%%%%%%%%%%%%%%%%%%%%%%%%%%%%%%%%%%%%%%%%%%%%%%%%%%%%%%%%%%%%%%%%%%%%%%%%%%%%%%%%%%%%%
\Im \, b_{0} &=& 
\frac{-52 - 168\,L + 44\,L^2 - 5\,L^3 - 172\,\zeta_2 + 60\,L\,\zeta_2 
        + 88\,\zeta_3}{r^2}
\nn \\ && + \frac{24 + 25\,L^2 - 72\,\zeta_2 + 4\,L\,\zeta_2 + 16\,\zeta_3}{r} 
\nn \\ && + 3 + \frac{L^2}{2} - \frac{7\,L^3}{6} + \zeta_2 + 
   L\,\left( -3 + 10\,\zeta_2 \right)  - 8\,\zeta_3 
\ ;
%%%%%%%%%%%%%%%%%%%%%%%%%%%%%%%%%%%%%%%%%%%%%%%%%%%%%%%%%%%%%%%%%%%%%%%%%%%%%%%%%%%%%%%%
\eea
\bea
%%%%%%%%%%%%%%%%%%%%%%%%%%%%%%%%%%%%%%%%%%%%%%%%%%%%%%%%%%%%%%%%%%%%%%%%%%%%%%%%%%%%%%%%
%%%% C_F*C_A
%%%%%%%%%%%%%%%%%%%%%%%%%%%%%%%%%%%%%%%%%%%%%%%%%%%%%%%%%%%%%%%%%%%%%%%%%%%%%%%%%%%%%%%%
\Im \, c_{-2} &=&  
\frac{11}{6\,r^2} + \frac{11}{12} 
\ ; \\
&& \nn \\
%%%%%%%%%%%%%%%%%%%%%%%%%%%%%%%%%%%%%%%%%%%%%%%%%%%%%%%%%%%%%%%%%%%%%%%%%%%%%%%%%%%%%%%%
\Im \, c_{-1} &=& 
 \frac{-47 + 18\,L - 9\,L^2 + 18\,\zeta_2}{9\,r^2}
  - \frac{67}{36}  + \frac{\zeta_2}{2} 
\ ; \\
&& \nn \\
%%%%%%%%%%%%%%%%%%%%%%%%%%%%%%%%%%%%%%%%%%%%%%%%%%%%%%%%%%%%%%%%%%%%%%%%%%%%%%%%%%%%%%%%
\Im \, c_{0} &=& 
  \frac{1}{216\,r^2}
\Big[-26477 + 4002\,L - 774\,L^2 + 108\,L^3 + 3240\,\zeta_2 \nn \\ && - 
      1728\,L\,\zeta_2 + 5184\,\zeta_3 \Big]
\nn \\ &&
+ \frac{-379 + 87\,L + 18\,\zeta_2}{9\,r} \nn \\ && + 
   \frac{-242 + 402\,L - 99\,L^2 - 108\,L\,\zeta_2 + 702\,\zeta_3}{108} 
\ ; 
%%%%%%%%%%%%%%%%%%%%%%%%%%%%%%%%%%%%%%%%%%%%%%%%%%%%%%%%%%%%%%%%%%%%%%%%%%%%%%%%%%%%%%%%
\eea
\bea
%%%%%%%%%%%%%%%%%%%%%%%%%%%%%%%%%%%%%%%%%%%%%%%%%%%%%%%%%%%%%%%%%%%%%%%%%%%%%%%%%%%%%%%%
%%%% C_F*T_R*N_f
%%%%%%%%%%%%%%%%%%%%%%%%%%%%%%%%%%%%%%%%%%%%%%%%%%%%%%%%%%%%%%%%%%%%%%%%%%%%%%%%%%%%%%%%
\Im \, d_{-2} &=&  
- \frac{2}{3\,r^2} - \frac{1}{3} 
\ ; \\
&& \nn \\
%%%%%%%%%%%%%%%%%%%%%%%%%%%%%%%%%%%%%%%%%%%%%%%%%%%%%%%%%%%%%%%%%%%%%%%%%%%%%%%%%%%%%%%%
\Im \, d_{-1} &=& 
\frac{10}{9\,r^2} + \frac{5}{9} 
\ ; \\
&& \nn \\
%%%%%%%%%%%%%%%%%%%%%%%%%%%%%%%%%%%%%%%%%%%%%%%%%%%%%%%%%%%%%%%%%%%%%%%%%%%%%%%%%%%%%%%%
\Im \, d_{0} &=& 
+ \frac{2\,\left( 556 - 174\,L + 9\,L^2 \right) }{27\,r^2} 
- \frac{8\,\left( -13 + 6\,L \right) }{9\,r}
\nn \\ && +  \frac{28 - 30\,L + 9\,L^2}{27} 
\ ;
%%%%%%%%%%%%%%%%%%%%%%%%%%%%%%%%%%%%%%%%%%%%%%%%%%%%%%%%%%%%%%%%%%%%%%%%%%%%%%%%%%%%%%%%
\eea
\bea
%%%%%%%%%%%%%%%%%%%%%%%%%%%%%%%%%%%%%%%%%%%%%%%%%%%%%%%%%%%%%%%%%%%%%%%%%%%%%%%%%%%%%%%%
%%%% C_F*T_R
%%%%%%%%%%%%%%%%%%%%%%%%%%%%%%%%%%%%%%%%%%%%%%%%%%%%%%%%%%%%%%%%%%%%%%%%%%%%%%%%%%%%%%%%
\Im \, e_{-2} &=& 0 
\ ; \\
&& \nn \\
%%%%%%%%%%%%%%%%%%%%%%%%%%%%%%%%%%%%%%%%%%%%%%%%%%%%%%%%%%%%%%%%%%%%%%%%%%%%%%%%%%%%%%%%
\Im \, e_{-1} &=& 0
\ ; \\
&& \nn \\
%%%%%%%%%%%%%%%%%%%%%%%%%%%%%%%%%%%%%%%%%%%%%%%%%%%%%%%%%%%%%%%%%%%%%%%%%%%%%%%%%%%%%%%%
\Im \, e_{0} &=& 
 \frac{2\,\left( -469 + 393\,L - 126\,L^2 + 9\,L^3 + 486\,\zeta_2 + 
        54\,L\,\zeta_2 - 324\,\zeta_3 \right) }{27\,r^2} 
\nn \\ && + 
   \frac{-76 + 24\,L - 3\,L^3 - 18\,L\,\zeta_2 + 108\,\zeta_3}{9\,r}
\nn \\ && 
+ \frac{56 - 30\,L + 9\,L^2}{27} 
\ . 
%%%%%%%%%%%%%%%%%%%%%%%%%%%%%%%%%%%%%%%%%%%%%%%%%%%%%%%%%%%%%%%%%%%%%%%%%%%%%%%%%%%%%%%%
\eea

\subsection{Two-loop Pseudoscalar Case}

\bea
%%%%%%%%%%%%%%%%%%%%%%%%%%%%%%%%%%%%%%%%%%%%%%%%%%%%%%%%%%%%%%%%%%%%%%%%%%%%%%%%%%%%%%%%
%%%% C_F^2
%%%%%%%%%%%%%%%%%%%%%%%%%%%%%%%%%%%%%%%%%%%%%%%%%%%%%%%%%%%%%%%%%%%%%%%%%%%%%%%%%%%%%%%%
\Re \, \bar b_{-2} &=&  
  \frac{-2\,\left( -1 + L \right) }{r} + 
   \frac{5 - 5\,L + 2\,L^2 - 12\,\zeta_2}{r^2} + 
   \frac{1 - 2\,L + L^2 - 6\,\zeta_2}{2}
\ ; \\
&& \nn \\
%%%%%%%%%%%%%%%%%%%%%%%%%%%%%%%%%%%%%%%%%%%%%%%%%%%%%%%%%%%%%%%%%%%%%%%%%%%%%%%%%%%%%%%%
\Re \, \bar b_{-1} &=& 
  \frac{-2\,L + 5\,L^2 - 32\,\zeta_2}{r} \nn \\ && + 
   \frac{1 - 39\,L + 21\,L^2 - 4\,L^3 - 136\,\zeta_2 + 80\,L\,\zeta_2}
    {2\,r^2}
   \nn \\ && + 
   \frac{2 - 2\,L + L^2 - L^3 - 8\,\zeta_2 + 20\,L\,\zeta_2}{2} 
\ ; \\
&& \nn \\
%%%%%%%%%%%%%%%%%%%%%%%%%%%%%%%%%%%%%%%%%%%%%%%%%%%%%%%%%%%%%%%%%%%%%%%%%%%%%%%%%%%%%%%%
\Re \, \bar b_{0} &=& 
  \frac{29}{4} - \frac{L^3}{6} + \frac{7\,L^4}{24} - 8\,\zeta_2 + 
   \frac{68\,{\zeta_2}^2}{5} - 
   \frac{3\,L^2\,\left( -1 + 8\,\zeta_2 \right) }{2} \nn \\ && - 14\,\zeta_3 + 
   L\,\left( -3 + \zeta_2 + 8\,\zeta_3 \right)  \nn \\ && + 
   \frac{1}{15\,r} 
   \Big[-30 - 180\,L - 30\,L^2 - 65\,L^3 + 240\,\zeta_2 + 
      1320\,L\,\zeta_2 \nn \\ && - 30\,L^2\,\zeta_2 - 360\,\ln(2)\,\zeta_2 + 
      78\,{\zeta_2}^2 - 90\,\zeta_3 - 240\,L\,\zeta_3 \Big]
   \nn \\ && + 
   \frac{1}{24\,r^2}
   \Big[1935 + 480\,L + 96\,L^2 - 224\,L^3 + 38\,L^4 - 24\,\zeta_2 \nn \\ && + 
      4512\,L\,\zeta_2 - 1536\,L^2\,\zeta_2 - 2304\,\ln(2)\,\zeta_2 + 
      1872\,{\zeta_2}^2 \nn \\ && - 624\,\zeta_3 - 192\,L\,\zeta_3 \Big]
\ ;
%%%%%%%%%%%%%%%%%%%%%%%%%%%%%%%%%%%%%%%%%%%%%%%%%%%%%%%%%%%%%%%%%%%%%%%%%%%%%%%%%%%%%%%%
\eea
\bea
%%%%%%%%%%%%%%%%%%%%%%%%%%%%%%%%%%%%%%%%%%%%%%%%%%%%%%%%%%%%%%%%%%%%%%%%%%%%%%%%%%%%%%%%
%%%% C_F*C_A
%%%%%%%%%%%%%%%%%%%%%%%%%%%%%%%%%%%%%%%%%%%%%%%%%%%%%%%%%%%%%%%%%%%%%%%%%%%%%%%%%%%%%%%%
\Re \,\bar c_{-2} &=&  
  \frac{-11\,\left( -1 + L \right) }{12} - 
   \frac{11\,\left( -3 + 2\,L \right) }{12\,r^2} + \frac{11}{6\,r}
\ ; \\
&& \nn \\
%%%%%%%%%%%%%%%%%%%%%%%%%%%%%%%%%%%%%%%%%%%%%%%%%%%%%%%%%%%%%%%%%%%%%%%%%%%%%%%%%%%%%%%%
\Re \, \bar c_{-1} &=& 
  \frac{-67 + 18\,\zeta_2}{18\,r} \nn \\ && + 
   \frac{-165 + 188\,L - 36\,L^2 + 12\,L^3 + 234\,\zeta_2 - 
      216\,L\,\zeta_2 - 72\,\zeta_3}{36\,r^2} \nn \\ &&  + 
   \frac{-49 + 67\,L + 18\,\zeta_2 - 18\,L\,\zeta_2 - 18\,\zeta_3}{36}
\ ; \\
&& \nn \\
%%%%%%%%%%%%%%%%%%%%%%%%%%%%%%%%%%%%%%%%%%%%%%%%%%%%%%%%%%%%%%%%%%%%%%%%%%%%%%%%%%%%%%%%
\Re \, \bar c_{0} &=& 
  \frac{628 + 1182\,L - 171\,L^2 + 594\,\zeta_2 - 108\,L\,\zeta_2 + 
      648\,\ln(2)\,\zeta_2 + 756\,\zeta_3}{54\,r} \nn \\ && + 
   \frac{1}{2160\,r^2} 
    \Big[-29055 + 114050\,L - 17130\,L^2 + 5460\,L^3 - 630\,L^4 \nn \\ && + 
      48480\,\zeta_2 - 97920\,L\,\zeta_2 + 32400\,L^2\,\zeta_2 + 
      103680\,\ln(2)\,\zeta_2 \nn \\ && - 81216\,{\zeta_2}^2 + 86040\,\zeta_3 - 
      34560\,L\,\zeta_3 \Big]
   \nn \\ && + 
   \frac{1}{540}
   \Big[-4345 + 1210\,L - 1005\,L^2 + 165\,L^3 + 8760\,\zeta_2 - 
      1980\,L\,\zeta_2 \nn \\ && + 270\,L^2\,\zeta_2 - 1701\,{\zeta_2}^2 + 
      4410\,\zeta_3 - 3510\,L\,\zeta_3 \Big]
\ ; 
%%%%%%%%%%%%%%%%%%%%%%%%%%%%%%%%%%%%%%%%%%%%%%%%%%%%%%%%%%%%%%%%%%%%%%%%%%%%%%%%%%%%%%%%
\eea
\bea
%%%%%%%%%%%%%%%%%%%%%%%%%%%%%%%%%%%%%%%%%%%%%%%%%%%%%%%%%%%%%%%%%%%%%%%%%%%%%%%%%%%%%%%%
%%%% C_F*T_R*N_f
%%%%%%%%%%%%%%%%%%%%%%%%%%%%%%%%%%%%%%%%%%%%%%%%%%%%%%%%%%%%%%%%%%%%%%%%%%%%%%%%%%%%%%%%
\Re \, \bar d_{-2} &=&  
\frac{-1 + L}{3} + \frac{-3 + 2\,L}{3\,r^2} - \frac{2}{3\,r}
\ ; \\
&& \nn \\
%%%%%%%%%%%%%%%%%%%%%%%%%%%%%%%%%%%%%%%%%%%%%%%%%%%%%%%%%%%%%%%%%%%%%%%%%%%%%%%%%%%%%%%%
\Re \, \bar d_{-1} &=& 
  \frac{-5\,\left( -1 + L \right) }{9} - 
   \frac{5\,\left( -3 + 2\,L \right) }{9\,r^2} + \frac{10}{9\,r}
\ ; \\
&& \nn \\
%%%%%%%%%%%%%%%%%%%%%%%%%%%%%%%%%%%%%%%%%%%%%%%%%%%%%%%%%%%%%%%%%%%%%%%%%%%%%%%%%%%%%%%%
\Re \, \bar d_{0} &=& 
  \frac{4\,\left( -10 - 30\,L + 9\,L^2 - 54\,\zeta_2 \right) }{27\,r} \nn \\
   && + 
   \frac{-3 - 880\,L + 204\,L^2 - 12\,L^3 - 1416\,\zeta_2 + 
      144\,L\,\zeta_2 - 144\,\zeta_3}{54\,r^2} \nn \\ && + 
   \frac{49 - 28\,L + 15\,L^2 - 3\,L^3 - 84\,\zeta_2 + 36\,L\,\zeta_2 - 
      36\,\zeta_3}{27}
\ ;
%%%%%%%%%%%%%%%%%%%%%%%%%%%%%%%%%%%%%%%%%%%%%%%%%%%%%%%%%%%%%%%%%%%%%%%%%%%%%%%%%%%%%%%%
\eea
\bea
%%%%%%%%%%%%%%%%%%%%%%%%%%%%%%%%%%%%%%%%%%%%%%%%%%%%%%%%%%%%%%%%%%%%%%%%%%%%%%%%%%%%%%%%
%%%% C_F*T_R
%%%%%%%%%%%%%%%%%%%%%%%%%%%%%%%%%%%%%%%%%%%%%%%%%%%%%%%%%%%%%%%%%%%%%%%%%%%%%%%%%%%%%%%%
\Re \, \bar e_{-2} &=& 0 
\ ; \\
&& \nn \\
%%%%%%%%%%%%%%%%%%%%%%%%%%%%%%%%%%%%%%%%%%%%%%%%%%%%%%%%%%%%%%%%%%%%%%%%%%%%%%%%%%%%%%%%
\Re \, \bar e_{-1} &=& 0
\ ; \\
&& \nn \\
%%%%%%%%%%%%%%%%%%%%%%%%%%%%%%%%%%%%%%%%%%%%%%%%%%%%%%%%%%%%%%%%%%%%%%%%%%%%%%%%%%%%%%%%
\Re \, \bar e_{0} &=& 
  \frac{407 - 56\,L + 15\,L^2 - 3\,L^3 - 252\,\zeta_2 + 36\,L\,\zeta_2}
    {27} 
    \nn \\ && 
    + \frac{1}{540\,r} 
      \Big[1760 - 240\,L - 360\,L^2 + 45\,L^4 + 720\,\zeta_2 - 
      540\,L^2\,\zeta_2 \nn \\ && - 3132\,{\zeta_2}^2 - 6480\,L\,\zeta_3\Big]
    \nn \\ && 
    + \frac{1}{270\,r^2}
        \Big[7920 + 5300\,L + 750\,L^2 - 1320\,L^3 + 45\,L^4 - 3240\,\zeta_2
        \nn \\ && + 
        14760\,L\,\zeta_2 - 540\,L^2\,\zeta_2 - 3132\,{\zeta_2}^2 + 
        6480\,\zeta_3 - 6480\,L\,\zeta_3\Big]
\ ;
%%%%%%%%%%%%%%%%%%%%%%%%%%%%%%%%%%%%%%%%%%%%%%%%%%%%%%%%%%%%%%%%%%%%%%%%%%%%%%%%%%%%%%%%
\eea

\bea
%%%%%%%%%%%%%%%%%%%%%%%%%%%%%%%%%%%%%%%%%%%%%%%%%%%%%%%%%%%%%%%%%%%%%%%%%%%%%%%%%%%%%%%%
%%%% C_F^2
%%%%%%%%%%%%%%%%%%%%%%%%%%%%%%%%%%%%%%%%%%%%%%%%%%%%%%%%%%%%%%%%%%%%%%%%%%%%%%%%%%%%%%%%
\Im \, \bar b_{-2} &=&  
1 - L + \frac{5 - 4\,L}{r^2} + \frac{2}{r}
\ ; \\
&& \nn \\
%%%%%%%%%%%%%%%%%%%%%%%%%%%%%%%%%%%%%%%%%%%%%%%%%%%%%%%%%%%%%%%%%%%%%%%%%%%%%%%%%%%%%%%%
\Im \, \bar b_{-1} &=& 
  1 - L + \frac{3\,L^2}{2} - \frac{2\,\left( -1 + 5\,L \right) }{r} + 
   \frac{39 - 42\,L + 12\,L^2 - 32\,\zeta_2}{2\,r^2} - 4\,\zeta_2
\ ; \\
&& \nn \\
%%%%%%%%%%%%%%%%%%%%%%%%%%%%%%%%%%%%%%%%%%%%%%%%%%%%%%%%%%%%%%%%%%%%%%%%%%%%%%%%%%%%%%%%
\Im \, \bar b_{0} &=& 
  3 + \frac{L^2}{2} - \frac{7\,L^3}{6} + \zeta_2 + 
   L\,\left( -3 + 10\,\zeta_2 \right)  - 8\,\zeta_3 \nn \\ && + 
   \frac{12 + 4\,L + 13\,L^2 - 36\,\zeta_2 + 4\,L\,\zeta_2 + 16\,\zeta_3}
    {r} \nn \\ && + \frac{-60 - 24\,L + 84\,L^2 - 19\,L^3 - 228\,\zeta_2 + 
      156\,L\,\zeta_2 + 24\,\zeta_3}{3\,r^2}
\ ;
%%%%%%%%%%%%%%%%%%%%%%%%%%%%%%%%%%%%%%%%%%%%%%%%%%%%%%%%%%%%%%%%%%%%%%%%%%%%%%%%%%%%%%%%
\eea
\bea
%%%%%%%%%%%%%%%%%%%%%%%%%%%%%%%%%%%%%%%%%%%%%%%%%%%%%%%%%%%%%%%%%%%%%%%%%%%%%%%%%%%%%%%%
%%%% C_F*C_A
%%%%%%%%%%%%%%%%%%%%%%%%%%%%%%%%%%%%%%%%%%%%%%%%%%%%%%%%%%%%%%%%%%%%%%%%%%%%%%%%%%%%%%%%
\Im \, \bar c_{-2} &=&  
\frac{11}{12} + \frac{11}{6\,r^2}
\ ; \\
&& \nn \\
%%%%%%%%%%%%%%%%%%%%%%%%%%%%%%%%%%%%%%%%%%%%%%%%%%%%%%%%%%%%%%%%%%%%%%%%%%%%%%%%%%%%%%%%
\Im \, \bar c_{-1} &=& 
  - \frac{67}{36}  + \frac{\zeta_2}{2} + 
   \frac{-47 + 18\,L - 9\,L^2 + 18\,\zeta_2}{9\,r^2}
\ ; \\
&& \nn \\
%%%%%%%%%%%%%%%%%%%%%%%%%%%%%%%%%%%%%%%%%%%%%%%%%%%%%%%%%%%%%%%%%%%%%%%%%%%%%%%%%%%%%%%%
\Im \, \bar c_{0} &=& 
  \frac{-197 + 57\,L + 18\,\zeta_2}{9\,r} \nn \\ && + 
   \frac{-11405 + 3426\,L - 1638\,L^2 + 252\,L^3 + 3240\,\zeta_2 - 
      3456\,L\,\zeta_2 + 3456\,\zeta_3}{216\,r^2} \nn \\ && + 
   \frac{-242 + 402\,L - 99\,L^2 - 108\,L\,\zeta_2 + 702\,\zeta_3}{108} 
\ ; 
%%%%%%%%%%%%%%%%%%%%%%%%%%%%%%%%%%%%%%%%%%%%%%%%%%%%%%%%%%%%%%%%%%%%%%%%%%%%%%%%%%%%%%%%
\eea
\bea
%%%%%%%%%%%%%%%%%%%%%%%%%%%%%%%%%%%%%%%%%%%%%%%%%%%%%%%%%%%%%%%%%%%%%%%%%%%%%%%%%%%%%%%%
%%%% C_F*T_R*N_f
%%%%%%%%%%%%%%%%%%%%%%%%%%%%%%%%%%%%%%%%%%%%%%%%%%%%%%%%%%%%%%%%%%%%%%%%%%%%%%%%%%%%%%%%
\Im \, \bar d_{-2} &=&  
- \frac{1}{3}  - \frac{2}{3\,r^2}
\ ; \\
&& \nn \\
%%%%%%%%%%%%%%%%%%%%%%%%%%%%%%%%%%%%%%%%%%%%%%%%%%%%%%%%%%%%%%%%%%%%%%%%%%%%%%%%%%%%%%%%
\Im \, \bar d_{-1} &=& 
\frac{5}{9} + \frac{10}{9\,r^2}
\ ; \\
&& \nn \\
%%%%%%%%%%%%%%%%%%%%%%%%%%%%%%%%%%%%%%%%%%%%%%%%%%%%%%%%%%%%%%%%%%%%%%%%%%%%%%%%%%%%%%%%
\Im \, \bar d_{0} &=& 
  \frac{28 - 30\,L + 9\,L^2}{27} + 
   \frac{2\,\left( 220 - 102\,L + 9\,L^2 \right) }{27\,r^2} - 
   \frac{8\,\left( -5 + 3\,L \right) }{9\,r}
\ ;
%%%%%%%%%%%%%%%%%%%%%%%%%%%%%%%%%%%%%%%%%%%%%%%%%%%%%%%%%%%%%%%%%%%%%%%%%%%%%%%%%%%%%%%%
\eea
\bea
%%%%%%%%%%%%%%%%%%%%%%%%%%%%%%%%%%%%%%%%%%%%%%%%%%%%%%%%%%%%%%%%%%%%%%%%%%%%%%%%%%%%%%%%
%%%% C_F*T_R
%%%%%%%%%%%%%%%%%%%%%%%%%%%%%%%%%%%%%%%%%%%%%%%%%%%%%%%%%%%%%%%%%%%%%%%%%%%%%%%%%%%%%%%%
\Im \, \bar e_{-2} &=& 0 
\ ; \\
&& \nn \\
%%%%%%%%%%%%%%%%%%%%%%%%%%%%%%%%%%%%%%%%%%%%%%%%%%%%%%%%%%%%%%%%%%%%%%%%%%%%%%%%%%%%%%%%
\Im \, \bar e_{-1} &=& 0
\ ; \\
&& \nn \\
%%%%%%%%%%%%%%%%%%%%%%%%%%%%%%%%%%%%%%%%%%%%%%%%%%%%%%%%%%%%%%%%%%%%%%%%%%%%%%%%%%%%%%%%
\Im \, \bar e_{0} &=& 
  \frac{56 - 30\,L + 9\,L^2}{27} 
  + \frac{4 + 12\,L - 3\,L^3 - 18\,L\,\zeta_2 + 108\,\zeta_3}{9\,r}
  \nn \\ &&  
  - \frac{2\,\left( 265 + 75\,L - 198\,L^2 + 9\,L^3 - 54\,\zeta_2 + 
        54\,L\,\zeta_2 - 324\,\zeta_3 \right) }{27\,r^2} 
\ . 
%%%%%%%%%%%%%%%%%%%%%%%%%%%%%%%%%%%%%%%%%%%%%%%%%%%%%%%%%%%%%%%%%%%%%%%%%%%%%%%%%%%%%%%%
\eea

%All the results of this Section can be obtained in an electronic form 
%by downloading the source of this manuscript from http://www.arxiv.org.

\section{Summary and Outlook}

If Higgs boson will be discovered, then the next important task will be
the experimental determination of its couplings to the other
known fundamental particles, in particular to heavy quarks $Q=c,b,t$. 
Therefore precise
predictions of Higgs boson decays into heavy quark-antiquark pairs are 
mandatory,
not only for the partial decay rates, but also for differential 
distributions.
\par
To put forward this aim we have computed the two-loop QCD corrections to
the decay amplitude of a neutral Higgs boson, with scalar and
pseudoscalar couplings  to fermions, into a heavy quark-antiquark pair, 
$h \to Q {\bar Q}$.
This amplitude is determined by two form factors, which we calculated  
with the
algebraic and analytic techniques employed in
\cite{RoPieRem1,RoPieRem2,RoPieRem3,Bernreuther:2004ih,Bernreuther:2005rw},
for arbitrary quark
mass $m$ and squared four-momentum $q^2$ of the spin-zero particle.
These scalar and pseudoscalar vertex functions have also
a number of other applications, including  resonant Higgs boson
production in high energetic muon collisions, $\mu^+ \mu^- \to Q {\bar 
Q} X$.
Applications to hadron physics include matrix elements  of
scalar and pseudoscalar quark currents, serving as interpolating fields for
heavy mesons,  between   heavy (anti)quark states.
Moreover, from the results presented here one obtains as a special case 
also the second order
QED corrections for the amplitudes $h \to f {\bar f}$, $f$ being a quark or 
a lepton.
\par
In addition, we  expanded the scalar and pseudoscalar form factors
near the $Q \bar Q$ threshold and in the regime $m^2/q^2\ll 1$. The 
former expansions may be
applied, for instance, to the decay of heavy nonstandard Higgs bosons into
$t\bar t$, the latter ones apply  to, e.g., $h \to b{\bar b}, c{\bar c}$.
\par
The renormalized form factors were expressed in terms of the
$\overline{\rm MS}$ coupling $\alpha_S$ for $(N_f+1)$ quark flavors and
the on-shell mass $m$ of the heavy  quark. They may be expressed
in straightforward fashion as functions of the running  
$\overline{\rm MS}$ mass
${\bar m}(\mu)$, too. It is expected that the use of
 this mass parameter  leads to a 
faster decrease of the coefficients of the perturbation series
\cite{Braaten:1980yq}.
However, this is no general rule (see, e.g., \cite{Harlander:1997xa}); 
the size of
the coefficients of the series when using $m$ or ${\bar m}$ must be 
studied for
each observable separately.
\par
The renormalized form factors contain soft and collinear divergences, which
in our IR-regularization scheme appear as poles in $\epsilon$.
These divergences have to be canceled against
the divergences arising from the real radiation of gluons and massless
quark-antiquark pairs, which in this paper were not
taken into account. The IR singularity structure
of QCD amplitudes involving massive quarks is so far not understood 
beyond the 1-loop level.
The presently known two-loop heavy-quark vertex functions
may provide insight into the generic singularity structure of the 
two-loop QCD amplitudes with massive quarks. As to the two-loop
scalar and pseudoscalar form factors given above we notice that the
infrared-divergent contributions proportional to the colour factor $C_F^2$
show Yenni-Frautschi-Suura exponentiation \cite{yfs}, which is 
characteristic
of Abelian gauge  theories. This is completely analogous to the case
of the two-loop QCD vector and axial vector vertex functions 
\cite{Bernreuther:2004ih}.
However the IR-divergent terms proportional to the other colour factors
cannot be explained by naive Abelian exponentiation. It remains an open 
problem whether an infrared factorization formula (as derived for massless 
QCD in~\cite{Catani:1998bh,Sterman:2002qn})
exists for the  generation of these
singularities beyond the one-loop level.

\par
As emphasized, the results of this paper are one building block towards
a differential  description of Higgs boson decays, $h \to Q {\bar Q}X$,
to second order QCD. 
% We plan to provide also the other pieces in the near future.

\section*{Acknowledgment}

This work was partially supported 
by Deutsche Forschungsgemeinschaft (DFG), 
SFB/TR9, by DFG-Graduiertenkolleg RWTH Aachen, by the Swiss National Science
Foundation  
(SNF) under contract 200021-101874, by the National 
Science Foundation under Grant No.\ PHY99-07949, and by the USA DoE under 
the grant DE-FG03-91ER40662, Task J.

%%%%%%%%%%%%%%%%%%%%%%%%%%%%%%%%%%%%%%%%%%%%%%%%%%%%%%%%%%%%%%%%%%%%%

\end{fmffile}

\end{document}